\documentclass[prb,twocolumn,preprintnumbers,amsmath,amssymb,superscriptaddress,showpacs]{revtex4}
\usepackage{graphicx}
\usepackage{latexsym}
\usepackage{amsmath}
\usepackage{graphics}
\usepackage{amssymb}
\usepackage{layout}
\usepackage{verbatim}
\usepackage{amsfonts,epsfig}
\usepackage{natbib}

\newcommand{\ket}[1]{\left|#1\right>}
\newcommand{\bra}[1]{\left< #1 \right|}
\newcommand{\beq}{\begin{equation}}
\newcommand{\eeq}{\end{equation}}
\newcommand{\bea}{\begin{eqnarray}}
\newcommand{\eea}{\end{eqnarray}}

\newcommand{\tr}{\hbox{Tr}}
\newcommand{\HH}{\hat{H}}
\newcommand{\mean}[1]{\langle{#1}\rangle{}}

\newcommand{\EZ}{E_{Z}}
\newcommand{\VST}{\hat{V}_{ST_{0}}}
\newcommand{\Th}{\hat{\theta}}
\newcommand{\EZT}{\mu_{\text{T}}}
\newcommand{\MA}{\mathcal{A}}
\newcommand{\sinc}{{\rm sinc}}

\begin{document}

\title{Hyperfine interaction induced dephasing of coupled spin qubits in semiconductor double quantum dots}
\author{Jo-Tzu Hung}
\affiliation{Department of Physics, University at Buffalo, SUNY, Buffalo, New York 14260-1500, USA}
\author{{\L}ukasz Cywi{\'n}ski}
%\email{lcyw@ifpan.edu.pl}
\affiliation{Institute of Physics, Polish Academy of Sciences, al.~Lotnik{\'o}w 32/46, PL 02-668 Warszawa, Poland}
\author{Xuedong Hu}
\affiliation{Department of Physics, University at Buffalo, SUNY, Buffalo, New York 14260-1500, USA}
\author{S. Das Sarma}
\affiliation{Condensed Matter Theory Center, Department of Physics, University of Maryland, College Park, MD 20742-4111, USA}
\date{\today}

\begin{abstract}
We investigate theoretically the hyperfine-induced dephasing of two-electron-spin states in a double quantum dot with a finite singlet-triplet splitting $J$. In particular, we derive an effective pure dephasing Hamiltonian, which is valid when the hyperfine-induced mixing is suppressed due to the relatively large $J$ and the external magnetic field.  Using both a quantum theory based on resummation of ring diagrams and semiclassical methods, we identify the dominant dephasing processes in regimes defined by values of the external magnetic field, the singlet-triplet splitting, and inhomogeneity in the total effective magnetic field.  We address both free induction and Hahn echo decay of superposition of singlet and unpolarized triplet states (both cases are relevant for singlet-triplet qubits realized in double quantum dots).  We also study hyperfine-induced exchange gate errors for two single-spin qubits.  Results for III-V semiconductors as well as silicon-based quantum dots are presented.
\end{abstract}

\maketitle

%%%%%%%%%%%%%%%%%%%%%
%%% INTRODUCTION
%%%%%%%%%%%%%%%%%%%%%
\section{Introduction \label{sec:introduction}}

% STATUS OF QD SPIN QUBITS
Spin qubits in quantum dots (QDs) or donors have been extensively investigated during the past decade.\cite{Hanson_RMP07, Liu_AP10, Morton_Nature11} Single spin qubits\cite{Loss_PRA98} have been initialized, manipulated, and read out both electrically in gated dots\cite{Elzerman_Nature04, Koppens_Nature06, Nowack_Science07, Koppens_PRL08, Pioro_NP08, Shaji_NP08, Morello_Nature10, Pla_Nature12} and optically in self-assembled dots.\cite{Press_Nature08,Press_NP10,Greilich_NP09}  Two-qubit exchange gates have been demonstrated recently in gated QDs,\cite{Brunner_PRL11, Nowack_Science11} so are two-electron\cite{Kim_NP11} and two-hole\cite{Greilich_NP11} couplings in a pair of vertically stacked self-assembled QDs.  Furthermore, a subspace formed by the singlet $S$ and the unpolarized triplet $T_{0}$ states of two electron spins has also been explored as a logical qubit,\cite{Levy_PRL02, Petta_Science05, Foletti_NP09, Maune_Nature12} and coupling between two such qubits\cite{VanWeperen_PRL11} and their entanglement\cite{Shulman_Science12} has been demonstrated recently.

% HF INTERACTION, SQDs
Slow decoherence relative to the control speed is one of the key criteria for a scalable quantum information processor.  In QDs made of III-V compounds or from natural Si, the dominant source of single-spin decoherence is the nuclear spin bath, coupled to the carrier spins by hyperfine (hf) interaction.\cite{Abragam, Optical_Orientation, Merkulov_PRB02, Hanson_RMP07}  The role of this hf coupling is clear: the energy splitting of an electron spin is affected by the fluctuating Overhauser field of the nuclear spins.  The nuclear spin bath is quasistatic due to its dynamics being much slower than the dynamics of the electron spins. Therefore, the strongest effect of the hf interaction is an inhomogeneous broadening in the distribution of electron spin splitting due to the random but static orientation of the nuclear spins, causing a decay of the electron free induction signal on the time scale of $T_{2}^{*}$.\cite{Merkulov_PRB02, Hanson_RMP07}  The slow nuclear spin dynamics within the bath causes homogeneous broadening, or pure dephasing of the electron spin qubit, which is measurable in a spin echo experiment. This nuclear spin dynamics is either due to hf interaction only at lower applied fields,\cite{Deng_PRB06, Yao_PRB06, Cywinski_PRL09, Cywinski_PRB09, Bluhm_NP10, Neder_PRB11} or, at higher fields, due to dipolar interaction between nuclear spins.\cite{deSousa_PRB03, Witzel_PRB06, Yao_PRB06}

For two uncoupled spins, spin product states such as $\ket{\uparrow\downarrow}$ and $\ket{\downarrow\uparrow}$ are the two-spin eigenstates.  If the two spins are initially prepared in a singlet state ($S$) $(\ket{\uparrow\downarrow} - \ket{\downarrow\uparrow})/\sqrt{2}$,\cite{Johnson_Nature05, Koppens_Science05, Petta_Science05} the random Overhauser field would strongly mix the $S$ and $T_{0}$ states, and the time scale on which this mixing leads to decay of the measured signal
is $T_{2}^{*}$, the same as that for a single-spin qubit.\cite{Johnson_Nature05, Koppens_Science05, Petta_Science05}  Similar to the single spin case,\cite{Koppens_PRL08} the application of a Hahn echo pulse sequence removes the influence of the quasistatic nuclear fluctuations, and reveals the much slower decoherence of two independent single spins.\cite{Petta_Science05,Bluhm_NP10}

% FINITE J
For two coupled electrons in a uniform effective magnetic field, the singlet and triplet states are the two-electron eigenstates.
At a finite exchange splitting $J$ in a double quantum dot (DQD), it was predicted\cite{Coish_PRB05} and then shown experimentally\cite{Laird_PRL06} that the hf-induced singlet-triplet mixing is suppressed, and the probability of the initialized singlet to remain in this state decays as a power-law towards a saturation value which approaches unity as $J$ is increased.
The decoherence effect of nuclear spin pair flips due to inter-nuclear dipolar interactions has also been analyzed in the limit of $J$ much smaller than the Zeeman splitting of the polarized triplet states.\cite{Yang_PRB08}
Recently, decoherence in the $S$-$T_{0}$ subspace has been investigated at finite $J$ in experiments on GaAs\cite{Dial_PRL13,Higginbotham_arXiv13} and InGaAs\cite{Weiss_PRL12} DQDs. In the former work the dominant role played by the charge noise was uncovered, while in the latter paper the effects of charge noise were minimized, and a lower bound on $T_{2}^{*}$ time due to interaction with nuclei was obtained.

% WHAT THE PAPER IS ABOUT
In this paper we systematically investigate the hf-induced dephasing of two-spin states at finite values of exchange splitting $J$.  We focus on the case of dephasing within the $S$-$T_{0}$ subspace, in which full control over the qubit state is possible because of the creation of a stationary Overhauser field gradient,\cite{Foletti_NP09} or a gradient of magnetic field generated by a proximal nanomagnet.\cite{Pioro_NP08,Petersen_PRL13}
We investigate both the limit of uniform effective field and of the finite effective field gradient, and we find that with an increasing magnitude of the gradient there is a smooth transition from strongly suppressed decoherence to decoherence that is similar to the case of single-spin qubits.  We also study the hf-induced decay in an Hahn echo experiment, which is possible with a controllable field gradient.\cite{Dial_PRL13} Lastly, we analyze the hf-induced exchange gate error, when the two spins in a DQD are treated as two single-spin qubits.

Our theoretical approach is based on first obtaining an effective pure dephasing Hamiltonian via an appropriate canonical transformation of the full hf Hamiltonian.\cite{Shenvi_scaling_PRB05, Yao_PRB06, Coish_PRB08, Yang_PRB08, Cywinski_PRL09, Cywinski_PRB09} The effective Hamiltonian $\tilde{H}$ is diagonal in the basis of the relevant states, allowing dephasing calculations for superpositions of these states. When dealing with terms in $\tilde{H}$ that are of second order in the {\it transverse} Overhauser field, we use the ring diagram theory\cite{Cywinski_PRL09,Cywinski_PRB09,Cywinski_PRB10} (RDT). The terms that are of first or second order in {\it longitudinal} Overhauser field are treated classically.  We also compare the RDT results with calculations based on semiclassical averaging over the quasistatic transverse Overhauser fields, underlining its connection to the short-time RDT calculations.\cite{Neder_PRB11,Cywinski_APPA11}

% OTHER DEPHASING SOURCES
For moderate values of singlet-triplet splitting, the shortest singlet-triplet dephasing times due to hf interaction derived in this paper are on the order of a microsecond (millisecond) for typical GaAs (Si) QDs.
It is important to point out that there are many other decoherence channels beyond hf coupling with nuclear spins for electrons in a DQD.  For example, at finite $J$, the orbital degree of freedom is not completely frozen, so that its fluctuations can lead to spin decoherence.  In particular, charge noise could be an important, even dominant, source of dephasing, as suggested by theory\cite{Coish_PRB05, Hu_PRL06, Culcer_APL09, Gamble_PRB12} and seen in experiment.\cite{Shulman_Science12,Dial_PRL13,Higginbotham_arXiv13}
In fact, the singlet-triplet dephasing time $T^{*}_{2} ~\! \approx \! 100$ ns due to charge noise in GaAs observed recently\cite{Dial_PRL13} is an order of magnitude smaller than the shortest hf-induced dephasing times predicted for this material in this paper.
Our results presented in this paper are most relevant in an experimental situation where charge noise is not dominant. However, despite the possibly critical role of charge noise in DQDs, we would like to stress that this noise can, in principle, be removed to a large degree (e.g.,~by different designs of samples or the gate circuitry), while the presence of nuclear spins is unavoidable in III-V materials, since all isotopes of the III-V elements have finite nuclear spins (unlike Si, where isotopic purification could, in principle, suppress the nuclear-induced spin decoherence).  Furthermore, it has recently been shown that $S$ and $T_{0}$ states of an optically controlled self-assembled QD molecule can be manipulated in a regime in which their splitting is insensitive (in the first order) to charge fluctuations.\cite{Weiss_PRL12} The existence of such ``sweet spots" was also predicted\cite{Stopa_NL08,Li_PRB10} in gated QDs.
Besides charge noise, the different charge distributions for the singlet and triplet states also allow electron-phonon interaction to cause dephasing in the $S$-$T_0$ subspace,\cite{Roszak_PRB09, Hu_PRB11, Gamble_PRB12} although this dephasing channel should be weak in a double dot if $J$ is not too large.\cite{Hu_PRB11}

% SPIN-ORBIT RELAXATION
In addition to pure dephasing, the $S$-$T_0$ qubit can undergo longitudinal relaxation through phonon emission, which is allowed by spin-orbit coupling\cite{Stano_PRL06,Raith_PRL12,Borhani_PRB12} and/or hyperfine mixing between the states.\cite{Raith_PRL12} Such dissipative relaxation processes lead to both transitions between $S$ and $T_{0}$, and leakages out of the qubit subspace (to $T_{\pm}$ states, for example). The characteristic time scales are most often much longer\cite{Borhani_PRB12} than the time scales considered in this paper, although recent calculations suggest that in some parameter regimes the $S$-$T_{0}$ transitions in GaAs DQDs can occur at the microsecond time scale.\cite{Raith_PRL12} One can, however, exploit the large anisotropy of the relaxation rates with respect to the direction of applied magnetic field, and suppress these processes by an appropriate choice of the in-plane $B$-field direction.\cite{Raith_PRL12}

We mention that although the single qubit decoherence is often theoretically studied in the literature in various contexts, there are few concrete analyses of multiqubit decoherence simply because the multiqubit decoherence problem is technically difficult due to the many possible decoherence channels for the entangled system when even a few qubits are coupled together.   Our current work demonstrates that the simplest spin system in which entanglement can occur, namely, a system of just two exchange-coupled electron spins weakly interacting with an environment of nuclear spins, is theoretically challenging, even though the full Hamiltonian for the problem is completely known.  Real experimental situations are obviously far more complex because it is unlikely that all the environmental influences would be completely known.  For example, as we mentioned above, for coupled spin qubits in semiconductor quantum dots, there would be, in addition to the nuclear-induced Overhauser noise, other decoherence mechanisms such as charge noise, stray fluctuating magnetic fields arising from random impurity spins in the semiconductor and from microwaves and other random fluctuations in the background.  However, the eventual construction of a fault-tolerant practical quantum computer necessarily requires understanding (and if possible, mitigating) all multiqubit decoherence mechanisms since the quantum error correction threshold is small, which implies that only the smallest amount of decoherence can be efficiently eliminated by the error correction protocols.  Our current work should be construed as a first step toward the goal of a comprehensive understanding of multiqubit decoherence in one of the most practical and widely studied quantum computer architecture proposals, namely, the spin quantum computer in semiconductor quantum dots.  It is somewhat sobering that even this first step of understanding Overhauser noise induced two-qubit decoherence in semiconductor quantum dots is already a very challenging problem.

The present paper is organized as follows. In Sec.~\ref{sec:DQD}, we describe the Hamiltonian of the two electrons in a DQD applicable in the regime of our interest. At finite $J$, the Hamiltonian of two electrons is given in the $\{S,T_{0}, T_{+}, T_{-}\}$ basis in Sec.~\ref{sec:H_4x4}.  The state of the nuclear bath is described in Sec.~\ref{sec:bath}.  In Sec.~\ref{sec:ST0}, we derive an effective Hamiltonian in the $S$-$T_0$ basis.  The dephasing of an $S$-$T_0$ qubit due to the various terms in the effective Hamiltonian is calculated for DQDs made of various materials in Sec.~\ref{sec:W_calculation_0}.  We identify the dominant dephasing mechanisms for various types of DQDs (GaAs, Si, or InGaAs) in Sec.~\ref{sec:dominant}, and discuss the existence of optimal value of $J$ at which the coherence time is predicted to be maximal.
In Sec.~\ref{sec:W_calculation_1}, we clarify the effect of a finite interdot field gradient, and analyze the decay of the Hahn echo signal.
The SWAP gate error due to the hf-induced dephasing processes is investigated in Sec.~\ref{sec:gate}, with a focus on the effect of inhomogeneous broadening.  A description of experimental protocols to measure $S$-$T_{0}$ decoherence is given in Appendix~\ref{app:experiment}.  Additional technical details are provided in Appendices~\ref{app:Hhf}--\ref{app:other_Heff}.

%%%%%%%%%%%%%%%%%%%%%
%%% DQD - HAMILTONIAN
%%%%%%%%%%%%%%%%%%%%%
\section{Two electron spins in a double quantum dot} \label{sec:DQD}

In this section we define the starting point of our calculations.  Specifically, we first identify the four relevant lowest-energy two-electron states in a DQD, then project the total Hamiltonian onto the basis that is a product of these four electronic states and the nuclear spin Hilbert space.  We also discuss the semiclassical description of the nuclear spin reservoir.  Starting from the general Hamiltonian, depending on the particular physical problems, we derive the effective Hamiltonian that couples the two levels that we are interested in to the nuclear spin reservoir.

\subsection{Low-Energy Two-Electron Hamiltonian in a DQD}\label{sec:H_4x4}

The system we consider in the current study consists of two electrons located in two (weakly) tunnel-coupled QDs, labeled left (L) and right (R), deep in the (1,1) charge configuration.\cite{Hu_PRA00, Petta_Science05, Coish_PRB05, Taylor_PRB07, Li_PRB12}  The total Hamiltonian for the coupled electron-nuclear spin system can be written as
\begin{equation}
\HH_{total} = \HH_{e} + \HH_{hf} + \HH_n \,,
\label{eq:H_total}
\end{equation}
where the three terms represent the electronic, hyperfine, and nuclear part of the Hamiltonian, respectively.  Below we discuss each of these terms in a two-electron DQD.

Deep in the (1,1) regime of the charge stability diagram, the four lowest-energy two-electron states, including one singlet $S(1,1)$ and three triplet states $T(1,1)$, are well approximated by the Heitler-London states. Specifically, $S(1,1) = \psi_{\rm S} \otimes (\ket{\uparrow\downarrow} - \ket{\downarrow\uparrow})/\sqrt{2}$ and $T(1,1) = \psi_{\rm AS} \otimes \ket{\uparrow\uparrow}, (\ket{\uparrow\downarrow} + \ket{\downarrow\uparrow})/\sqrt{2}$, and $\ket{\downarrow\downarrow}$.  The orbital parts $\psi_{\rm S/AS}$ are symmetric and antisymmetric combinations of $\Psi_{L}(\mathbf{r})$ and $\Psi_{R}(\mathbf{r})$ states, which are the single-electron ground state orbital of the potentials for the L and R dots.  States in the doubly occupied $(2,0)$ and $(0,2)$ configurations have significantly higher energies, and are not included explicitly in our consideration.
The only role they play is to lower the energy of the singlet with respect to $T_{0}$ by the exchange splitting $J$ ($J \! > \! 0$ for typical values of magnetic fields used in experiments), when the DQD has a finite tunnel coupling.\cite{Petta_Science05, Koppens_Science05}

In the presence of a magnetic field, which has both a uniform $B \! =\! (B_{L}+B_{R})/2$ and a gradient $\Delta B  \! = \!  (B_{L}-B_{R})/2$ component, the electronic Hamiltonian in the $\{S,T_{0},T_{+},T_{-}\}$ basis is
\beq
\HH_{e} = \left( \begin{array}{cccc} -J & \Delta E_{Z} & 0 & 0 \\  \Delta E_{Z} & 0 & 0 & 0 \\ 0 & 0 & -\EZ & 0\\ 0 & 0 & 0 & \EZ \end{array} \right ) \,\, , \label{eq:H0}
\eeq
where $\EZ = -g\mu_{B}B$ and $\Delta\EZ = -g\mu_{B}\Delta B$ (the sign convention is such that the positive $B$ field lowers the energy of $T_{+}$ in GaAs, where the effective g-factor is negative).

The two electrons couple to the environmental nuclear spins through the contact hf interaction, which takes the general form
\beq
\HH_{hf} = \sum_{i} \MA_{\alpha[i]} \mathbf{S}_{1} \cdot \mathbf{I}_{i}\nu_{0}\delta(\mathbf{r}_{1} - \mathbf{R}_{i}) + \sum_{i} \MA_{\alpha[i]} \mathbf{S}_{2} \cdot \mathbf{I}_{i}\nu_{0}\delta(\mathbf{r}_{2} - \mathbf{R}_{i}) \,\, ,  \label{eq:H1hf}
\eeq
where $\mathbf{S}_{1,2}$ are the spin operators of the two electrons at positions $\mathbf{r}_{1,2}$, $\mathbf{I}_{i}$ are the spin operators of nuclei at site $\mathbf{R}_{i}$, and $\MA_{\alpha[i]}$ is the hf constant corresponding to the species $\alpha[i]$ (e.g.,~$^{69}$Ga, $^{71}$Ga, and $^{75}$As in GaAs, or $^{29}$Si in Si) of the nucleus at site $i$. The values of hf constants for relevant nuclei are given in Table \ref{tab:parameters}. $\nu_{0}$ is the volume of the primitive unit cell, and using a single-electron wave function $\phi(\mathbf{r}) \! = \! \frac{1}{\sqrt{\nu_{0}}} \Psi(\mathbf{r})$, where $\Psi(\mathbf{r})$ is the envelope function, the hf-interaction energy (i.e.,~the Knight shift) of the $i$-th nucleus interacting with one electron is $A_{i} = \MA_{\alpha[i]} |\Psi(\mathbf{r}_{i})|^{2}$.

Projecting Hamiltonian~(\ref{eq:H1hf}) onto the $\{S,T_{0},T_{+},T_{-}\}$ basis (see Appendix \ref{app:Hhf} for details), we obtain\cite{Coish_PRB05, Taylor_PRB07, Sarkka_PRB08}
\beq
\HH_{hf} = \left( \begin{array}{cccc}
0 & \sum_{i}B_{i}I^{z}_{i} & -\sum_{i}\frac{B_{i}}{\sqrt{2}}I^{+}_{i} & \sum_{i}\frac{B_{i}}{\sqrt{2}}I^{-}_{i} \\
\sum_{i}B_{i}I^{z}_{i} & 0 & \sum_{i}\frac{C_{i}}{\sqrt{2}}I^{+}_{i} &  \sum_{i}\frac{C_{i}}{\sqrt{2}}I^{-}_{i} \\
-\sum_{i}\frac{B_{i}}{\sqrt{2}}I^{-}_{i} & \sum_{i}\frac{C_{i}}{\sqrt{2}}I^{-}_{i} & \sum_{i}C_{i}I^{z}_{i} & 0\\
\sum_{i}\frac{B_{i}}{\sqrt{2}}I^{+}_{i} & \sum_{i}\frac{C_{i}}{\sqrt{2}}I^{+}_{i} & 0 & -\sum_{i}C_{i}I^{z}_{i} \end{array} \right ) \,,  \label{eq:Hhf}
\eeq
where
\bea
B_{i} & = & \frac{1}{2} ( A^{L}_{i} - A^{R}_{i} ) \,\, , \\
C_{i} & = & \frac{1}{2} ( A^{L}_{i} + A^{R}_{i} ) \,\,.
\eea
Here, $A^{L(R)}_{i} = \MA_{\alpha[i]}|\Psi_{L(R)}(\mathbf{R}_{i})|^{2}$ denotes the hf coupling between an electron in the L(R) orbital and a nuclear spin at $\mathbf{R}_{i}\,$.  Here we have neglected the overlap between the L and R orbital wavefunctions.  Keeping the finite overlaps amounts to small quantitative corrections to the matrix elements of $\HH_{hf}$, and neglecting them does not cause any qualitative change (for the general form of $\hat{H}_{hf}$ and its derivation, see Appendix \ref{app:Hhf}).
The same holds for the corrections to two-electron hf interaction terms brought by tunneling-induced admixture of $(0,2)$ singlet to $(1,1)$ singlet state [we assume interdot bias, or detuning, to be such that $(0,2)$ charge state is the more strongly coupled doubly-charged state]. The main influence of these corrections is to diminish the hf interaction between the singlet and the nuclei as the amplitude of $S(1,1)$ state, which is hf-coupled to other states, decreases. As long as this decrease is small, i.e.,~we are far enough from anticrossing of $(1,1)$ and $(0,2)$ singlet states, the modifications brought by this effect are not qualitative.

The diagonal terms for the two polarized triplet states in $\HH_{hf}$ are the longitudinal Overhauser field along the external field direction $z$.  On the time scale of interest to the present study, it is quasistatic.\cite{Merkulov_PRB02,Taylor_PRB07}  We therefore express the field operator in terms of its ensemble average and fluctuations:
\begin{eqnarray}
\hat{\mu} & \equiv &\sum_{i}C_{i}I^{z}_{i} =  \mu_{\text{O}} + \delta\hat{\mu} \;,\\
\mu_{\text{O}} & \equiv & \sum_{i}C_{i} \mean{I^{z}_{i}}\;,
\end{eqnarray}
where $\mean{...}$ denotes an average over the nuclear bath.

In Hamiltonian $\HH_{hf}$, the Overhauser field difference between the two dots, $\sum_{i}B_{i}I^{z}_{i}$, couples $S$ and $T_0$ states.   It can be an important control for universal manipulation of an $S$-$T_0$ qubit.\cite{Foletti_NP09}  Again we split this term into the average and the fluctuations\cite{Yang_PRB08}:
\beq
\hat{\theta} \equiv \sum_{i}B_{i}I^{z}_{i} \! = \! \theta_{\text{O}} + \delta\hat{\theta} \,\, .
\eeq
Here the mean field $\theta_{\text{O}} \! \equiv \! \mean{\hat{\theta}}$ can be built up through dynamical nuclear spin polarization (DNP),\cite{Foletti_NP09} while the fluctuation $\delta\hat{\theta} \! \equiv \! \hat{\theta} - \theta_{\text{O}}$ can be reduced with respect to its ``natural'' high-temperature value during the DNP process.\cite{Bluhm_PRL10}

In experiments where the Overhauser field in the DQD is prepared by DNP through multiple sweeps across the $S$-$T_{+}$ anticrossing,\cite{Petta_PRL08, Foletti_NP09, Bluhm_PRL10} both finite $\theta_{\text{O}}$ and $\mu_{\text{O}}$ are established.  Typically both $\theta_{\text{O}}$ and $\mu_{\text{O}}$ are of the order $100$ mT (or $2.5$ $\mu$eV) for GaAs.  In Ref.~\onlinecite{Foletti_NP09}, $\theta_{\text{O}}$ reaches above $200$ mT, with $\mu_{\text{O}}$ reaching approximately $100$ mT. Note that an equivalent role can be played by an external magnetic field gradient $\Delta B_{z}$ (leading to finite $\Delta E_{Z}$), which could be established using a nanomagnet located close to the DQD,\cite{Tokura_PRL06,Pioro_NP08,Petersen_PRL13} with reported values of $\sim \! 10$ mT field difference between the two dots.

The total Hamiltonian for the two electron spins and the hyperfine interaction now takes the form
\begin{eqnarray}
\HH_e + \HH_{hf} & = & \left( \begin{array}{cccc} -J & \theta_{\text{T}} & 0 & 0 \\  \theta_{\text{T}} & 0 & 0 & 0 \\ 0 & 0 & -\mu_{\text{T}} & 0\\ 0 & 0 & 0 & \mu_{\text{T}}  \end{array} \right ) \nonumber \\
&& \hspace*{-0.9in} + \left( \begin{array}{cccc}
0 & \delta\hat{\theta} & -\sum_{i}\frac{B_{i}}{\sqrt{2}}I^{+}_{i} & \sum_{i}\frac{B_{i}}{\sqrt{2}}I^{-}_{i} \\
\delta\hat{\theta} & 0 & \sum_{i}\frac{C_{i}}{\sqrt{2}}I^{+}_{i} &  \sum_{i}\frac{C_{i}}{\sqrt{2}}I^{-}_{i} \\
-\sum_{i}\frac{B_{i}}{\sqrt{2}}I^{-}_{i} & \sum_{i}\frac{C_{i}}{\sqrt{2}}I^{-}_{i} & \delta \hat{\mu} & 0\\
\sum_{i}\frac{B_{i}}{\sqrt{2}}I^{+}_{i} & \sum_{i}\frac{C_{i}}{\sqrt{2}}I^{+}_{i} & 0 & -\delta \hat{\mu} \end{array} \right )\!\!. \;\;\;\;\;\;\;  \label{eq:HH}
\end{eqnarray}
Here we have combined the quasistatic mean-field Overhauser terms with the external magnetic field, with $\Delta \EZ$ replaced by
\begin{equation}
\theta_{\text{T}} \! \equiv \! \theta_{\text{O}} + \Delta \EZ \,\, , \label{eq:thetaT}
\end{equation}
and $\EZ$ replaced by
\beq
\mu_{\text{T}} \! \equiv \! \EZ - \mu_{\text{O}} \,\, . \label{eq:muT}
\eeq

\begin{table}\centering
\caption{The hf constants and the nuclear Zeeman energies for $B=1$ T. The parameters for Ga, As and In are taken from Ref. \onlinecite{Cywinski_PRB09}. For $^{29}$Si, the value of $\MA_{\alpha}$ comes from Ref. \onlinecite{Assali_PRB11}. For comparison, $-\EZ/g \approx 57.8$ $\mu$eV at $B=1$ T.}
\label{tab:parameters}
\begin{tabular}{ccc}
	\hline\hline
	Nuclear species $\alpha$&\hspace{0.5in}	$\MA_{\alpha}$ ($\mu$eV) &\hspace{0.5in}$\omega_{\alpha}$ (neV) \\
	\hline	
	$^{69}$Ga	&\hspace{0.5in}	$35.9$		&\hspace{0.5in}$-42.1$\\
	$^{71}$Ga	&\hspace{0.5in}	$45.9$		&\hspace{0.5in}$-53.6$\\
	$^{75}$As		&\hspace{0.5in}	$42.9$		&\hspace{0.5in}$-30.1$\\
	$^{113}$In	&\hspace{0.5in}	$55.8$		&\hspace{0.5in}$-38.4$\\
	$^{115}$In	&\hspace{0.5in}	$56.0$		&\hspace{0.5in}$-38.5$ \\
	$^{29}$Si		&\hspace{0.5in}	$2.15$		&\hspace{0.5in}$34.8$ \\
	\hline\hline
\end{tabular}
\end{table}

The last term in $\HH_{total}$ is the nuclear Zeeman energy:
\beq
	\HH_{n} = \sum_i \omega_{\alpha[i]} I^{z}_{i} \,\,,
\eeq
where $\omega_{\alpha[i]}$ is the Zeeman splitting of the nucleus of species $\alpha$ at site $i$. Typically, these splittings are smaller than the electronic one by three orders of magnitude (see Table \ref{tab:parameters}).
Note that including the finite values of $\omega_{\alpha}$ in the case of multiple isotopes (as is the case in III-V based QDs) is crucial for description of Hahn echo decay of a single electron spin\cite{Cywinski_PRL09,Cywinski_PRB09,Bluhm_NP10} (or an $S$-$T_0$ qubit\cite{Neder_PRB11} at $J\! = \! 0$, which is equivalent to two independent single spins), while the nuclear Zeeman energies generally have much smaller influence on dephasing during the free evolution of the qubit. Below, we will show that these statements also hold for the $S$-$T_{0}$ decoherence at finite $J$.

%%%%%%%%%%%%%%%%%%%%%%%%%%
%%% THE NUCLEAR BATH
%%%%%%%%%%%%%%%%%%%%%%%%%%
\subsection{The nuclear bath and its semiclassical description}  \label{sec:bath}

As we have discussed at the end of the previous section, one of the key features of a coupled electron-nuclear-spin system is the smallness of the intrinsic nuclear energy scales (both the Zeeman energies and the dipolar interactions among the nuclei).  Consequently, at experimentally realistic temperatures, the equilibrium nuclear density operator is proportional to unity, $\hat{\rho}_{I} \propto \mathbf{1}$.
When nuclear spins are dynamically polarized,\cite{Petta_PRL08, Vink_NP09, Latta_NP09, Xu_Nature09, Foletti_NP09, Bluhm_PRL10, Frolov_arXiv12, Petersen_PRL13} the direction of the polarization in each dot is along the applied field ($z$) direction. The components of the nuclear spins transverse to this direction are randomized on a time scale of $\sim \! 100$ $\mu$s due to intra-nuclear dipolar interactions,\cite{Merkulov_PRB02,Khaetskii_PRL02} so that for experiments in which the total data acquisition time is much longer than this time scale, the appropriate nuclear density matrix is diagonal in the basis of eigenstates of $I^{z}_{i}$.  A semiclassical description of the nuclear reservoir is thus valid for at least some situations.\cite{Neder_PRB11}  Here we discuss some of the most important characteristics of the nuclear reservoir.

The Overhauser field is defined as ${\mathbf h} = \sum_i A_i \mathbf{I}_i$.  The maximal value of the Overhauser field (as felt by a single electron in a given orbital) in a fully polarized nuclear bath is
\beq
\mathcal{A}_{M} = \sum_{i}A_{i}I_{i} = \sum_{\alpha} n_{\alpha} I_{\alpha}\mathcal{A}_{\alpha} \,\, ,
\eeq
where $i$ denotes the nuclear sites, $I_{i}$ is the $i$-th nuclear spin, $\alpha$ denotes the nuclear species, and $n_{\alpha}$ is the average number of nuclei of this species in the unit cell (i.e.,~in both III-V compounds and in Si, we have $\sum_{\alpha}n_{\alpha}\! = \! 2$), and $\mathcal{A}_{\alpha}$ are the hf couplings of nuclei of $\alpha$ species given in Table \ref{tab:parameters}. With the envelope functions $\Psi_{L,R}$ normalized as $\int |\Psi_{L,R}(\mathbf{r})|^2 \text{d}^{3}r \! = \! \nu_{0}$, where $\nu_{0}$ is the volume of the Wigner-Seitz unit cell, we have then $A^{L,R}_{i} \! = \! \mathcal{A}_{\alpha[i]} |\Psi_{L,R}(\mathbf{r}_{i})|^2$, as stated before. We also define the number of unit cells $N_{\Psi}$, in which the probability of finding an electron described by wavefunction $\Psi(\mathbf{r})$ has appreciable magnitude
\beq
N_{\Psi} \equiv \frac{\int |\Psi(\mathbf{r})|^2 \text{d}^{3}r}{\int |\Psi(\mathbf{r})|^4 \text{d}^{3}r} \,\, .  \label{eq:NT}
\eeq
This definition implies that
\beq
\sum_{i \in \alpha} A_{i}^2 \approx n_{\alpha} \mathcal{A}^{2}_{\alpha}  \sum_{u} |\Psi(\mathbf{r}_{u})|^4 = \frac{n_{\alpha}\mathcal{A}^{2}_{\alpha}}{N_{\Psi}} \,\, ,
\eeq
where the sum over $u$ is over all the Wigner-Seitz unit cells (we assume that the envelope function is practically constant within each cell).

On a time scale on which the nuclear spins can be considered static, we can replace the quantum averages over the nuclear bath by classical averages over the values of the static Overhauser field $\mathbf{h}$ described by a Gaussian probability distribution,\cite{Merkulov_PRB02,Khaetskii_PRL02}
\beq
\text{Tr}\Big[ \hat{\rho}_{I} f \big (\sum_{i}A_{i}\mathbf{\hat{I}}_{i} \big ) \Big ] \approx \int P(\mathbf{h}) f(\mathbf{h})  \text{d}^{3}h
\eeq
where
\beq
P(\mathbf{h}) = \frac{1}{2\pi \sigma^{2}_{\perp}} e^{-h^{2}_{\perp}/2\sigma^{2}_{\perp,\Psi}} \frac{1}{\sqrt{2\pi}\sigma_{z}} e^{-(h_{z} - \mean{h_{z}})^{2} / 2\sigma^{2}_{z,\Psi}} \,\, ,  \label{eq:PO}
\eeq
where $\mathbf{h}_{\perp}$ is the transverse component of $\mathbf{h}$, and $\mean{h_{z}}$ is the average value of the longitudinal Overhauser field (in the $z$ direction).
The width of the distribution of $h_{z}$ is given by
\beq
\sigma^{2}_{z,\Psi} = \sum_{i} A^{2}_{i}\left [ \mean{(I^{z}_{i})^{2}}-\mean{I^{z}_{i}}^2 \right ] \,\, . \label{eq:sigmazphi}
\eeq
Under realistic experimental conditions, the nuclear spin bath is in the thermal state at the high-temperature limit, $\hat{\rho}_I \propto \mathbf{1}$. This state is isotropic, so that the variance of the $z$ projection of any given spin is $\mean{(I^{z}_{i})^{2}}-\mean{I^{z}_{i}}^2 \! = \! I_{\alpha}(I_{\alpha}+1)/3$ for $i \in \alpha$. A \emph{narrowed} state of the bath, in which the variance is reduced from this ``high temperature'' value, is also often considered, and it can be created in experiments.\cite{Barthel_PRL09,Bluhm_PRL10,Vink_NP09,Latta_NP09,Xu_Nature09, Frolov_arXiv12} We account for the possibility of narrowing of the distribution of $h^{z}$ fields by introducing a narrowing factor $n_{F} \! < \! 1$, defined as the ratio between actual $\sigma_{z}$ and its high temperature value given by the above expression. We thus have
\beq
\sigma^{2}_{z,\Psi} = n^{2}_{F} \frac{\sum_{\alpha} n_{\alpha} I_{\alpha}(I_{\alpha}+1) \mathcal{A}^{2}_{\alpha}}{3N_{\Psi}} \equiv n^{2}_{F} \frac{\mathcal{A}^{2}}{N_{\Psi}} \,\, , \label{eq:sigmaz}
\eeq
where we have defined the ``typical'' energy of hf interaction $\mathcal{A}$. For GaAs $\mathcal{A}$ is of the same order of magnitude as the maximal Overhauser field $\mathcal{A}_{M}$.  For silicon with a fraction $f$ of the spinful $^{29}$Si nuclei, $\mathcal{A}\! \propto \! \mathcal{A}_{M}/\sqrt{f}$.

For typical achievable values of nuclear polarization, the width of the distribution of the transverse Overhauser field is given by an analogous formula, albeit without the $n_{F}$ factor.  In other words,
\beq
\sigma_{\perp,\Psi} = \frac{\sigma_{z,\Psi}}{n_F} \, \, . \label{eq:sigmaperp}
\eeq
In III-V compounds, the number of nuclei interacting appreciably with an electron is $N_{S} \! \equiv \! 2N_{\Psi}$ (the factor of $2$ appears because there are two nuclei per unit cell), and the typical value $\sigma_{\perp}$ for a GaAs QD with $N_{S} \! \approx \! 10^{6}$ spins is a few mT ($\sigma_{\perp} \! \lesssim \! 0.1 \mu$eV).
For silicon, the result depends also on the concentration of the spin-$1/2$ $^{29}$Si nuclei, given by $f \! \equiv \! n_{\text{Si}}/2$ (with $f\!= \! 0.047$ for natural silicon). With $I_{\text{Si}} \! = \! 1/2$, we obtain
\beq
\sigma_{\perp,\Psi}^{\text{Si}}(f) =  \sqrt{f} \frac{\mathcal{A}_{\text{Si}}}{\sqrt{2 N_{\Psi}}} = \frac{\mathcal{A}_{\text{Si}}f}{\sqrt{N_{S}}} \,\,,
\eeq
where the value of $\mathcal{A}_{\text{Si}}$ is given in Table \ref{tab:parameters} (note that we are using here a different definition for $\mathcal{A}_{\text{Si}}$ compared to Ref.~\onlinecite{Assali_PRB11}, where an $f$-dependent quantity was used).

In the following we will be mostly interested in the distribution of the \emph{difference} of the Overhauser fields between the two dots,
\beq
\theta = \frac{h^{z}_{L} - h^{z}_{R}}{2} \,\, ,
\eeq
which could have a finite average and/or a narrowed distribution. Using the values of $\sigma_{z,L/R}$ for the two dots (L and R), we introduce
\beq
\sigma^{2}_{z} = n^{2}_{F}(\sigma^{2}_{z,L} +  \sigma^{2}_{z,R} ) = n^{2}_{F} \frac{\mathcal{A}^2}{N_{D}} \,\, , \label{eq:sigmaz_DQD}
\eeq
where
\beq
\frac{1}{N_{D}} = \frac{1}{N_{L}} + \frac{1}{N_{R}} \,\, ,
\eeq
and we have taken the natural (non-narrowed) values for $\sigma_{z,L/R}$, {\em{}i.e.,} we have used Eq.~(\ref{eq:sigmazphi}) with $\Psi_{L,R}(\mathbf{r})$. The $n_{F}$ factor accounts now for possibly reduced standard deviation of the difference of the Overhauser field in the two QDs. It should be noted that such a narrowing can be achieved by enforcing a correlation between the values of $h^{z}_{L}$ and $h^{z}_{R}$, that is by modifying the joint probability distribution for the two fields,  without affecting the distribution of each one of them considered separately. The state narrowing obtained in experiments on singlet-triplet qubits in DQDs is of this nature.\cite{Bluhm_PRL10}
The standard deviation of the transverse components of the Overhauser field difference, $\sigma_{\perp}$, is, analogous to the single-electron case from Eq.~(\ref{eq:sigmaperp}), defined as $\sigma_{\perp} \! = \! \sigma_{z}/n_{F}$.

%%%%%%%%%%%%%%%%%%%%%%%%%%%%%%%%%
%%% DERIVATIONS OF S-T0 DEPHASING HAMILTIONIAN
%%%%%%%%%%%%%%%%%%%%%%%%%%%%%%%%%

\subsection{Effective Hamiltonian in the $S$-$T_{0}$ subspace} \label{sec:ST0}
One focus of the present paper is the decoherence of $S$-$T_{0}$ qubits. Here we derive the effective Hamiltonian in the basis of $\ket{S}$ and $\ket{T_{0}}$ states in the presence of a large external magnetic field.  It is directly applicable to experiments on $S$-$T_{0}$ qubits whenever exchange splitting $J$ is large enough.

\begin{figure}[t]
	\includegraphics[width=0.8\linewidth]{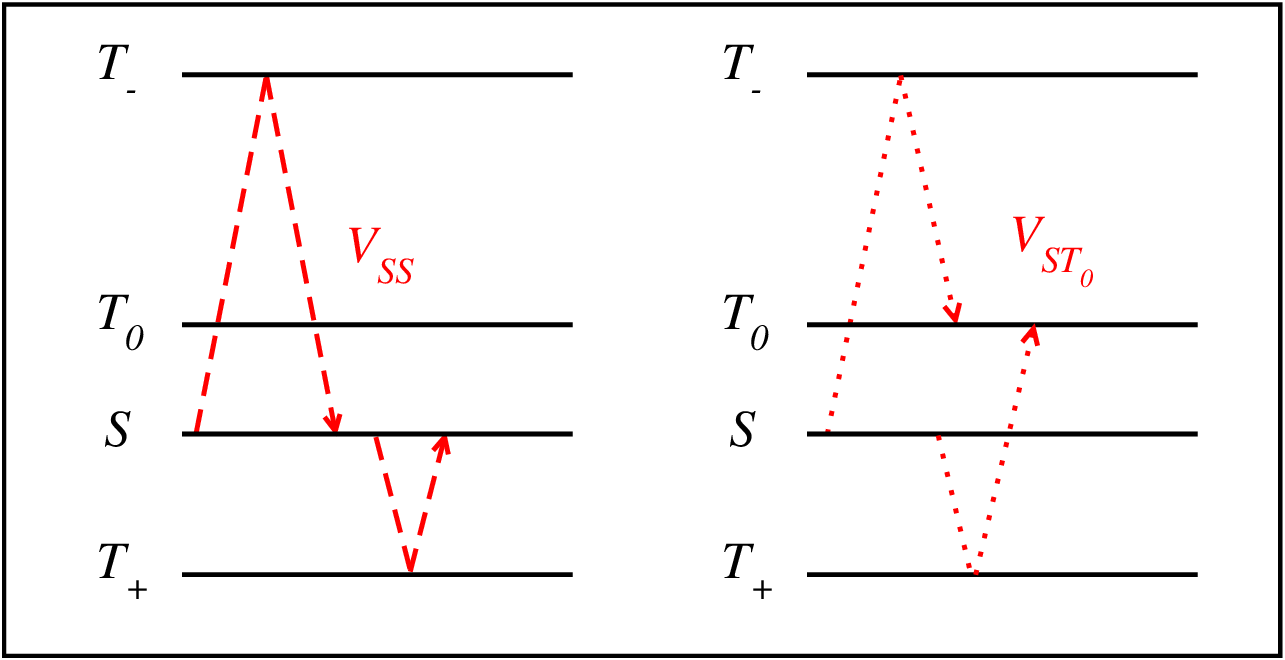}	
	\caption{(Color online) The virtual transitions that lead to the second-order effective interactions $\hat{V}_{SS}$ and $\VST$, as described in text.}\label{fig:VssVst}
\end{figure}

As shown in Eq.~(\ref{eq:HH}), $\ket{S}$ and $\ket{T_{0}}$ states are coupled to the polarized triplet states $\ket{T_{\pm}}$ by the transverse Overhauser field $\sum_{i} A^{L/R}_{i} I^{\pm}_{i}$.  In a finite external magnetic field, such that $\mu_{\rm T} \gg \sigma_{\perp}$, dephasing between $S$ and $T_{0}$ states can be faithfully described by an effective Hamiltonian in the subspace of these two states, treating the coupling to the $T_{\pm}$ states perturbatively.\cite{Winkler}  With the zeroth-order Hamiltonian given by the first matrix in Eq.~(\ref{eq:HH}), 
the condition for the perturbative treatment is
\beq
	\sigma_{\perp} \ll |J \pm \mu_{\text{T}}|\, , \,\, |\mu_{\text{T}}| \,\, . \label{eq:assumption1}
\eeq
The effective Hamiltonian in the $\{S,T_{0}\}$ subspace is then:
\bea
	\tilde{H}_{ST_0} & = & \left ( \begin{array}{cc} -J + \hat{V}_{SS} & \VST+\theta_{\text{T}} + \delta\hat{\theta} \\
	\VST^{\dagger} + \theta_{\text{T}} + \delta\hat{\theta}  & 0 \end{array} \right ) \nonumber \\
	& & + \sum_i \omega_{\alpha[i]} I^{z}_{i} + \left ( \begin{array}{cc}  \HH_{B} &  0 \\ 0  & \HH_{C} \end{array} \right )\,\,, \label{eq:Hfirst}
\eea
which contains second-order effective interactions among the nuclei.  In particular, $\hat{V}_{SS}$ comes from the virtual flip flops between $S$ and $T_{\pm}$:
\bea
	\hat{V}_{SS} & = & \frac{J}{\EZT^2-J^2} \sum_{i, j} B_{i} B_j I^{+}_{i}I^{-}_{j}  \nonumber \\
 & & \hspace*{-0.8in} = v_{ss} \sum_{i, j} (A^{L}_{i}A^{L}_{j}+A^{R}_{i}A^{R}_{j} - A^{L}_{i}A^{R}_{j}-A^{R}_{i}A^{L}_{j}    ) I^{+}_{i}I^{-}_{j} \,, \label{eq:VSS}
\eea
with $v_{ss}= J/4(\EZT^2-J^2)$.  $\VST$ represents flip flops between $S$ and $T_{0}$ via virtual transitions through $T_{\pm}$.  It consists of a Hermitian $\hat{V}_{H}$ and an anti-Hermitian $\hat{V}_{AH}$ part:
\beq
	\VST  =  \hat{V}_{H} + \hat{V}_{AH}    \,\, , \label{eq:VST}\\
\eeq
where
\bea
	\hat{V}_{H} & = & -\frac{1}{4} \left(\frac{1}{\EZT} + \frac{\EZT}{\EZT^2-J^2} \right) \sum_{i,j} B_{i}C_{j} (I^{+}_{i}I^{-}_{j}+I^{-}_{i}I^{+}_{j}) \nonumber \\
& = & v_{H} \sum_{i,j} (A^{L}_{i}A^{L}_{j}-A^{R}_{i}A^{R}_{j}) (I^{+}_{i}I^{-}_{j}+I^{-}_{i}I^{+}_{j}) \\
\hat{V}_{AH} & = & -\frac{1}{4}\frac{J}{\EZT^2-J^2} \sum_{i,j} B_{i}C_{j} (I^{+}_{i}I^{-}_{j} - I^{-}_{i}I^{+}_{j}) \nonumber \\
& = & v_{AH} \sum_{i\neq j} (A^{L}_{i}A^{R}_{j}-A^{R}_{i}A^{L}_{j}) I^{+}_{i}I^{-}_{j} \,\, ,
\eea
with
\begin{eqnarray*}
v_{H} & = & -\frac{1}{16} \left(\frac{1}{\EZT} + \frac{\EZT}{\EZT^2-J^2} \right) \\
v_{AH} & = & -\frac{1}{8}\frac{J}{\EZT^2-J^2} \,.
\end{eqnarray*}
Fig. \ref{fig:VssVst} gives a cartoon that depicts the virtual flip-flop transitions contributing to $\hat{V}_{SS}$ and $\VST$.

Hamiltonian $\HH_{ST_0}$ in Eq.~(\ref{eq:Hfirst}) contains terms that are linear in the nuclear spin operators $I^{z}_{i}$ but are second-order in the hyperfine coupling strength (within our approximation of neglecting the orbital overlap when calculating the hf interactions, $B^{2}_{i} = C^{2}_{i}= A^{2}_{i}/4$):
\bea
	\HH_{B} & = & \frac{1}{\EZT + J}\sum_{i}B^{2}_{i}I^{z}_{i} = \frac{1}{4(\EZT + J)}\sum_{i}A^{2}_{i}I^{z}_{i} \,\, , \nonumber \\
	\HH_{C} & = & \frac{1}{\EZT}\sum_{i}C^{2}_{i}I^{z}_{i} = \frac{1}{4\EZT}\sum_{i}A^{2}_{i}I^{z}_{i} \,\,. \label{eq:HbHc}
\eea
They influence the $S$-$T_{0}$ coherence in a way identical to what the longitudinal Overhauser field does to a single spin.  Averaging over a thermal distribution of the Overhauser field felt by a single spin, $h^{z}=\sum_i A_i I_i^z$, leads to a strong inhomogeneous broadening\cite{Merkulov_PRB02} and a dephasing time of $T_{2}^{*} \! \approx \! 1/\sigma_{z}$, where $\sigma_{z}$ is the spread of the values of the longitudinal Overhauser field (see Sec.~\ref{sec:bath} for precise definition). This time is of the order of 10 ns in GaAs QDs for a single electron spin, while dephasing due to these $I^{z}$-linear terms is strongly suppressed in the case of an $S$-$T_0$ qubit, because the interaction strength is significantly reduced, {\em{}i.e.,}~the qubit-nuclei couplings in Eq.~(\ref{eq:HbHc}) are $\sim A^{2}_{i}/\mu_{\rm T} \! \ll \! A_{i}$ for the values of $\mu_{\rm T}$ considered here.

Note that $\hat{V}_{SS}$, $\hat{V}_{ST_0}$, and $\delta \hat{\theta}$ are all fluctuations due to nuclear spins, and average to zero in a thermal nuclear spin reservoir.  In the absence of these fluctuations, the universal control of the $S$-$T_0$ qubit can be achieved via tuning of $J$ and $\theta_{\rm T}$.\cite{Foletti_NP09}  In the context of decoherence when $J$ is finite (more precisely, when $\sqrt{J^2 + 4\theta_{\rm T}^2} \gg \sigma_z, \sigma_\perp$), we have two interesting regimes to consider.  The first is when $\theta_{\rm T} = 0$, i.e., the DQD is in a uniform total field.  In this limit, the DQD has left-right symmetry, so that the singlet and triplet states are system eigenstates.  Since $S$ and $T_0$ are $S_z = 0$ states, with no magnetic moment in either quantum dots, they are not directly affected by the Overhauser field $h_z$.  The $S$-$T_0$ qubit made up from these two states should thus have a significantly longer inhomogeneous broadening time ($T_2^*$) than a single spin.  We will study this regime in detail in Sec.~\ref{sec:W_calculation_0}.  The other regime is when $\theta_{\rm T} \gg \sigma_{z}$, i.e., a magnetic field gradient is present (whether due to nuclear spin polarization or applied externally).  Now the left-right symmetry of the DQD is broken, and the true eigenstates of the system are superpositions of $S$ and $T_0$ states.  The electron spin densities in the two dots do not vanish anymore, so that $h_z$ can affect the two-spin coherence directly, and the system acquires single-spin qubit characteristics.  This regime will be studied in Sec.~\ref{sec:W_calculation_1}.

%%% J=0
An extreme case is when $J \! \ll \! \sigma_{z},\sigma_{\perp}$, which is a regime already investigated in existing experimental \cite{Petta_Science05, Koppens_Science05, Johnson_Nature05, Bluhm_NP10, Maune_Nature12} and theoretical \cite{Witzel_PRB08,Neder_PRB11} studies.
Here, the exchange coupling is effectively turned off.  The dynamics of the two independent electron spins are determined by the Overhauser fields in the respective dots.  The interdot nuclear spin flip flops are completely suppressed, so that all interactions involving such flip flops, $\hat{V}_{SS}$ and $\hat{V}_{AH}$, vanish in this limit.  In addition, now $\HH_{B} = \HH_{C}$ so that they do not affect the two-spin dynamics.  The remaining hf-mediated interaction is due to intradot flip flops:
\beq
\lim_{J\to 0}\hat{V}_{H} = -\frac{1}{8\EZT}\sum_{i,j} (A^{L}_{i}A^{L}_{j}-A^{R}_{i}A^{R}_{j}) (I^{+}_{i}I^{-}_{j} + I^{-}_{i}I^{+}_{j})\,\,. \label{eq:VH0}
\eeq

When $J\! = \! 0$, the two-spin eigenstates are spin product states.  Within the $S_z = 0$ subspace, it is more convenient to consider the Hamiltonian in the basis of $\ket{\pm X} \! = \! \frac{1}{\sqrt{2}}(\ket{S} \pm \ket{T_{0}}) = \{|\uparrow\downarrow\rangle, |\downarrow \uparrow\rangle \}$ states.  The resulting Hamiltonian is of pure dephasing form in this product basis:
\beq
\hat{H}_{J=0} \approx (\hat{V}_{H} + \theta_{\text{T}} + \delta \hat{\theta}) ( \ket{+X}\bra{+X} - \ket{-X}\bra{-X} ) \,\, .
\eeq
In a free evolution experiment with an initial singlet state, the ensemble coherence will now decay in $T_{2}^{*} \! \approx \! 1/\sigma_{z}$ due to the $\delta \hat{\theta}$ term.\cite{Petta_Science05, Bluhm_PRL10} On the other hand, in a Hahn echo experiment,\cite{Petta_Science05, Bluhm_NP10} the influence of $\delta \hat{\theta}$ is removed, and the signal decay is due to $\hat{V}_H$ from Eq.~(\ref{eq:VH0}).  Since this interaction is a sum of two commuting terms from two uncoupled dots, the appropriately defined $S$-$T_{0}$ decoherence function is a product of the two single-dot decoherence functions.\cite{Bluhm_NP10, Neder_PRB11}
This observation establishes a one-to-one correspondence between single-spin Hahn echo decay\cite{Koppens_PRL08} due to hf-mediated interactions considered theoretically in Refs.~\onlinecite{Yao_PRB06, Cywinski_PRL09, Cywinski_PRB09, Cywinski_PRB10}, and the $J\! = \!0$ singlet-triplet Hahn echo decay.\cite{Petta_Science05, Bluhm_NP10}

%%%%%%%%%%%%%%%%%%%%%%%%%%%%%%%%%%%%%%%
%%% CALCULATIONS OF DEPHASING FUNCTIONS
%%%%%%%%%%%%%%%%%%%%%%%%%%%%%%%%%%%%%%%
\section{Calculations of $S$-$T_{0}$ dephasing in the absence of interdot effective field gradient} \label{sec:W_calculation_0}
In this section, we study two-spin decoherence within the $S$-$T_{0}$ subspace in the absence of interdot magnetic field gradient, i.e.,~$\theta_{\rm T} = 0$ (practically, this is true as long as $\theta_{\rm T} < \sigma_z$).  With $J \gg \sigma_z, \sigma_\perp$ (note that the magnitudes of $\hat{V}_{SS}$ and $\VST$ are negligible compared to $\sigma_{z}$, as shown in Appendix \ref{app:classical}), we can perform a second canonical transformation to diagonalize Hamiltonian (\ref{eq:Hfirst}), treating the off-diagonal terms as a perturbation.
We now obtain the final form of the effective Hamiltonian in a uniform effective magnetic field,
\bea
	\tilde{H}_{\theta_{\rm T} = 0} &=&\nonumber \left ( \begin{array}{cc}
		-J + \hat{V}_{SS} + \hat{H}_{A} + \hat{V}_{\delta\theta} & 0 \\
		0  & -\hat{H}_{A}-\hat{V}_{\delta\theta} \end{array} \right ) \\
		&&+ \sum_i \omega_{\alpha[i]} I^{z}_{i}
			+ \left ( \begin{array}{cc}  \HH_{B} &  0 \\
			0  &  \HH_{C} \end{array} \right ) \,\,, \label{eq:Hfinal_thetaT=0}
\eea
where
\bea
	\HH_A &=&  -\frac{1}{J} \sum_{i,j} B_{i}B_{j} I^{z}_{i} I^{z}_{j} = -\frac{\delta\hat{\theta}^2}{J} \,\,, \label{eq:Ha}\\
	\hat{V}_{\delta\theta} &=& -\frac{1}{J} \left( \{\hat{V}_{H}, \delta\hat{\theta} \} + [\hat{V}_{AH},\delta\hat{\theta}] \right ) \,\,.\label{eq:exact_Vtheta0}
\eea
To evaluate $\hat{V}_{\delta \theta}$, we use a $1/N$ type approximation where we neglect the commutators of nuclear operators.  The commutator in the above equation thus vanishes, so that
\beq
	\hat{V}_{\delta\theta} \approx
		v_{\theta} \, \delta \hat{\theta} \sum_{i,j} (A^{L}_{i}A^{L}_{j}-A^{R}_{i}A^{R}_{j}) (I^{+}_{i}I^{-}_{j} + I^{-}_{i}I^{+}_{j}) \,\,, \label{eq:Vtheta0}
\eeq
where
$$v_{\theta}=\frac{1}{8J} \left( \frac{1}{\EZT}+\frac{\EZT}{\EZT^{2}-J^{2}} \right) \,.$$

With $\theta_{\rm T} = 0$, $S$ and $T_0$ are two-spin eigenstates.  The dynamics of the $S$-$T_{0}$ coherence is quantified by the following decoherence function
\beq
W^{ST_{0}}(t) \equiv  \frac{\rho_{_{ST_{0}}}(t)}{\rho_{_{ST_{0}}}(0)} = \tr_{I}\left ( \hat{\rho}_{I} e^{i\HH_{T_{0}}t}e^{-i\hat{H}_{S}t} \right ) \,\, . \label{eq:W(t)}
\eeq
Here $\hat{\rho}_{I}(0)$ is the initial nuclear density operator, while $\HH_{S}$ and $\HH_{T_{0}}$ are the operators appearing on the diagonal in the effective Hamiltonian from Eq.~(\ref{eq:Hfinal_thetaT=0}). In Appendix \ref{app:experiment} we discuss how the above quantity can be measured in electrically controlled $S$-$T_0$ qubits.  Some of the information contained in $W_{ST_{0}}(t)$ function may also be indirectly inferred from the optical spectra of two-electron states in coupled self-assembled InGaAs quantum dots.\cite{Weiss_PRL12}

Our main goal is to quantify each of the dephasing processes as we vary $J$ and $\mu_{\rm T}$.  There are three kinds of hf-related terms appearing in $\tilde{H}_{\theta_{\rm T} = 0}$: (1) $\HH_{B}$ and $\HH_{C}$, the $I^{z}$-linear terms, (2) $\HH_{A}$, the square of the longitudinal Overhauser field, and (3) $\hat{V}_{SS}$ and $\hat{V}_{\delta\theta}$, which are second order in transverse Overhauser field.  Below we discuss their individual contributions to $S$-$T_{0}$ dephasing dynamics.  Such a treatment makes sense when the time scales on which various terms operate are very different.  If the importance of these interactions is comparable, treating various interactions as independent would introduce a quantitative error
when the commutator of the two competing terms is non-negligible.
In the following, we will present results for the envelope $W(t)$, in which we have removed the fast oscillating part of $W_{ST_{0}}(t)$ from Eq.~(\ref{eq:W(t)}):
\beq
W(t) \equiv e^{-iJt}W_{ST_{0}}(t) \,\, .
\eeq

%%%
%%% HB and HC
%%%
\subsection{Dephasing due to $\HH_{B}$ and $\HH_{C}$} \label{sec:HBC}
\begin{figure}[t]
	\includegraphics[width=0.9\linewidth]{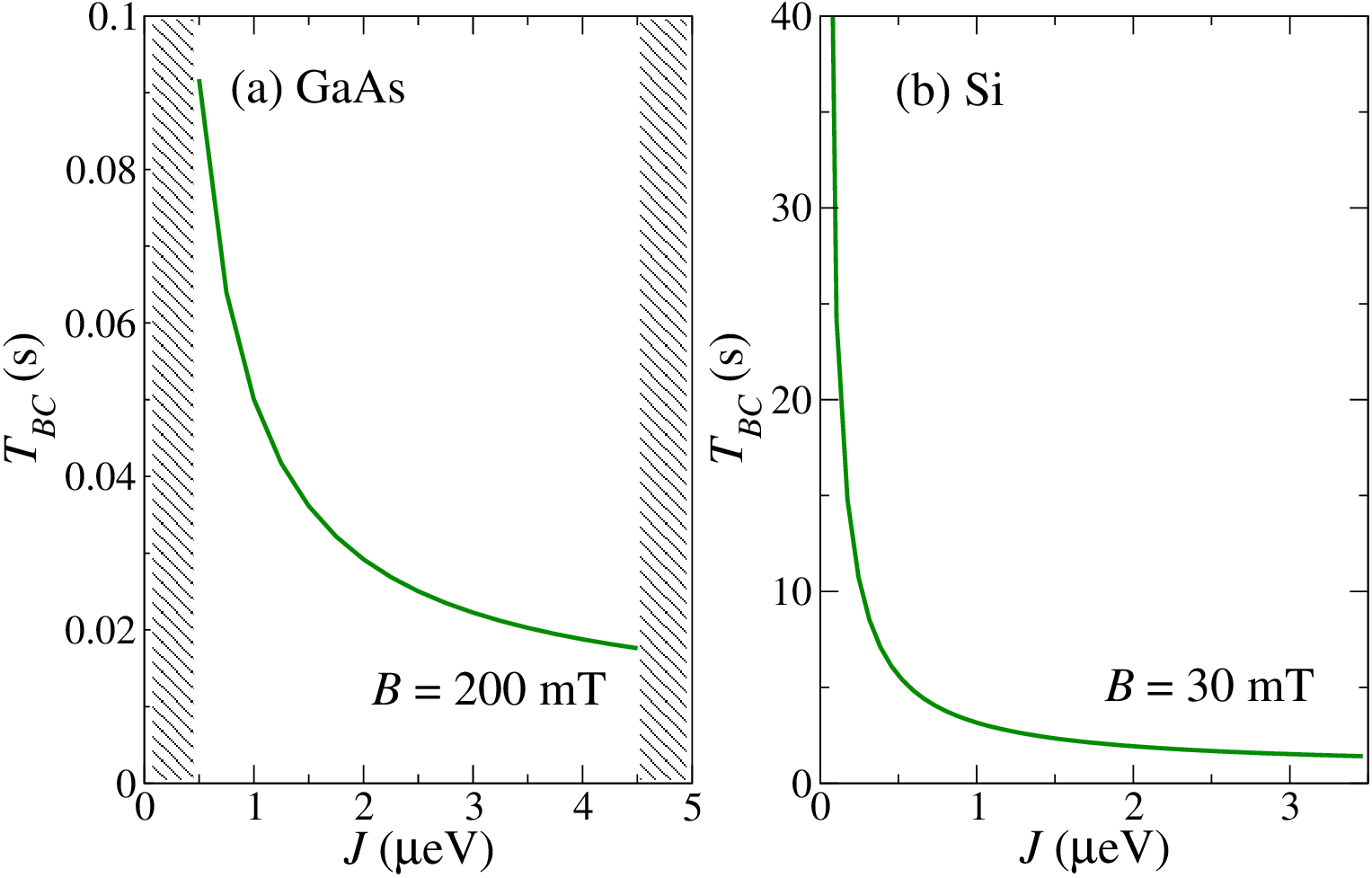}	
	\caption{(Color online) The times cale $T_{BC}$ of $S$-$T_{0}$ coherence decay due to $J^{z}$-linear terms (see Eq.~(\ref{eq:TBC})) for (a) GaAs DQD with $N_{L}\! = \! N_{R} \! =\! 10^{6}$ at $B\! = \! 200$ mT ($|\EZT| \approx 5.0~\mu$eV), and (b) DQD based on natural silicon with $N_{L}\! = \! N_{R} \! =\! 10^{5}$ at $B\! = \! 30$ mT ($|\EZT| \approx 3.5~\mu$eV). The regions in which  $\sigma_{\perp}\ll |J\pm \EZT|,|\EZT|$ is not fulfilled are hatched.}\label{fig:T2*}
\end{figure}

While $\HH_B$ and $\HH_C$ leads to dephasing that is completely analogous to the hf-induced inhomogeneous broadening of a single spin free evolution, the magnitude of the dephasing here is strongly suppressed because of the reduced coupling between the nuclei and the qubit.  From Eq.~(\ref{eq:W(t)}), keeping only $\HH_{B}$ and $\HH_{C}$ in the Hamiltonian, we obtain
\beq
W_{B,C}(t) =\text{Tr}_{\text{I}}\left ( \hat{\rho}_{I}(0) e^{i(\HH_{C}-\HH_{B})t} \right ) \,\, . \label{eq:WBC_average}
\eeq
Neglecting the wavefunction overlap, $\HH_{C}-\HH_{B} \! = \! \gamma ( \sum_{i\in L} A^{2}_{i}I^{z}_{i} + \sum_{i\in R} A^{2}_{i}I^{z}_{i})$, with
\beq
\gamma = \frac{J}{4\EZT(\EZT+J)}  \,\, .
\eeq
For a large number of nuclei ($N_{S} \gg 1$), as discussed in Sec.~\ref{sec:bath}, tracing over the nuclear spin density matrix in Eq.~(\ref{eq:WBC_average}) can be approximated by averaging over a classical Gaussian distribution of energy splittings with variance
\begin{eqnarray}
\sigma^{2}_{B,C} & = & \gamma^{2}\sum_{i}A_{i}^{4}\sigma^{2}_{i} \nonumber \\
& = & \frac{\gamma^{2}}{3} \sum_{\alpha} n_{\alpha} I_{\alpha}(I_{\alpha}+1) \mathcal{A}^{4}_{\alpha} \sum_{u} |\Psi(\mathbf{r}_{u})|^{8} \,\, .
\end{eqnarray}
where the variance for the $i$-th spin, $\sigma_{i}^{2} = \mean{(I^{z}_{i})^2} - \mean{I^{z}_{i}}^2$, is approximated by its value of $\frac{1}{3}I(I+1)$ at vanishing nuclear polarization $p$ (this is a good approximation at small $p$, since corrections are of the order $p^2$).
The value of $\sum_{i}A^{4}_{i}$ depends on the distribution of the nuclear couplings, i.e.,~the shape of electron wavefunction, but it can be roughly estimated as $1/N_{\Psi}^3$, as shown in Appendix \ref{app:rho}. The resulting coherence decay is then given by
\beq
W_{BC}(t) = e^{-(t/T_{2}^{*})^2} \,\, ,
\eeq
with
\beq
T_{BC} \approx \frac{\sqrt{6}}{|\gamma|n_{F} \sqrt{\sum_{\alpha}n_{\alpha} I_{\alpha}(I_{\alpha}+1) \mathcal{A}^4_{\alpha}}} \frac{1}{\sqrt{N^{-3}_{L}+N^{-3}_{R}}}  \,\, , \label{eq:TBC}
\eeq
where $N_{L}$ and $N_{R}$ are the numbers of unit cells in the L and R dots (as defined in Eq.~(\ref{eq:NT})).
The typical values of this time for GaAs and Si are shown in Fig.~\ref{fig:T2*}.

%%% FIGURE: T_BC

The very long $T_{BC}$ shown in Fig.~\ref{fig:T2*} is a clear illustration of the strongly suppressed electron-nuclear spin interaction, which is a consequence of the highly symmetric nature of the singlet and unpolarized triplet states.
In the following we generally neglect the contributions from $\HH_B$ and $\HH_C$. Other consequences of the reduced effective Knight shifts experienced by the nuclei will be discussed at the end of Sec.~\ref{sec:VSS}.

%%%
%%% H_A dephasing
%%%
\subsection{Dephasing due to the second-order longitudinal Overhauser field $\HH_{A}$} \label{sec:HA}

\begin{figure}[h]
	\includegraphics[width=0.9\linewidth]{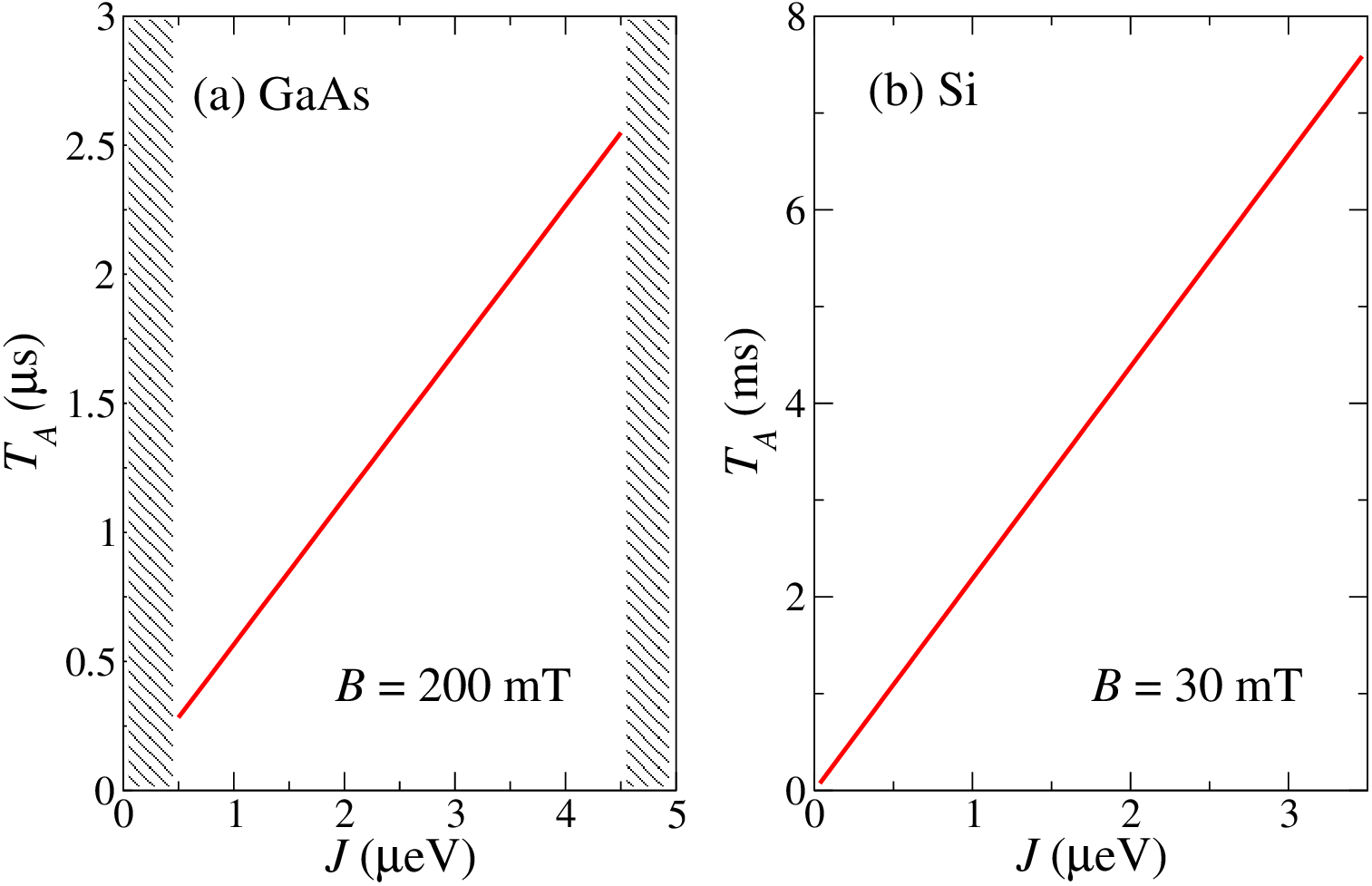}	
	\caption{(Color online) The time scale $T_{A}$ of $S$-$T_{0}$ coherence decay.
(a) GaAs DQD with $N_{L}\! = \! N_{R} \! =\! 10^{6}$ at $B\! = \! 200$ mT ($|\EZT| \approx 5.0~\mu$eV), and (b) natural Si DQD with $N_{L}\! = \! N_{R} \! =\! 10^{5}$ at $B\! = \! 30$ mT ($|\EZT| \approx 3.5~\mu$eV). In both cases the effective field gradient $\theta_{\rm T} \! = \! 0$. The hatched regions are where $\sigma_{\perp}\ll |J\pm \EZT|,|\EZT|$ is not fulfilled. No narrowing is assumed ($n_{F} \! = \! 1$).}\label{fig:TA}
\end{figure}

$\HH_{A}$-induced dephasing can be calculated classically because of the quasistatic nature of the longitudinal Overhauser field.  We can rewrite $\HH_{A}$ in terms of the classical Overhauser field $\theta = (h^{z}_{L}-h^{z}_{R})/2$, where $\mathbf{h}_{L/R} = \sum_{i} A^{L/R}_{i} \mathbf{I}_i$,
\beq
	H_{A} = -\frac{\theta^2}{J} \,.\label{eq:Ha_classical}
\eeq
As in Sec.~\ref{sec:bath}, we treat $\theta$ as a Gaussian random variable, and obtain the relevant decoherence function $W_{A}(t)$ by evaluating the Gaussian integral:
\beq
W_{A}(t) = \int P(\theta) \, e^{2 i \theta^2 t/J} d\theta \,\, ,
\eeq
where
\beq
P(\theta) = \frac{1}{\sqrt{2\pi} \sigma_{\theta}} e^{-\frac{\theta^{2}}{2\sigma^{2}_{\theta}}} \,\, ,
\eeq
with a distribution width $\sigma_{\theta} = \sigma_{z}/2$, where $\sigma_{z}$ is defined in Eq.~(\ref{eq:sigmaz_DQD}).  We then obtain (see also Ref.~\onlinecite{Yang_PRB08})
\beq
	W_{A}(t) = \frac{e^{\frac{i}{2}\arctan(\eta_{A} t)}}{\left(1+\eta_{A}^{2}t^{2} \right)^{1/4}} \,, \label{eq:WA_0}
\eeq
where we have defined $\eta_{A} \!=\! 4\sigma^{2}_{\theta}/J\!=\!\sigma^{2}_{z}/J$.  The characteristic decay time scale $T_{A}$ is defined by $|W_{A}(T_{A})|=1/e$, giving us
\beq
	T_{A} = \frac{e^{2}J}{\sigma^{2}_{z}} = \frac{e^{2}N_{D}J}{n^{2}_{F}\mathcal{A}^2} \,\,. \label{eq:TA0}
\eeq	
In Fig.~\ref{fig:TA} we show the typical values of this dephasing time for GaAs and Si.  These $T_A$ dephasing times are much shorter than $T_{BC}$ from the previous section, but are significantly longer than the single-spin inhomogeneous broadening dephasing time $T_2^*$, which is in the order of 10 ns in a GaAs QD.  As such, $T_A$ represents the inhomogeneous broadening of the pure $S$-$T_0$ two-level system at finite $J$ when $\theta_{\rm T} = 0$.  Qualitatively, $T_A$ is long because $\HH_A$ is of the second order in the Overhauser field fluctuations $\delta\theta$, while for single spins the inhomogeneous broadening is of the first order in $\delta\theta$.  The very long $T_A$, as compared with the short single-spin $T_2^*$, means that narrowing of the nuclear bath may not be necessary for $S$-$T_0$ qubits if $\theta_{\rm T} = 0$, though it is worth noting that $T_{A}$ does increase as $n_{F}^{-2}$ when the state of nuclear bath is narrowed (more precisely, when the distribution of $\delta \theta$ is narrowed).

%%% T_A figure

%%%
%%% V_SS dephasing
%%%
\subsection{Dephasing due to $\hat{V}_{SS}$}  \label{sec:VSS}

$\HH_A$, $\HH_B$, and $\HH_C$ lead to inhomogeneous broadening because longitudinal Overhauser fields are quasistatic.  On the other hand, $\hat{V}_{SS}$ and $\hat{V}_{\delta\theta}$ are due to transverse Overhauser fields, the dynamics of which cannot be neglected completely, and the influence of which cannot be fully removed by a Hahn echo (see Sec.~\ref{sec:echo}).  In this section we study $S$-$T_0$ dephasing induced by $\hat{V}_{SS}$.

According to our general approach, in which we treat each of the dephasing-inducing terms in the full Hamiltonian separately, we write the effective Hamiltonian as $\tilde{H}= (\HH_{Z} + \hat{V}_{SS})  \ket{S}\bra{S} + \HH_{Z}\ket{T_{0}}\bra{T_{0}}$. The Zeeman term is kept here, because while it does not lead to any dephasing by itself, it modifies the dynamics caused by $\hat{V}_{SS}$, as we will show below.

An important question is why we can neglect $\HH_{A}$, $\HH_{B}$, and $\HH_{C}$ terms in the current calculation.  These terms, apart from leading to $S$-$T_{0}$ dephasing in ways described above, in principle also modify dephasing caused by $\hat{V}_{SS}$ by giving different effective Knight shifts to nuclear spins at different locations in a quantum dot.  By neglecting these terms, the operators in $\hat{V}_{SS}$ are expressed in the interaction picture only with respect to the ``noninteracting'' part of the Hamiltonian.  Specifically, the nuclear operators in the interaction picture are:
\beq
I^{\pm}_{k}(t) = I^{\pm}_{k} e^{ \pm i \omega_{k} t } \,\,, \label{eq:Jpm}
\eeq
where $\omega_k$ are nuclear Zeeman energy of the $k$th nucleus.  The $\HH_{A}$, $\HH_{B}$, and $\HH_{C}$ terms would have modified this time dependence by changing the nuclear spin frequency (in particular, keeping the nonlinear in $I^{z}_{k}$ term $\HH_{A}$ would make the exact calculations much more complicated).  However, it turns out that the time scale on which the dependence of $\hat{I}^{\pm}_{k}(t)$ becomes visibly modified by the presence of one of these terms is typically much longer than the time scale of coherence decay due to either these terms by themselves or due to $\hat{V}_{SS}$. For example, including $\HH_{B}$ and $\HH_{C}$ in the interaction picture would lead to corrections in time-dependence of transverse nuclear operators on a time scale of $t_{B} \! \sim \! 4 N_{D}^{2}(\EZT+J)/\mathcal{A}^2$, which will turn out to be much larger than $T_{SS}$ (derived below) unless the number of nuclei is very small (which could happen, for example, in small QDs made of isotopically purified silicon). Similarly, one can quickly estimate that the time scale on which the corrections due to $\HH_{A}$ appear is much larger than $T_{A}$ given in Eqs.~(\ref{eq:TA0}), unless $N_{S}$ or $n_{F}$ are so small that $n_{F}\sqrt{N_{S}} \! < \! 1$.  In short, neglecting the $I^{z}$-dependent hf terms does not lead to any qualitative change in our calculations below, while it dramatically reduces the complexity (and therefore increases the transparency) of the calculation.

We now outline our calculation of dephasing due to $\hat{V}_{SS}$.  The decoherence function is
\beq
W_{SS}(t) = \tr_{J}\left\{ \hat{\rho}_{I} \mathcal{T}\left[e^{-i\int^{t}_{0}d\tau \mathcal{V}_{SS}(\tau)}\right] \right\}\,\,,  \label{eq:WSS_initial}
\eeq
where $\mathcal{T}$ is the time-ordering operator, and $\mathcal{V}_{SS}(\tau)$ is $\hat{V}_{SS}$ in the interaction picture with respect to $\hat{H}_{Z}$, so that the nuclear spin operators are given by Eq.~(\ref{eq:Jpm}).  The calculation is performed using the $1/N$-expansion-based RDT from Refs.~\onlinecite{Cywinski_PRL09,Cywinski_PRB09}.  In the interaction picture we have
\beq
	\mathcal{V}_{SS}(t) = v_{ss} \sum_{k, l} \phi_{k}\phi_{l}A_{k}A_{l} I^{+}_{k}(t) I^{-}_{l}(t)\,\, .
\eeq
where $\phi_{i} \! = \! 1$ for $i \in L$ and $\phi_{i} \! = \! -1$ for $i \in R$.  When averaging over the nuclear operators, we contract the nuclear indices in pairs, so that the sign factors drop out ($\phi_{i}^{2} \! = \! 1$).  The $T$-matrix defined in Ref.~\onlinecite{Cywinski_PRB09}, after taking into account the slightly different structure of the averaged exponent, is then given by
\beq
	T_{kl} = v_{ss}  2\sqrt{a_{k}a_{l}} A_{k}A_{l} \, e^{i\omega_{kl}t/2} \, \frac{ \sin \frac{\omega_{kl}\,t}{2}}{\omega_{kl}} \,\,, \label{eq:Tkl}
\eeq
where $\omega_{kl}=\omega_{k}-\omega_{l} $, and $a_{k} = {\langle I^{+}_{k}I^{-}_{k} \rangle}_{0} = \frac{2}{3}I_{k}(I_{k}+1)$ for an unpolarized nuclear spin bath. Applying the RDT, we re-sum the linked-ring terms from the perturbative expansion and obtain a general formula\cite{Cywinski_PRB09}:
\beq
	W_{SS}(t) = \prod^{N}_{m} \frac{e^{-i \arctan(\lambda_{m}(t))}}{\sqrt{1+(\lambda_{m} (t))^{2}}} \,\,,
\eeq
where $\lambda_{m}(t)$ are the eigenvalues of $T_{kl}$.  As discussed in Ref.~\onlinecite{Cywinski_PRB09}, we can simplify the calculation by introducing a ``coarse-grained'' $\tilde{T}$-matrix of $N_{I} \times N_{I}$ dimension, with $N_{I}$ being the number of distinct nuclear species $\alpha$, each of them having a distinct value of the Zeeman splitting $\omega_{\alpha}$. Note that within the used approximations (the neglect of $\HH_{A,B,C}$ terms) this step is an exact transformation, leading to
\beq
\tilde{T}_{\alpha\beta} = 2 v_{ss} \sqrt{a_{\alpha}a_{\beta}} \sqrt{n_{\alpha}n_{\beta}} \frac{\mathcal{A}_{\alpha}\mathcal{A}_{\beta}}{N_{D}} \, e^{i\omega_{\alpha\beta}t/2} \, \frac{\sin \frac{\omega_{\alpha\beta}\,t}{2}}{\omega_{\alpha\beta}}  \,\, . \label{eq:Tab}
\eeq

For a system with multiple species in the short-time limit ($t \! \ll \! 1/\omega_{\alpha\beta}$, when $\tilde{T}_{\alpha\beta} \! \sim \! \delta_{\alpha\beta}$), or for a system with a single nuclear species, we obtain
\beq
	W_{SS}(t \! \ll \! 1/\omega_{\alpha\beta}) \approx 	 \frac{e^{-i \arctan(\eta_{_{SS}} t)}}{\sqrt{1+(\eta_{_{SS}} t)^{2}}} \,\,, \label{eq:WSS_short}
\eeq
where
\begin{eqnarray}
\eta_{_{SS}} & = & |v_{ss}|  (\sum_{k \in L}a_{k} A^{2}_{k}+\sum_{k \in R}a_{k} A^{2}_{k}) \,\, \nonumber\\
& = & 2|v_{ss}| (\sigma^{2}_{\perp,L} + \sigma^{2}_{\perp,R}) \equiv 2|v_{ss}|\sigma^{2}_{\perp} \,\, .
\end{eqnarray}
The result of Eq.~(\ref{eq:WSS_short}) can also be obtained using a semiclassical quasistatic bath approximation, as shown in Appendix \ref{app:classical}, where one can see how  this expression follows from a product of two Gaussian averages (over the $h_x$ and $h_y$ components of the transverse Overhauser fields) of phase factors $\sim \exp(-i \xi h_{x,y}^2)$.
If the characteristic decay time $T_{_{SS}}$ defined by $|W_{SS}(T_{_{SS}})|\! = \! 1/e$ falls in this short-time regime, we obtain
\beq
	T_{SS}= \frac{\sqrt{e^{2}-1}}{2v_{ss}} \frac{1}{\sigma^{2}_{\perp}} = 2\sqrt{e^2-1}\frac{|\mu_{\rm T}^2 - J^2|}{J\sigma^{2}_{\perp}}
	%\sim \frac{\EZT^2-J^2}{J} \frac{N}{\mathcal{A}^2}
	\,\, . \label{eq:Tss}
\eeq

For a system with multiple nuclear species in the long-time limit, $t \! \gg \! 1/\omega_{\alpha\beta}$, the heteronuclear spin contributions in Eq.~(\ref{eq:Tkl}) become negligible compared to the homo-nuclear ones. Then $W_{SS}(t)$ can be approximated by a product of functions describing decoherence caused by each spin species treated separately:
\beq
	W_{SS}(t \! \gg \! 1/\omega_{\alpha\beta}) \approx \prod_{\alpha} W_{SS,\alpha}(t) \,\,, \label{eq:WSS_long}
\eeq
where the homonuclear contribution with the nuclear species $\alpha$ is given by
\beq
	W_{SS,\alpha}(t \! \gg \! 1/\omega_{\alpha\beta}) \approx \frac{e^{-i \arctan(\eta_{{SS},\alpha} t)}}{\sqrt{1+(\eta_{{SS},\alpha}t)^2}}	\,\,.
\eeq
with $\eta_{_{SS},\alpha} \! = \!2| v_{ss}| n_{\alpha}\mathcal{A}^{2}_{\alpha}/N_{D}$.  In GaAs, where all $\eta_{\alpha}$ have similar values, the estimate of the characteristic decay time (when it indeed falls in the $t \! \gg \! 1/\omega_{\alpha\beta}$ regime) is $T_{SS} \! \approx \! 1/\eta_{\alpha}$.  The asymptotic decay of the coherence function (for $t \! \gg \! \text{max}_{\alpha} \eta_{_{SS},\alpha}^{-1}$) depends then on the number of nuclear species, e.g., $W^{\text{long}}_{SS}(t) \sim t^{-3}$ in GaAs, while $W^{\text{long}}_{SS}(t) \sim t^{-1}$ in Si.
The dependence of the power-law character of the decay on the number of species in the long-time limit can be most easily understood from the classical averaging approach presented in Appendix \ref{app:classical}. According to the RDT calculation of free evolution, at long times, the nuclear flip flops between spins of different species can be neglected, and we can treat each species separately. For $K$ species, we can perform the $2K$-fold integration over independent Gaussian variables ($x$ and $y$ components of the Overhauser fields due to $K$ species), and each integral contributes one factor of $1/\sqrt{t}$ to the asymptotic behavior. This can be compared, for example, with the case of decay of Rabi oscillations of a single spin coupled to a nuclear bath,\cite{Dobrovitski_PRL09} in which an analogous average over a \emph{single} component of the Overhauser field lead to $\sim \! 1/\sqrt{t}$ asymptotic decay.

%%% TSS Figure
\begin{figure}[h]
	\includegraphics[width=0.9\linewidth]{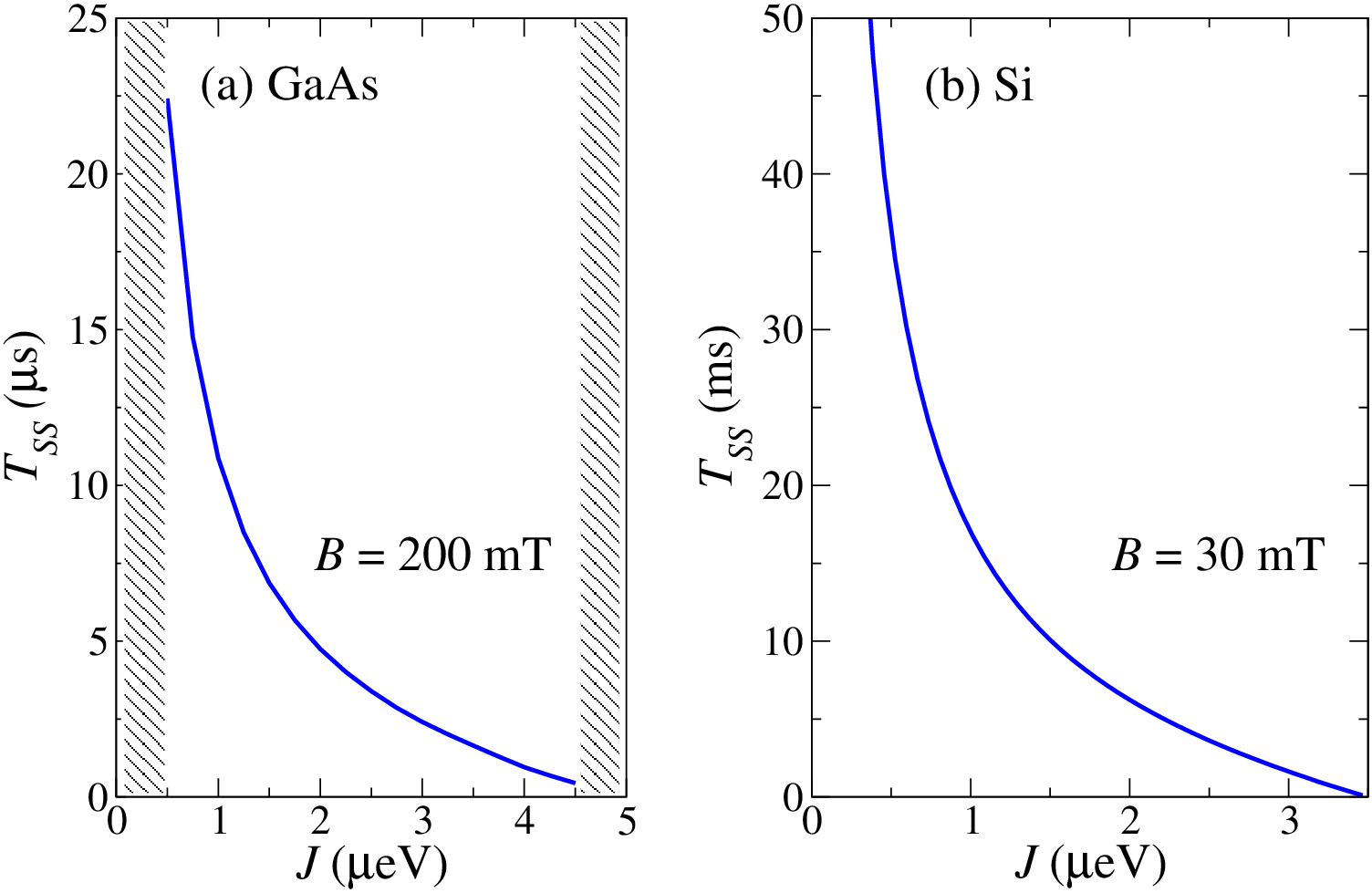}	
	\caption{(Color online) The time scale $T_{SS}$ of $S$-$T_{0}$ coherence decay due to the $\hat{V}_{SS}$ term for (a) GaAs DQD with $N_{L}\! = \! N_{R} \! =\! 10^{6}$ at $B\! = \! 200$ mT ($|\EZT| \approx 5.0~\mu$eV), and (b) DQD based on natural silicon with $N_{L}\! = \! N_{R} \! =\! 10^{5}$ at $B\! = \! 30$ mT ($|\EZT| \approx 5.0~\mu$eV). The diagonal hatching regions are where $\sigma_{\perp}\ll |J\pm \EZT|,|\EZT|$ is not fulfilled.}\label{fig:TSS}
\end{figure}

In Fig. \ref{fig:TSS} we plot the dephasing time $T_{SS}$ due to $\hat{V}_{SS}$ for GaAs and Si as a function of the exchange splitting $J$. This dephasing mechanism is enhanced by larger $J$ because of the $J$-dependence of the $v_{ss}$ coupling, which in turn reflects the fact that the $\hat{V}_{SS}$ term originates from the $J$-induced asymmetry between the two virtual processes shown in Fig. \ref{fig:VssVst}.  Numerically, $T_{SS}$ is longer than $T_A$ for smaller $J$, but shorter at larger $J$ values, so that it can become the dominant dephasing channel for an $S$-$T_0$ qubit in the large-$J$ regime.  We will discuss the crossover in more detail later in this section.

The suppression of Knight shifts experienced by the nuclei in a DQD (compared to a single spin) has implications beyond simplifying the dephasing calculation in this section.  For example, under certain conditions the transverse components of the Overhauser field can be treated by simply averaging the system's evolution over the distribution (\ref{eq:PO}) of classical Overhauser fields (see Appendix \ref{app:classical}).  For a single spin in a QD, this approach is valid when the evolution time $t$ is shorter than the inverse of the typical spread of nuclear spin Knight shifts,\cite{Merkulov_PRB02} i.e.,~$t \! \ll \! N_{\Psi}/\mathcal{A}$.\cite{Liu_NJP07,Cywinski_PRB09,Coish_PRB10}
Thus in this short-time limit, the exact shape of the electron's wavefunction is irrelevant.  On the other hand, when $t \! \gg \! N_{\Psi}/\mathcal{A}_{\alpha}$, this wavefunction shape has to be taken into account.  The qualitative form of the decay then changes, and the semiclassical approximation (with the nuclei replaced by a classical field or a precessing large spin) fails.\cite{Liu_NJP07,Cywinski_PRB09,Coish_PRB10}  However, for an $S$-$T_0$ qubit in a DQD with a finite $J$ and no $\theta_{\rm T}$, the strongly reduced effective Knight shift for the nuclei leads to a strong enhancement of the time scale on which the shape of the wavefunction is irrelevant.

Another consequence of the suppressed Knight shift is the diminished importance of the bath dynamics induced by dipolar interactions among the nuclear spins. For a single spin qubit this dynamics leads to
fluctuations of the qubit's energy splitting by the longitudinal Overhauser field operator $\sum_{i}A_{i}I^{z}_{i}$.  This dephasing mechanism causes a narrowed-state free induction decay (FID) on a time scale of $T_{\text{dip}} \geq 10$ $\mu$s in a typical GaAs QD\cite{Yao_PRB06, Witzel_PRB06, Witzel_AHF_PRB07, Witzel_PRB08} (while these predictions have not been yet verified, the theory given in these references correctly accounts for spin-echo decay measurements in GaAs\cite{Bluhm_NP10} and silicon\cite{Tyryshkin_JPC06, Abe_PRB10}).  For an $S$-$T_{0}$ qubit with a finite $J$, the time scale of coherence decay due to dipolar dynamics of nuclei is expected to be much longer than $T_{\text{dip}}$, which makes this mechanism irrelevant in the case of pure $S$-$T_0$ dephasing.

%%%
%%% V_Theta dephasing
%%%
\subsection{Dephasing due to $\hat{V}_{\delta\theta}$}  \label{sec:VTheta}

The Hamiltonian governing the $\hat{V}_{\delta\theta}$-induced dephasing takes the form $\HH \! = \! \hat{V}_{\delta\theta}(\ket{S}\bra{S} - \ket{T_{0}}\bra{T_{0}})$. According to Eq.~(\ref{eq:Vtheta0}), $\hat{V}_{\delta\theta}$ consists of two independent (commuting) terms related to the two dots in the DQD, thus the total decoherence function is a product of the single-dot decoherence functions: $W_{\delta\theta}(t)\! = \! W_{\delta\theta,L}(t) \times W_{\delta\theta,R}(t)$.

$\hat{V}_{\delta\theta}$ contains both transverse and longitudinal components of the Overhauser field, so that the full RDT calculation is rather cumbersome (the semiclassical calculation is also more involved than in the case of $\hat{V}_{SS}$, as shown in Appendix \ref{app:classical}). On the other hand, it turns out that in most cases dephasing caused by this term is much slower than the previously considered mechanisms (due to $\HH_{A}$ and $\hat{V}_{SS}$), and an exact calculation of coherence decay due to $\hat{V}_{\delta\theta}$ is of little practical relevance. Below, we present a calculation of a lower bound for the coherence time $T_{\delta\theta}$. This lower bound is obtained by replacing the $\delta\hat{\theta}$ operator in Eq.~(\ref{eq:Vtheta0}) with $\sigma_{z}$.  The decoherence function in this case is given by,
\beq
	W_{\delta\theta}(t) = \tr_{I}\left\{ \hat{\rho}_{I}\,
		\mathcal{\bar{T}}\left[e^{i\int^{t}_{0} \hat{V}_{\delta\theta}(\tau) d\tau}\right]
		\mathcal{T}\left[e^{-i\int^{t}_{0} \hat{V}_{\delta\theta}(\tau) d\tau}\right] \right\} \\
	%&=& \tr_{I}\left\{ \hat{\rho}^{I}(0)\, \mathcal{T}_{c} \left[e^{-i\int_{c} \mathcal{V}_{\theta}(\tau) d\tau}\right] %\right\} \,\,,	
\eeq
where $\mathcal{T} (\mathcal{\bar{T}})$ is the time (anti-)ordering operator, and $\hat{V}_{\delta\theta}(\tau)$ is in the interaction picture, with the nuclear spin operators given by Eq.~(\ref{eq:Jpm}).

Following a procedure analogous to the one described in the previous section, we obtain, for example,
\bea
	T^{L}_{kl}(t) \approx 4\,v_{\theta}\sigma_z \sqrt{a_{k}a_{l}} A_{k}A_{l}\, e^{i\omega_{kl}t/2}\, \sinc{\left(\frac{\omega_{kl} t}{2}\right)\,t}
		\label{eq:Tkl_theta}  \,\, ,
\eea
where the nuclear indices $k$ and $l$ refer to the nuclei in dot L and $\sinc{(x)}=(\sin{x})/x$. The formula for $ T^{R}_{kl}$ is analogous, albeit with a negative sign.  The total decoherence function can then be obtained as
\beq
	W_{\delta\theta}(t) = \prod^{N_{L}}_{m} \frac{e^{-i \arctan\left(\lambda^{L}_{m}(t)\right)}}{\sqrt{1+\left(\lambda^{L}_{m}(t)\right)^{2}}}
	\prod^{N_{R}}_{n}	\frac{e^{-i \arctan\left(\lambda^{R}_{n}(t)\right)}}{\sqrt{1+\left(\lambda^{R}_{n}(t)\right)^{2}}}  \,\,,
\eeq
where $\lambda^{L(R)}_{m}$ are the eigenvalues of $T^{L(R)}_{kl}$ matrices.

When $\omega_{kl}t \ll 1 $, or for a homo-nuclear system, $T^{L}_{\theta, kl} \approx 4v_{\theta}\sigma_z \sqrt{a_{k}a_{l}}A_{k}A_{l}t$, and the decoherence function can be simplified:
\beq
	W_{\delta\theta}\left(t\! \ll \! \omega_{\alpha\beta}^{-1} \right) \approx \frac{e^{-i \arctan\left(\eta^{L}_{\delta\theta} t \right)}}{\sqrt{1+ \left( \eta^{L}_{\delta\theta} t \right)^{2}}} \frac{e^{i \arctan \left( \eta^{R}_{\delta\theta} t \right)}}{\sqrt{1 + \left(\eta^{R}_{\delta\theta} t \right)^{2}}} \,\,, \label{eq:Wtheta}
\eeq
where $\eta^{L/R}_{\delta\theta}= 4v_{\theta}\sigma_z a_{k} \sum_{k\in L/R}A^{2}_{k} = 8v_{\theta}\sigma_{z} \sigma^{2}_{\perp,L/R}$.  For simplicity we assume a symmetric DQD ($\sigma_{\perp,L}=\sigma_{\perp,R}=\sigma_{\perp} /\sqrt{2}$), such that
\beq
	W_{\delta\theta}\left(t\! \ll \! \omega_{\alpha\beta}^{-1} \right) \approx \frac{1}{1+ \left(\eta_{\delta\theta}t \right)^{2}} \,\,,\label{eq:Wthetashort}
\eeq
where $\eta_{\delta\theta} = 4v_{\theta}\sigma_z\sigma^{2}_{\perp}$.  The characteristic decay time is then given by
\beq
T_{\delta\theta} \gtrsim  \frac{2J|\mu_{\rm T}|}{\sigma_{z}\sigma^{2}_{\perp}} \left | \frac{1-(J/\EZT)^{2}}{2-(J/\EZT)^{2}}\right | \,\,. \label{eq:Ttheta0}
\eeq

\begin{figure}[h]
	\includegraphics[width=0.9\linewidth]{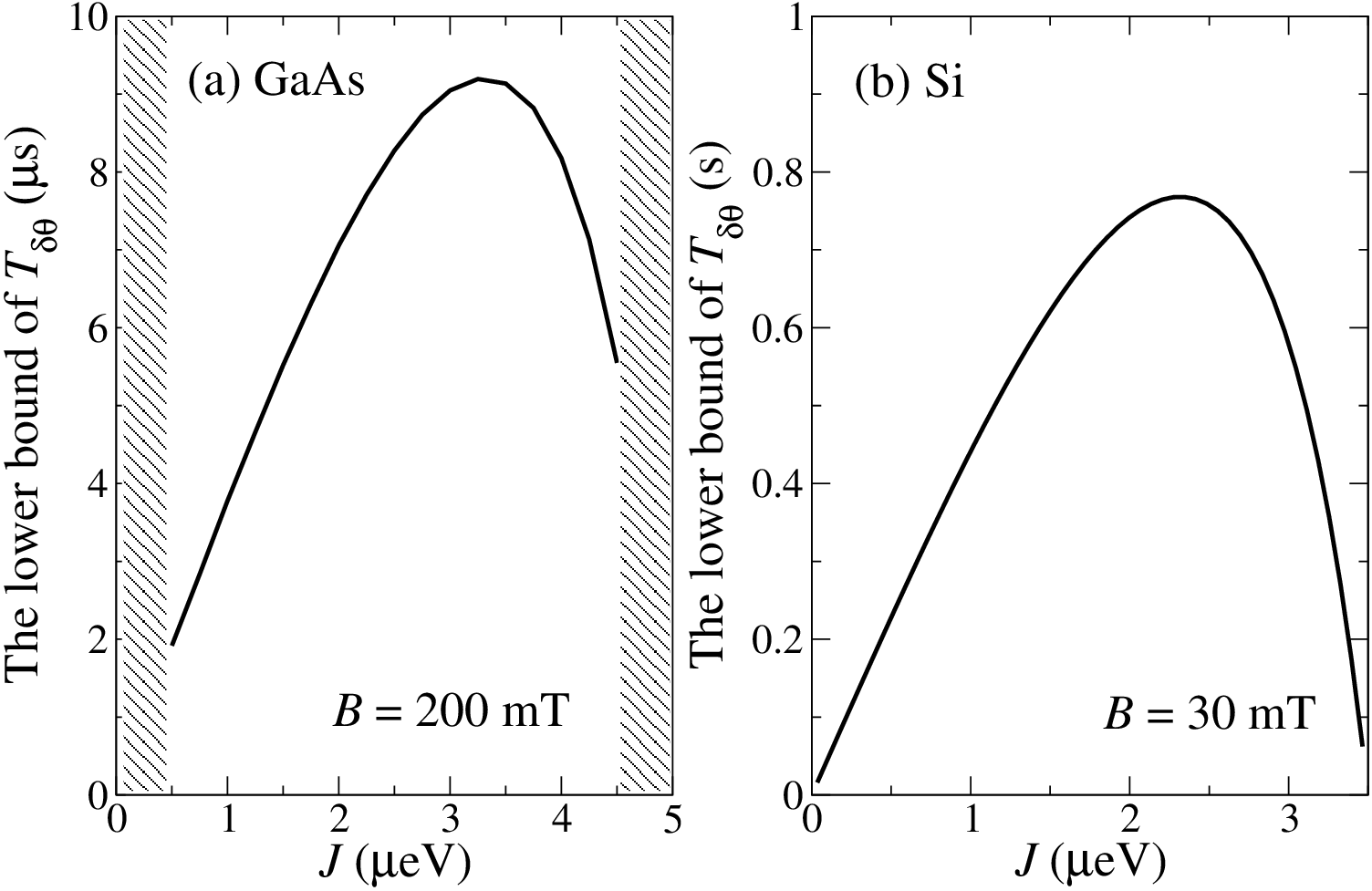} %{figure/t_theta.eps} as a function of J/\EZT
	\caption{Lower bound for $T_{\delta\theta}$ as a function of $J$ for (a) GaAs with $N_{L}=N_{R}=10^{6}$ at $B=200$ mT ($\EZT\approx 5.0~\mu$eV) and (b) natural Si with $N_{L}=N_{R}=10^{6}$ at $B=30$ mT ($|\EZT| \approx 3.5~\mu$eV). The diagonal hatching regions are where $\sigma_{\perp}\ll |J\pm \EZT|,|\EZT|$ is not fulfilled. No narrowing is assumed ($n_{F} \! = \! 1$).}\label{fig:Ttheta0}
\end{figure}

Although Eq.~(\ref{eq:Ttheta0}) is derived assuming a low $B$-field in a hetero-nuclear material (so that $T_{\delta\theta}$ is shorter than $\omega^{-1}_{\alpha\beta}$), it remains a good estimate of the characteristic decay time scale at higher fields in III-V materials. In Fig.~\ref{fig:Ttheta0} we show results of example calculations of $T_{\delta\theta}$ decay times for GaAs and Si.  The nonmonotonic behavior of $T_{\delta\theta}$ as a function of $J$ is due to dependence of the coupling strength $v_{\theta}$ on $J$. As shown in Fig.~\ref{fig:VssVst}, there are two terms contributing to $\hat{V}_{\delta\theta}$: one associated with virtual transitions involving $T_{-}$, the other involving $T_{+}$. When $J$ increases, the former is suppressed while the latter is enhanced, and a maximum in $T_{\delta\theta}(J)$ appears as a result of this competition.

%%%%%%%%%%%%%%%%%%%%%%%%%%%%%%%%%%%%%%%%%%%%%%%%%
%%% IDENTIFICATION OF DOMINANT DEPHASING CHANNELS
%%%%%%%%%%%%%%%%%%%%%%%%%%%%%%%%%%%%%%%%%%%%%%%%%
\subsection{Identification of the dominant dephasing mechanisms in GaAs and Si}  \label{sec:dominant}
Now that we have investigated the individual dephasing channels for an $S$-$T_0$ qubit by the hyperfine interaction, we would like to identify the dominant mechanisms in various regimes defined by the value of $B$ field $\mu_{\rm T}$, singlet-triplet splitting $J$, degree of nuclear reservoir narrowing $n_F$, dot sizes etc.  Below, we first list the approximate formulas for the characteristic dephasing times due to each of the above-considered mechanisms. For the sake of clarity, in these formulas we assume $N_{L} \! = \! N_{R} \! = \! N$, so that $N_{D} \! = \! N/2$.
Defining $r \! = \! J/\mu_{\rm T}$, we have
\begin{eqnarray}
T_{BC} & \approx & 4\sqrt{6}\left| \frac{1+r}{r} \right | \frac{|\mu_{\rm T}|}{\sigma_{z}} \frac{2N_{D}}{\mathcal{A}}  \,\,, \\
T_{A} & \approx & \frac{e^2 J}{\sigma_{z}^{2}} \,\, ,  \label{eq:TAsimple} \\
T_{SS} & \approx & 2\sqrt{e^2-1}\frac{\mu^{2}_{T}}{J\sigma^{2}_{\perp}}|1-r^2| \,\, ,  \label{eq:TSSsimple}  \\
T_{\delta\theta} & \gtrsim & \frac{2J|\mu_{\rm T}|}{\sigma_{z}\sigma^{2}_{\perp}}\left | \frac{1-r^2}{2-r^2} \right | \,\, .
\end{eqnarray}
Notice that while $T_{BC}$ decreases with increasing $J$, even at $r\approx 0.9$ it is still longer by a factor of $\sim \! |mu_{T}|/\sigma_{z} \! \gg\! 1$ than $N_{D}/\mathcal{A}$, which is $\approx \! 10$ $\mu$s ($\approx 1$ ms) in a typical GaAs (Si) dot with $N \! \approx \! 10^6$ ($10^{5}$).  We thus do not include $T_{BC}$ in the discussion below.

\begin{figure}[h]
	\includegraphics[width=0.9\linewidth]{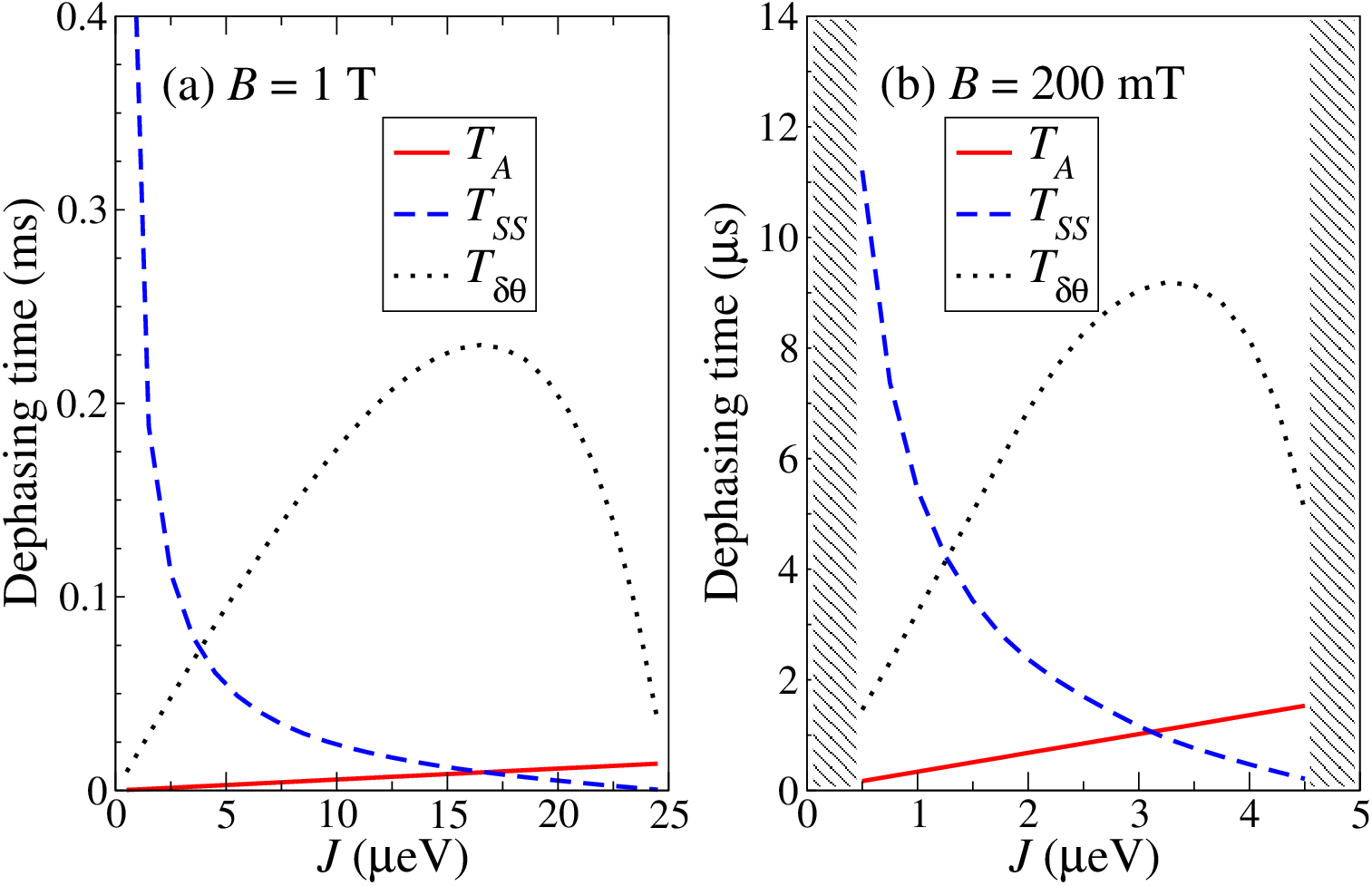}	
	\caption{(Color online) Dephasing times induced by $\HH_{A}$ (solid red line), $\hat{V}_{SS}$ (dashed blue line) and $\hat{V}_{\delta\theta}$ (dotted black line) as a function of $J/\EZT$ for GaAs at (a) $B=1$ T ($|\EZT| \approx 25~\mu$eV) and (b) $B=200$ mT ($|\EZT| \approx 5.0~\mu$eV), both with $N_{L}=N_{R}=10^{6}$, resulting in $\sigma_{z}=\sigma_{\perp} \approx 0.1$ $\mu$eV with no narrowing ($n_{F} \! = \!1$).}\label{fig:T_GaAs}
\end{figure}

\begin{figure}[h]
	\includegraphics[width=0.8\linewidth]{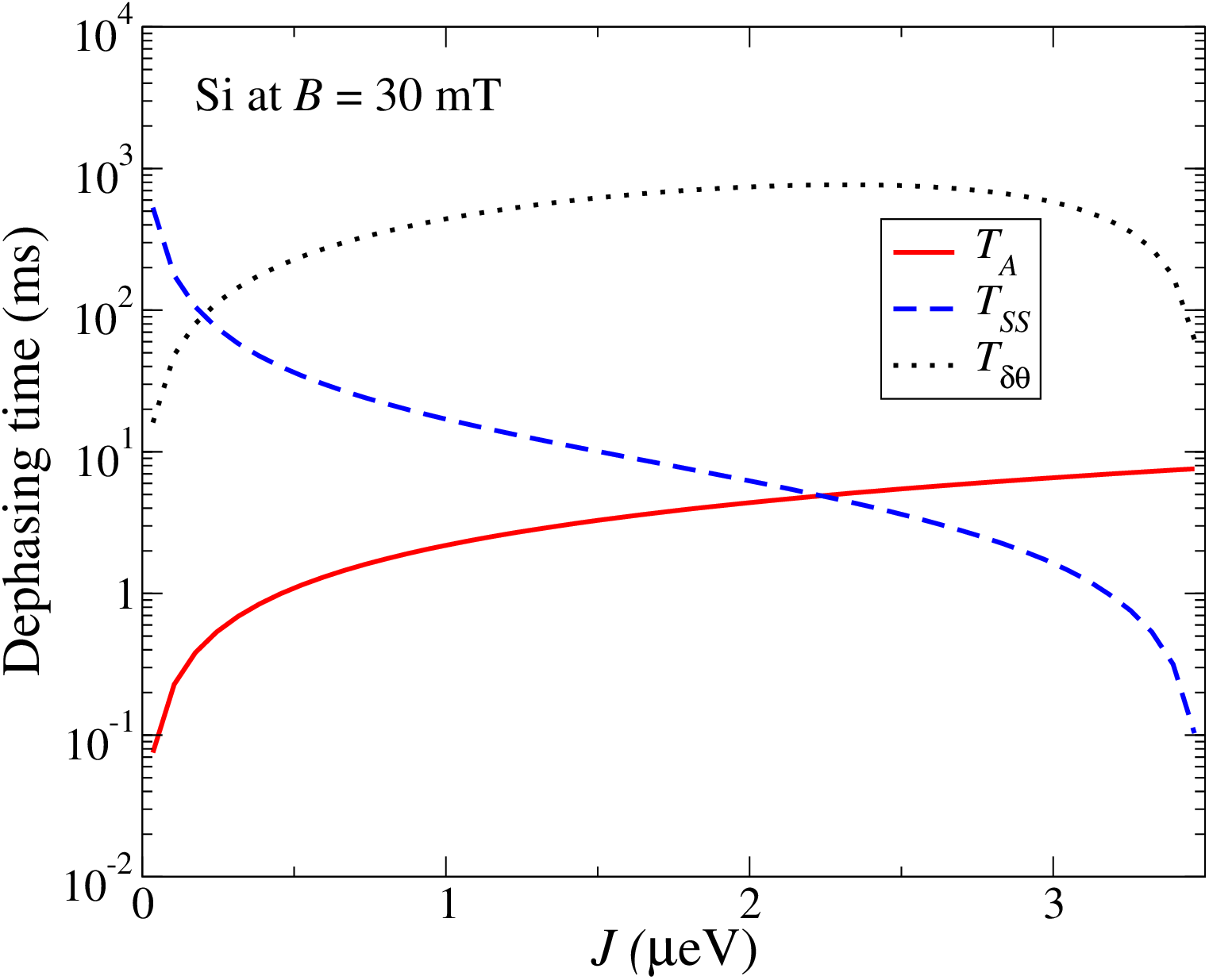}	
	\caption{(Color online) Dephasing times induced by $\HH_{A}$ (solid red line), $\hat{V}_{SS}$ (dashed blue line) and $\hat{V}_{\delta\theta}$ (dotted black line) as a function of $J/\EZT$ for Si at $B=30$ mT ($|\EZT| \approx 3.5~\mu$eV), both with $N_{L}=N_{R}=10^{5}$, resulting in $\sigma_{z}=\sigma_{\perp}\approx 1.5$ neV. No narrowing is assumed ($n_{F} \! = \! 1$). }\label{fig:T_Si}
\end{figure}

Without nuclear state narrowing, there is a competition between $\HH_{A}$ and $\hat{V}_{SS}$ induced dephasing.  At small enough $J$, dephasing due to $\HH_{A}$ dominates, while $T_{SS}$ becomes the shortest time scale at large $J$.  We show examples of calculations of all these times for GaAs and Si in Figs.~\ref{fig:T_GaAs} and \ref{fig:T_Si}, respectively.  From Eqs.~(\ref{eq:TAsimple}) and (\ref{eq:TSSsimple}) we can obtain $J_{\text{max}}$ for which the two mechanisms give the same dephasing time:
\beq
J_{\text{max}}  \approx 0.64 \mu_{\rm T}\,\,.
\eeq
The approximate sign here can be replaced by an equal sign in the case of a single-isotope material (or when decay occurs at a time scale shorter than the Larmor precession period of the nuclei). At $J_{\text{max}}$ the dephasing time is maximal: we anticipate a nonmonotonic $J$ dependence for the observed $T_{2}$ time.  The decoherence function should be well approximated by Eq.~(\ref{eq:WA_0}) for $J \! \ll \! J_{\text{max}}$, and by Eqs.~(\ref{eq:WSS_short}) and (\ref{eq:WSS_long}) for $J \! \gg \! J_{\text{max}}$.


When the nuclear state is narrowed with $n_{F}<1$, $T_{A}$ and $T_{\delta\theta}$ are extended by factors of $n_{F}^{-2}$ and $n_{F}^{-1}$, respectively.  With strong enough narrowing, the $\hat{V}_{SS}$-induced dephasing, i.e.,~dephasing due to hf-induced virtual transitions between $S$ and $T_{\pm}$ states, becomes the dominant source of hf dephasing in an $S$-$T_{0}$ qubit.

\subsection{Dephasing in an InGaAs DQD}
\begin{figure}[h]
	\includegraphics[width=0.9\linewidth]{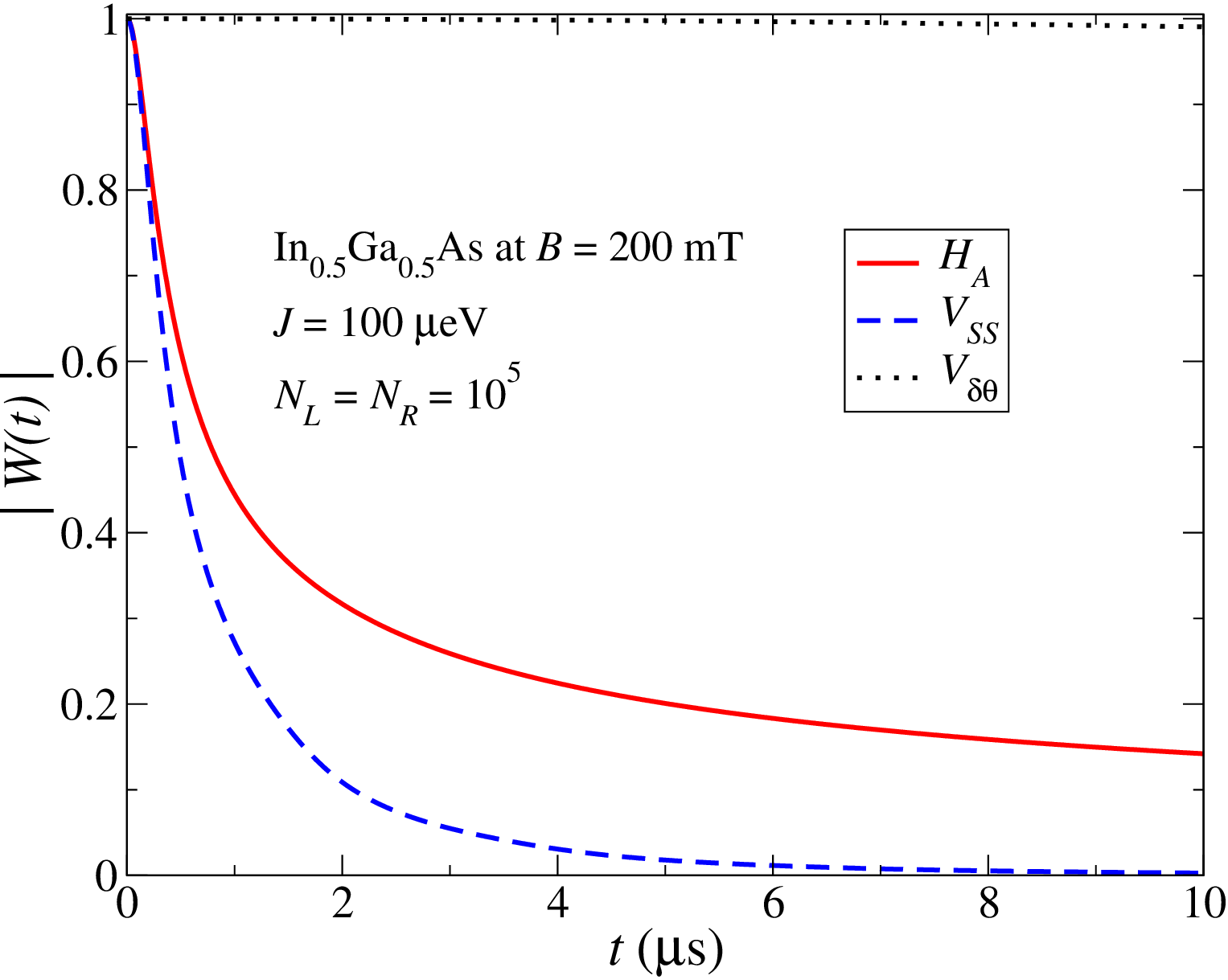}	
	\caption{(Color online) The decay of the decoherence function, calculated separately for each of the mechanisms, for In$_{0.5}$Ga$_{0.5}$As DQD with two electrons at $B \! = \! 200$ mT (assuming the $g$-factors of the electrons in both dots to be $\approx\! 0.5$), with $J\! = \! 100 $  $\mu$eV and for $N_{L} \!= \! N_{R} \! =10^{5}$. Here $|\EZT| \approx 5.8~\mu$eV, and $n_{F} \! =\! 1$.}\label{fig:InGaAs}
\end{figure}

We can apply the theory presented in this section to two electrons in a vertically stacked self-assembled InGaAs DQD, for which an estimate of the $S$-$T_{0}$ coherence time has been experimentally determined recently.\cite{Weiss_PRL12} Due to the strong interdot tunneling achievable in these structures, the values of $J$ are much larger than those in gated GaAs DQDs.  Most interestingly, one can investigate the $S$-$T_{0}$ coherence at the value of interdot detuning where $J$ is to first-order insensitive to the fluctuations of the electric bias (i.e.,~the charge noise), while still satisfying $J \gg \sigma_{z}$. In Ref.~\onlinecite{Weiss_PRL12}, at such an ``optimal point'' with respect to charge noise, $J_{\text{op}} \! \approx \! 100$ $\mu$eV, which is in fact larger even than the $\mu_{\rm T}$ splitting due to the applied $B$ field. While we have confined ourselves so far to the regime where $J \! < \! |\mu_{\rm T}|$, all our results are applicable for $J > |\mu_{\rm T}|$ (as long as we avoid the region of strong hyperfine-induced $S$-$T_{\pm}$ mixing --- in other words, $|J \pm \mu_{\rm T}| \gg \sigma_{\perp}$ has to be satisfied). In Fig.~\ref{fig:InGaAs}, we present the decoherence functions $W_A(t)$, $W_{SS}(t)$, and $W_{\delta\theta}(t)$ for a In$_{0.5}$Ga$_{0.5}$As DQD.  The dot parameters are $N_{L} \! = \! N_{R} \! = \! 10^{5}$, $J \! = \! 100$ $\mu$eV, and $B\! = \! 200$ mT (the electron $g$-factor is approximated by $g_{\text{eff}} \! \approx \! 0.5$).  For these parameters $H_{A}$ and $V_{SS}$ are of similar importance, and $V_{SS}$ dominates the long-time decay.  The calculated coherence time is of the order of a microsecond, which is in qualitative agreement with the lower bound of $200$ ns given in Ref.~\onlinecite{Weiss_PRL12}.

%%%%%%%%%%%%%%%%%%%%%%%%%%%%%%%%%%%%%%%%%%%%%%%%%%%%%%%%%%%%%%%%%%%%%%%
%%% CALCULATIONS OF DEPHASING FUNCTIONS IN THE PREsENCE OF THE GRADIENT
%%%%%%%%%%%%%%%%%%%%%%%%%%%%%%%%%%%%%%%%%%%%%%%%%%%%%%%%%%%%%%%%%%%%%%%
\section{Calculations of $S$-$T_{0}$ dephasing in the presence of an interdot magnetic field gradient} \label{sec:W_calculation_1}
In the presence of a finite field gradient $\theta_{\rm T} \gg \sigma_z$ (either from a nanomagnet, or due to a previously prepared Overhauser field gradient), one needs to obtain the new eigenstates that account for the $\theta_{\text{T}}$-induced mixing of $S$ and $T_{0}$ states, and then to re-derive the pure dephasing Hamiltonian in the new eigen-basis.
Nevertheless, the physical picture is quite clear here.  The mixing of $S$ and $T_{0}$ states means that in the new eigenstates, electron spin density in each dot does not vanish anymore.  As such the linear longitudinal Overhauser field $\delta \hat{\theta}$ would lead to dephasing between the eigenstates, similar to what happens to single spin qubits.  Indeed, if $\theta_{\rm T} \gg J$, the eigenstates would approach the product states again, so that we should recover dephasing of two independent spins exactly.  In this section, we focus on this transition from reduced dephasing in the highly-symmetric $S$-$T_0$ two-level system to the single-spin dephasing in the large-field-gradient limit.

\subsection{$S$-$T_{0}$ free-induction decay dephasing in the presence of an interdot magnetic field gradient}

At finite $\theta_{\rm T}$, the eigenstates of the first matrix in Eq.~(\ref{eq:HH}) are
\bea
\ket{S'} & = & \cos \gamma \ket{S} + \sin\gamma \ket{T_{0}} ,\nonumber \\
\ket{T_{0}'} & = & -\sin \gamma \ket{S} + \cos\gamma \ket{T_{0}}, \nonumber
\eea
where the mixing angle $\gamma$ is defined by
\beq
\tan 2\gamma = -\frac{2\theta_{\rm T}}{J} \,\,.
\eeq
The full range of $\gamma$ is from 0 to $\pi/4$, corresponding to limits of $\theta_{\rm T} \ll J$ and $\theta_{\rm T} \gg J$, respectively.  The corresponding eigenenergies are
\begin{eqnarray}
E_{S'} = -\frac{J}{2} -\frac{1}{2}\sqrt{J^{2}+4\theta^{2}_{\rm T}} = -\frac{J}{2}\left(1+\frac{1}{\cos 2\gamma} \right) \,, && \\
E_{T_{0}'} = -\frac{J}{2} +\frac{1}{2}\sqrt{J^{2}+4\theta^{2}_{\rm T}} = -\frac{J}{2}\left(1-\frac{1}{\cos 2\gamma} \right ) \,. &&
\end{eqnarray}

The complete dephasing Hamiltonian in the basis of $S'-T_0'$ is then
\begin{widetext}
\beq
	\tilde{H}^{'}_{ST_0} = \left ( \begin{array}{cc} E_{S'} & 0  \\ 0 & E_{T'_{0}} \end{array} \right )
	+ \left( \begin{array}{cc} \delta\theta \sin{2\gamma} &  \delta\theta\cos{2\gamma} \\
		\delta\theta\cos{2\gamma} &  -\delta\theta\sin{2\gamma} \end{array} \right)
	+ \left( \begin{array}{cc}  \!\!\hat{V}_{SS}\cos^{2}{\gamma} + \hat{V}_{\text{H}}\sin{2\gamma} \!\!&\!\!
			-  \hat{V}_{SS}\frac{\sin{2\gamma}}{2} + \hat{V}_{\text{H}}\cos{2\gamma}+\hat{V}_{\text{AH}} \!\!\\
			\!\!-\hat{V}_{SS}\frac{\sin{2\gamma}}{2} +  \hat{V}_{\text{H}}\cos{2\gamma}-\hat{V}_{\text{AH}} \!\!&\!\!
			\hat{V}_{SS}\sin^{2}{\gamma} - \hat{V}_{\text{H}}\sin{2\gamma} \!\!\end{array} \right),
%	&& + \left[ \begin{array}{cc} \HH_{B}\cos^{2}{\gamma}+\sin^{2}{\gamma}\HH_{C} & (\HH_{C}-\HH_{B})  \frac{\sin{2\gamma}}{2} \\
%			 (\HH_{C}-\HH_{B})\frac{\sin{2\gamma}}{2} &  \sin^{2}{(\gamma}{2}\HH_{C}+ \HH_{B}\sin^{2}{\gamma} \end{array} \right] \,\,.	
\label{eq:H'}		
\eeq
\end{widetext}
where we have neglected $\HH_{B}$ and $\HH_{C}$ since their influence on dephasing is very weak, as discussed before.

The influence of the off-diagonal terms in $\tilde{H}^{'}_{ST_{0}}$ could be included by performing another canonical transformation. However, the off-diagonal terms related to the transverse Overhauser fields generally only give subleading corrections to the diagonal terms, except for the case of $\theta_{\rm T} \gg J$ (when the diagonal $\hat{V}_{SS}$ term vanishes) or $\theta_{\rm T} \ll J$ (when the diagonal $\delta \theta$ and $\hat{V}_H$ terms vanish) limits.
Keeping only the lowest order terms (up to second order in hf coupling strength), we obtain the pure dephasing Hamiltonian in the $S'-T_0'$ space as
\begin{widetext}
\bea
	\tilde{H}^{'}_{ST_0} & \approx & \left ( \begin{array}{cc} E_{S'} & 0  \\ 0 & E_{T'_{0}} \end{array} \right )
	+ \left( \begin{array}{cc} \delta\theta \sin{2\gamma} - \frac{\delta\theta^2}{J} \cos^3{2\gamma} & 0 \\
		0 &  -\delta\theta\sin{2\gamma} + \frac{\delta\theta^2}{J} \cos^3{2\gamma} \end{array} \right) \nonumber \\
	& & + \left( \begin{array}{cc}  \!\!\hat{V}_{SS}\cos^{2}{\gamma} + \hat{V}_{\text{H}}\sin{2\gamma} & 0  \\
			0 & \hat{V}_{SS}\sin^{2}{\gamma} - \hat{V}_{\text{H}}\sin{2\gamma} \!\!\end{array} \right).
%	&& + \left[ \begin{array}{cc} \HH_{B}\cos^{2}{\gamma}+\sin^{2}{\gamma}\HH_{C} & (\HH_{C}-\HH_{B})  \frac{\sin{2\gamma}}{2} \\
%			 (\HH_{C}-\HH_{B})\frac{\sin{2\gamma}}{2} &  \sin^{2}{(\gamma}{2}\HH_{C}+ \HH_{B}\sin^{2}{\gamma} \end{array} \right] \,\,.	
\label{eq:H'_dephase}		
\eea
\end{widetext}
We can define $\HH_{A}' \! = \! -\delta\theta^{2}\cos^{3}{2\gamma}/J$, which becomes $\HH_{A}$ when $\theta_{\rm T}\! =\! 0$.

\begin{figure}[h]
	\includegraphics[width=0.9\linewidth]{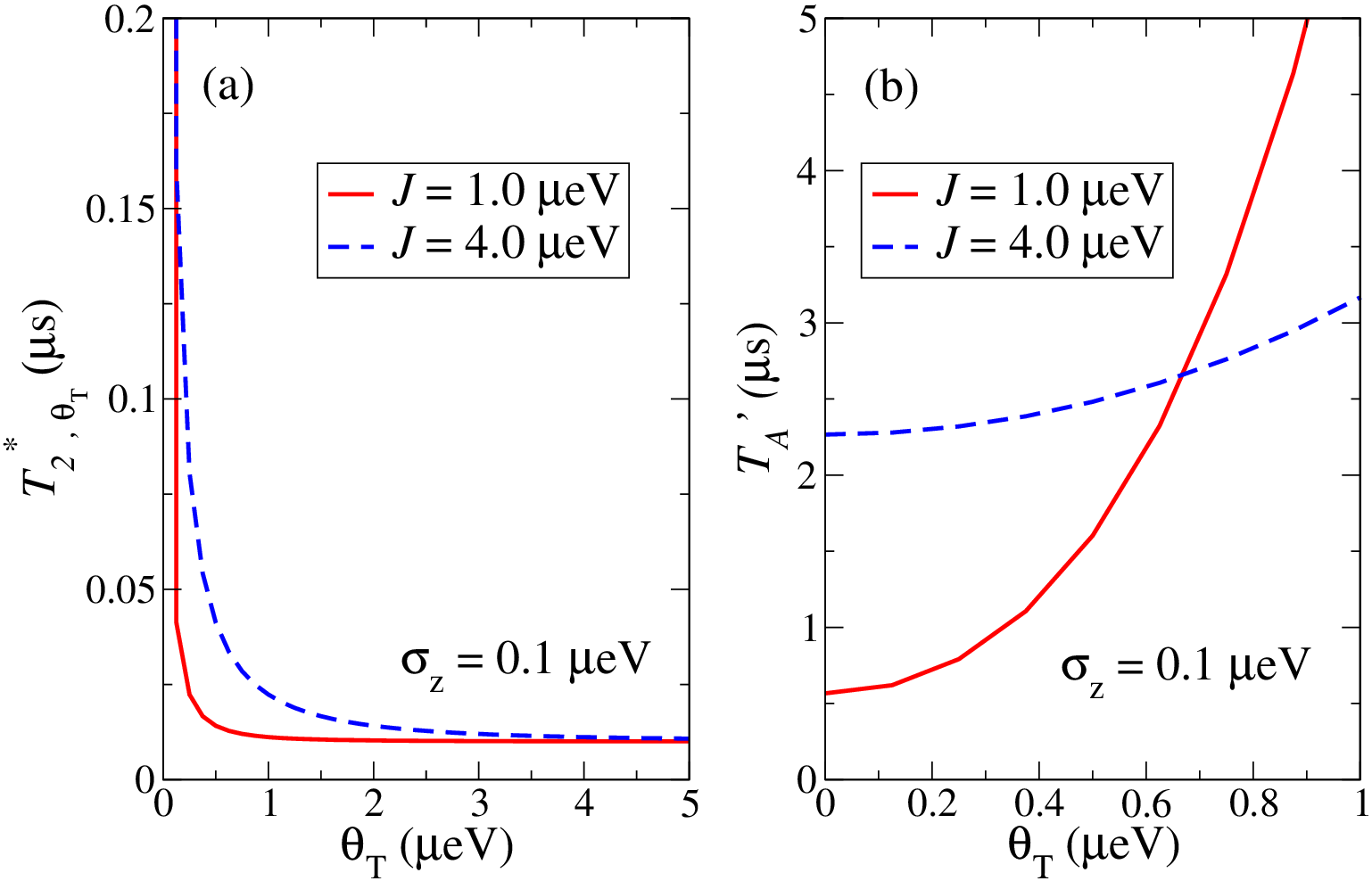}	
	\caption{(Color online) The dephasing time (a) $T^{*}_{2,\theta_{\rm T}}$ due to the $\delta \hat{\theta}$-linear terms and (b) $T'_{A}$ from the second canonical transformation in GaAs as a function of $\theta_{\rm T}$ at two fixed values of $J$. The rms of the Overhauser field difference between the dots is $\sigma_{z}\! = \! 0.1$ $\mu$eV. For $\theta_{\rm T} \gg J$, $T^{*}_{2,\theta_{\rm T}}$ is saturated at $\sim 10$ ns.}\label{fig:finite_gradient_z}
\end{figure}

The key new feature here is the re-appearance of the $\delta \hat{\theta}$-linear terms in the dephasing Hamiltonian.  The $\delta \hat{\theta}$-induced inhomogeneous broadening that plagues single-spin qubits is now back in action, though its effect is reduced when the field gradient $\theta_{\rm T}$ is small compared to the exchange splitting $J$ ({\em{}i.e.,} when the mixing angle $\gamma$ is small). The inhomogeneous broadening dephasing time due to the $\delta \hat{\theta}$-linear terms is given by
\beq
	T^{*}_{2,\theta_{\rm T}} = \frac{1}{|\sin{2\gamma}|}\frac{\sqrt{2}}{\sigma_{z}} \approx \frac{\sqrt{2}J}{4\sigma_{z}\theta_{\rm T}} \,\,, \label{eq:T2starDQD}
\eeq
where the approximate formula holds when $\gamma \! \ll \! 1$ (but for $\theta_{T}$ still larger than $\sigma_z$).  When $\gamma$ approaches $\pi/4$ (i.e.,~$\theta_{\rm T} \! \gg \! J$), $T^{*}_{2,\theta_{\rm T}}$ approaches the $T^{*}_{2} \sim 1/\sigma_z$ for a single spin in a QD, which is about $10$ ns for a typical GaAs QD (see Fig.~\ref{fig:finite_gradient_z}). On the other hand, for $\theta_{\rm T} \! \ll \! J$, $T^{*}_{2,\theta_{\rm T}} \propto J/\theta_{\rm T}\sigma_{z}$. As $\theta_{\rm T}$ decreases, $T^{*}_{2,\theta_{\rm T}}$ grows, until we reach the regime $\theta_{\rm T} \lesssim \sigma_z$, when dephasing due to the quadratic $\HH_A'$ term becomes more important.  In general, $\HH_A'$ leads to a characteristic dephasing time of
\beq
	T'_{A} = \frac{e^{2}}{\sigma^{2}_{z}}\frac{J}{\cos^{3}{2\gamma}} \,\,.
\eeq
When $\theta_{\rm T} \sim \sigma_z \ll J$ we have $\gamma \ll 1$, so that $\cos 2\gamma \sim 1$, and $T_A' \approx T_A$.
In Fig. \ref{fig:finite_gradient_z}, we plot $T^{*}_{2,\theta_{\rm T}}$ and $T'_{A}$ due to the longitudinal Overhauser fields as a function of $\theta_{\rm T}$ at fixed values of $J$. A transition of $T^{*}_{2, \theta_{\rm T}}$ from two-spin to single-spin coherence dynamics is shown in Fig. \ref{fig:finite_gradient_z}(a) as $\theta_{\rm T}$ approaches $J/2$.  When $\theta_{\rm T} \lesssim \sigma_{z}$, the resulting two-spin dephasing is essentially the same as in the previously discussed case of $\theta_{\rm T} = 0$. Instead of $T^{*}_{2}$, $T'_{A} \sim T_A$ now represents the longitudinal Overhauser field induced inhomogeneous broadening.

\begin{figure}[h]
	\includegraphics[width=0.9\linewidth]{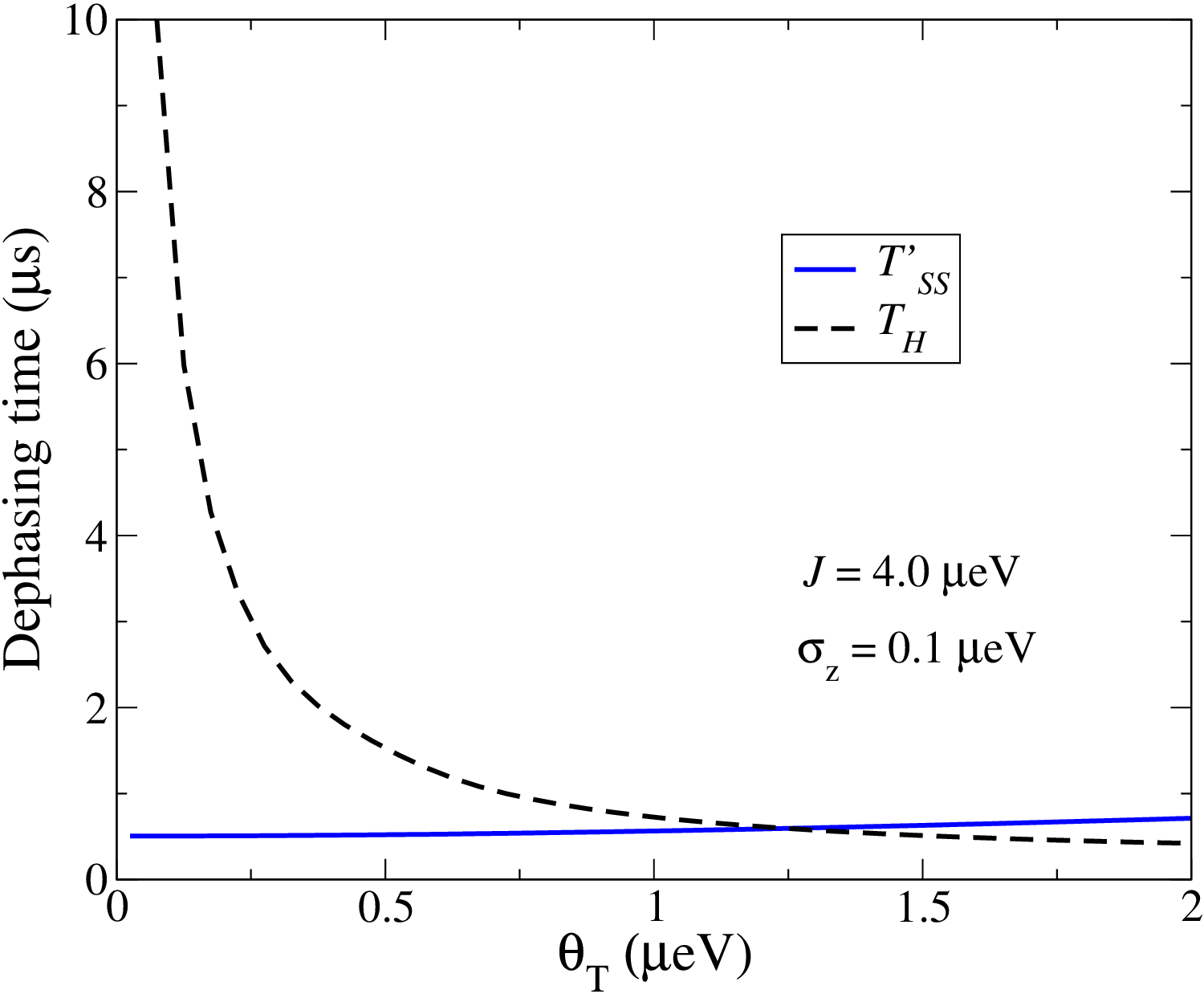}	
	\caption{(Color online) The dephasing times $T'_{SS}$ from Eq.~(\ref{eq:TSSprime}) and $T_{H}$ from Eq.~(\ref{eq:TH}), due to the terms given in Eqs.~(\ref{eq:VSSprime}) and (\ref{eq:VHtheta}), respectively, for GaAs DQD with $N_{L}\!=\! N_{R} \! = 10^{6}$ at $B \! = \! 200$ mT ($|\EZT| \approx 5.0~\mu$eV).
	}  \label{fig:finite_gradient_xy}
\end{figure}

The remaining terms on the diagonal are related to the previously discussed $\hat{V}_{SS}$ and $\hat{V}_{\delta\theta}$ terms. The dephasing due to the first of them,
\beq
 \left( \begin{array}{cc} \hat{V}_{SS}\cos^{2}\gamma &  0  \\
		0 &  \hat{V}_{SS}\sin^{2}\gamma \end{array} \right) \,\, , \label{eq:VSSprime}
\eeq
is calculated in the same way as in Sec.~\ref{sec:VSS} (compare also with the Hahn echo calculation in the next section), only with $\tilde{T}_{\alpha\beta}$ term in Eq.~(\ref{eq:Tab}) multiplied by $\cos 2\gamma$. The resulting dephasing time is
\beq
T'_{SS} = \frac{T_{SS}}{\cos 2 \gamma} \,\, , \label{eq:TSSprime}
\eeq
which shows that the dephasing due to the transverse Overhauser fields described by $\hat{V}_{SS}$ is suppressed by the presence of the field gradient $\theta_{\rm T}$. On the other hand, the influence of the interactions previously appearing in $\hat{V}_{\delta\theta}$ term is enhanced by the finite $\theta_{\rm T}$, similar to the inhomogeneous broadening due to $\delta \theta$.  The dephasing due to the term
\beq
\left( \begin{array}{cc} \hat{V}_{H}\sin 2\gamma &  0  \\
		0 & -\hat{V}_{H}\sin 2\gamma \end{array} \right) \,\, , \label{eq:VHtheta}
\eeq
is calculated in a way analogous to the one in Sec.~\ref{sec:VTheta}, only with $\sigma_{z}$ in Eq.~(\ref{eq:Tkl_theta}) replaced by $\theta_{\rm T}$. The coherence dynamics is thus described by Eqs.~(\ref{eq:Wtheta}) and (\ref{eq:Wthetashort}) in which $\eta_{\delta\theta}$ is replaced by $\eta_{H} \! = \! -\eta_{\delta\theta}J\sin 2\gamma/2\sigma_{z}$.  The resulting dephasing time is given by
\beq
T_{H} = \frac{2\sigma_{z}}{J}\frac{T_{\delta\theta}}{|\sin 2\gamma |}  \,\,.  \label{eq:TH}
\eeq
When $\theta_{\rm T} \! \ll \! J$, $T_{H} \! \approx \! \frac{\sigma_{z}}{\theta_{\rm T}} T_{\delta\theta}$, with $T_{\delta\theta}$ given by Eq.~(\ref{eq:Ttheta0}).  Clearly, this dephasing channel is strongly enhanced when $\theta_{\rm T} \! \gg \! \sigma_{z}$. As $\theta_{\rm T}$ increases, so that $\theta_{\rm T} \gg J \gg \sigma_z$, $T_H \rightarrow 2T_{\delta\theta}\sigma_{z}/J \! \ll \! T_{\delta\theta}$. In Fig.~\ref{fig:finite_gradient_xy}, we show examples of calculated values of $T'_{SS}$ and $T_{H}$. However, one can see that with exception of the regime of small $\theta_{T}$, these decay times due to the transverse Overhauser fields are much longer than the $T^{*}_{2,\theta_{T}}$ time from Eq.~(\ref{eq:T2starDQD}) shown in Fig. \ref{fig:finite_gradient_z}(a).

%%%%%%%%%%%%%%%%%%%%%
%%%%%%% Hahn Spin echo %%%%%%
%%%%%%%%%%%%%%%%%%%%%
\subsection{Spin echo for the $S$-$T_{0}$ qubit in the presence of an effective magnetic field gradient} \label{sec:echo}

A finite field gradient (i.e.,~$\theta_{\rm T} > \sigma_{z}$) can produce controlled rotations around the $x$ axis of the $S$-$T_{0}$ Bloch sphere when $J = 0$.  Such rotations allow faster preparation of superposition of $S$ and $T_{0}$ states (see Appendix \ref{app:experiment}), but more importantly, they allow the Hahn-echo experiment,\cite{Dial_PRL13} in which a $\pi$-rotation about the $x$ axis [from here on we will label a rotation of angle $\phi$ around the $\alpha$-axis as a $(\phi)_\alpha$ pulse] is performed in the middle of a free evolution (i.e.,~at $\tau \! \equiv \! t/2$).  If a qubit is initialized in the $S$ state, the full Hahn echo (HE) sequence would consist of a $(\pi/2)_x$-pulse (to prepare a $S$-$T_{0}$ superposition), free evolution for time $t/2$ in the finite-$J$ regime, then a $\pi_x$ pulse, another free evolution period of $t/2$, and finally another $(\pi/2)_x$ pulse, which prepares the qubit for the final projective measurement on the $S$ state. After simple manipulations (similar to the ones shown in Appendix \ref{app:experiment}), we arrive at the singlet probability at the end of the sequence:
\beq
P_{S}^{HE}(2\tau) = \frac{1}{2} - \frac{1}{4}\text{Tr}\left( \hat{\sigma}_{y}e^{-i\hat{H}\tau} \hat{\sigma}_{x}e^{-i\hat{H}\tau}\hat{\sigma}_{y}\hat{\rho}_{I}e^{i\hat{H}\tau}\hat{\sigma}_{x}e^{i\hat{H}\tau} \right ) \,\, ,
\eeq
where $\hat{H}$ is the total Hamiltonian in the $S$-$T_{0}$ subspace.  Now we rotate to the $S'-T_0'$ basis.  As we did previously, we again make a pure dephasing approximation for the Hamiltonian $\hat{H}'$ in the rotated basis, such that
\begin{eqnarray}
P_{S}^{HE}(2\tau) & = & \frac{1}{2} + \frac{1}{2}\cos^{2}2\gamma \,\text{Re}W^{HE}(2\tau) + \nonumber\\
& &  \frac{1}{2}\sin^{2}2\gamma\, \text{Re}W^{FID}(2\tau) \,\, ,
\end{eqnarray}
where
\beq
W^{HE}(2\tau) = \text{Tr}_{I} \left ( \hat{\rho}_{I} e^{i\tilde{H'}_{T_{0}}\tau}e^{i\tilde{H'}_{S}\tau} e^{-i\tilde{H'}_{T_{0}}\tau}e^{-i\tilde{H'}_{S}\tau}\right ) \,\, , \label{eq:WHE}
\eeq
is the truly ``echoed'' part of the signal, while
\beq
W^{FID}(2\tau) = \text{Tr}_{I} \left ( \hat{\rho}_{I} e^{2i\tau\tilde{H'}_{T_{0}}} e^{-2i\tau\tilde{H'}_{S}} \right ) \,\,
\eeq
is the component of the signal which decays just like the free evolution signal (Free induction decay, or FID). The presence of this term is due to the fact that the field gradient used for rotations about the $x$ axis is ``always on'' in the current-generation experiments.\cite{Dial_PRL13}   Here $\tilde{H}'_{S}$ and $\tilde{H}'_{T_0}$ are the diagonal terms of the $S'-T_0'$ pure dephasing Hamiltonian.  Using a rotated coordinate system on the Bloch sphere (with $+z'$ direction corresponding to a rotated $S'$ state), all the rotations are slightly tilted away from $x'$ axis.  The resulting pulse error leads to an appearance of a part of the signal which corresponds to evolution unaffected by the $\pi$ pulse.

Note that for $|2\theta_{\rm T}/J| \ll 1$, $2\gamma \ll 1$, and the FID-like part of the signal is just a small correction. Only at $2\,\theta_{\rm T} \approx J$ the two components contributing to $P_{S}(t)$ would be comparable. Since it is expected that $W^{HE}(t)$ exhibits slower decay than $W^{FID}(t)$, in this regime the echo decay time will be close to the FID decay time calculated previously. Below we focus on the $J \gg 2\,|\theta_{\rm T}|$ regime, and calculate the time dependence of $W^{HE}(t)$.

\subsubsection{Echo decay due to longitudinal Overhauser fields}
The Hahn echo sequence cancels completely any contribution to dephasing of $\tilde{H}'_{B,C}$ terms. Furthermore, the phase accumulated during the evolution due to the diagonal $\delta\hat{\theta}\,\sin 2\gamma$ and $\HH_A'$ terms is completely canceled.

\subsubsection{Echo decay due to the transverse Overhauser fields}
There are two terms in $\tilde{H}'$ due to transverse Overhauser fields.  The first is the modified $\hat{V}_{SS}$ interaction term:
\bea
\tilde{V}_{SS}' & = & \hat{V}_{SS} \left ( \begin{array}{cc} \cos^{2}\gamma & 0 \\ 0 & \sin^{2}\gamma   \end{array}  \right )  \nonumber \\
& =& \frac{\hat{V}_{SS}}{2} \left( \hat{I}' + \cos2\gamma \hat{\sigma}_z' \right)\,\, , \label{eq:VSSp}
\eea
where $\hat{I}'$ and $\hat{\sigma}_z'$ are identity and Pauli-$z$ matrix in the $S'-T_0'$ subspace.  The identity part of the interaction does not cause decoherence between the $S'$ and $T_0'$ states, and will be dropped from the following considerations.  The second is the large-$\theta_{\rm T}$ analogue of $\hat{V}_{\delta\theta}$ term considered previously:
\beq
\tilde{V}'_{\theta} = \hat{V}_{H} \sin 2\gamma \hat{\sigma}_{z}'  \approx - \frac{2\theta_{\rm T}}{J}\hat{V}_{H}\hat{\sigma}_{z}' \,\, . \label{eq:Vtp}
\eeq
Below we will consider their independent contributions to dephasing.

The calculations in both cases are straightforward modifications of spin echo theory of Refs.~\onlinecite{Cywinski_PRL09,Cywinski_PRB09,Neder_PRB11}. Following the derivation of Ref.~\onlinecite{Cywinski_PRB09}, the $W^{HE}(t)$ function from Eq.~(\ref{eq:WHE}) can be written as
\beq
W^{HE}(t) = \text{Tr}_{I} \left[ \hat\rho_{I} \mathcal{T}_{C} \exp \left( -i\int_{C}f(t'_{c})\tilde{\mathcal{V}}'(t'_{c}) \text{d}t'_{c}   \right )  \right ] \,\, ,
\eeq
where $\mathcal{T}_{C}$ is the operator ordering $\tilde{V}'(t_{c})$ on the closed time-loop contour,\cite{Cywinski_PRB09} $\tilde{V}'(t_{c})$ is the respective interaction term [from Eq.~(\ref{eq:VSSp}) or Eq.~(\ref{eq:Vtp})], $t_{c} \! = \! (t,c)$ is the time variable on the contour (with $c\! = \! \pm$), and $f(t_{c})$ is the temporal Hahn echo filter function defined on the contour.

As discussed previously, $\tilde{\mathcal{V}}'_{SS}(t'_{c})$ interaction is given by
\beq
\mathcal{V}'_{SS}(t'_{c}) = \frac{c}{2} \cos2\gamma \, v_{ss} \sum_{k, l} \phi_{k}\phi_{l}A_{k}A_{l} I^{+}_{k} I^{-}_{l} e^{i\omega_{kl}t'}\,\, .
\eeq
where $\phi_{i} \! = \! 1$ for $i \in L$ and $\phi_{i} \! = \! -1$ for $i \in R$. The corresponding filter function is
\beq
f(t'_{c}) = \Theta(t-t')\Theta(t'-t/2) - \Theta(t/2-t')\Theta(t')   \,\, , \label{eq:filter}
\eeq
where $\Theta(x)$ is the Heaviside function. The resulting $\tilde{T}$ matrix is
\beq
\tilde{T}_{\alpha\beta} = v_{ss}\cos 2\gamma \sqrt{a_{\alpha}a_{\beta}}\sqrt{n_{\alpha}n_{\beta}}\frac{\mathcal{A}_{\alpha}\mathcal{A}_{\beta}}{N_{D}} \frac{4ie^{i\omega_{\alpha\beta}t/2}}{\omega_{\alpha\beta}} \sin^{2} \frac{\omega_{\alpha\beta}t}{4} \,\, ,
\eeq
where $N_{D} = 1/(N_{L}^{-1}+N_{R}^{-1})$. Following Ref.~\onlinecite{Cywinski_PRB09}, we obtain
\beq
W^{HE}_{SS}(t) = \frac{1}{1+\frac{1}{2}R_{SS}(t) } \,\, , \label{eq:WHESS}
\eeq
with
\beq
R_{SS}(t) = 16 v^{2}_{ss} \cos^{2}2\gamma \sum_{\alpha \neq \beta} a_{\alpha}a_{\beta} n_{\alpha}n_{\beta} \frac{\mathcal{A}^{2}_{\alpha}\mathcal{A}^{2}_{\beta}}{N_{D}^{2}}  \frac{\sin^{4} \frac{\omega_{\alpha\beta}t}{4}}{\omega^{2}_{\alpha\beta}} \,\, . \label{eq:RSS}
\eeq
We thus see that the Hahn echo $S$-$T_{0}$ coherence decay due to $\hat{V}_{SS}$ interaction is of the same form as the echo decay of a single spin coherence (cf.~Refs.~\onlinecite{Cywinski_PRL09,Cywinski_PRB09}).

\begin{figure}[h]
	\includegraphics[width=0.9\linewidth]{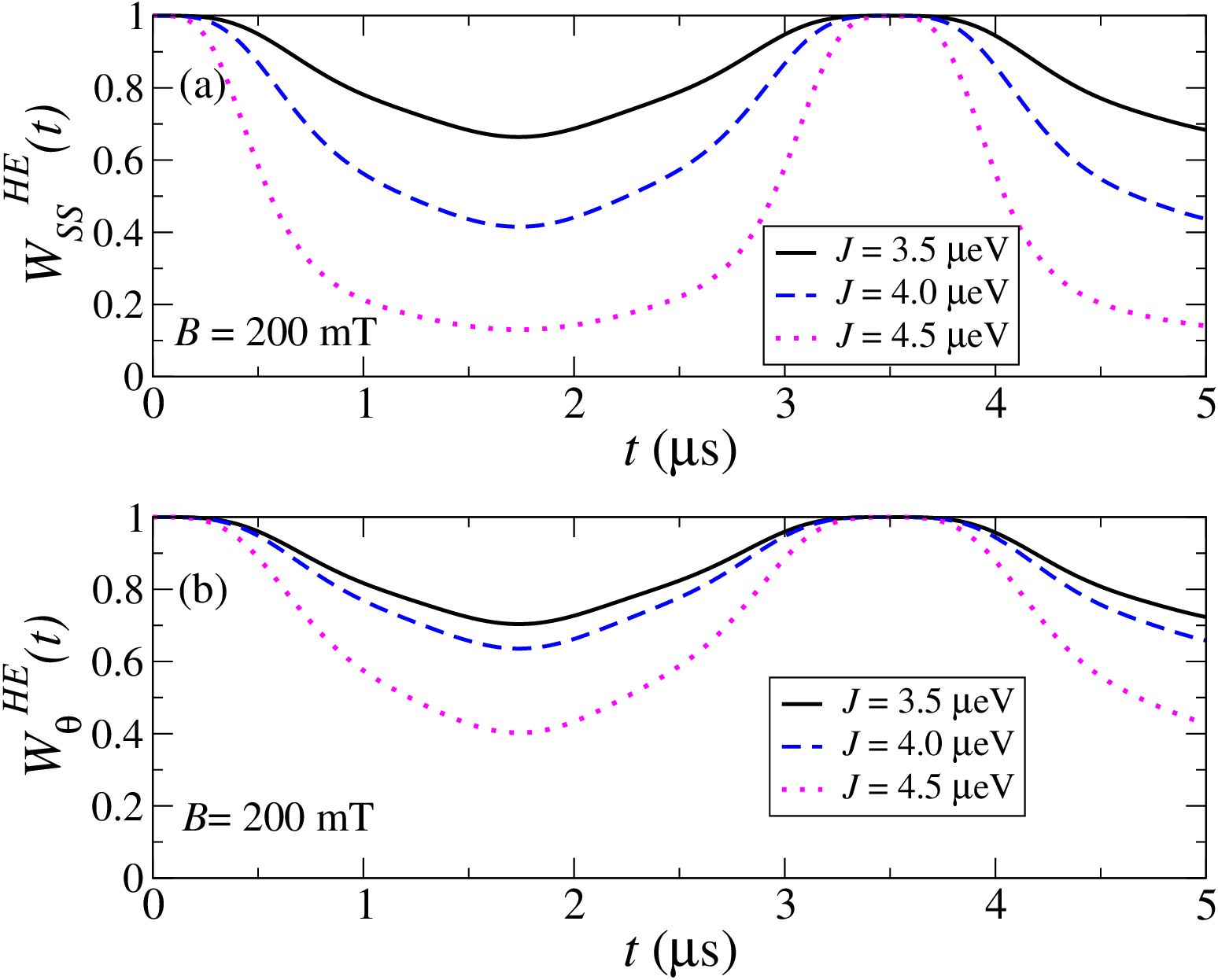}
	\caption{(Color online) (a) $W^{HE}_{SS}(t)$ and (b) $W^{HE}_{\theta}(t)$ for GaAs DQD with $N_{L}\!=\!N_{R}\!=\!10^{6}$ at $B=200$ mT and $\theta_{\rm T}=40$ mT with various values of $J$. Here, $|\EZT|\approx 5.0~\mu$eV.}\label{fig:echoB02}
\end{figure}

\begin{figure}[h]
	\includegraphics[width=0.9\linewidth]{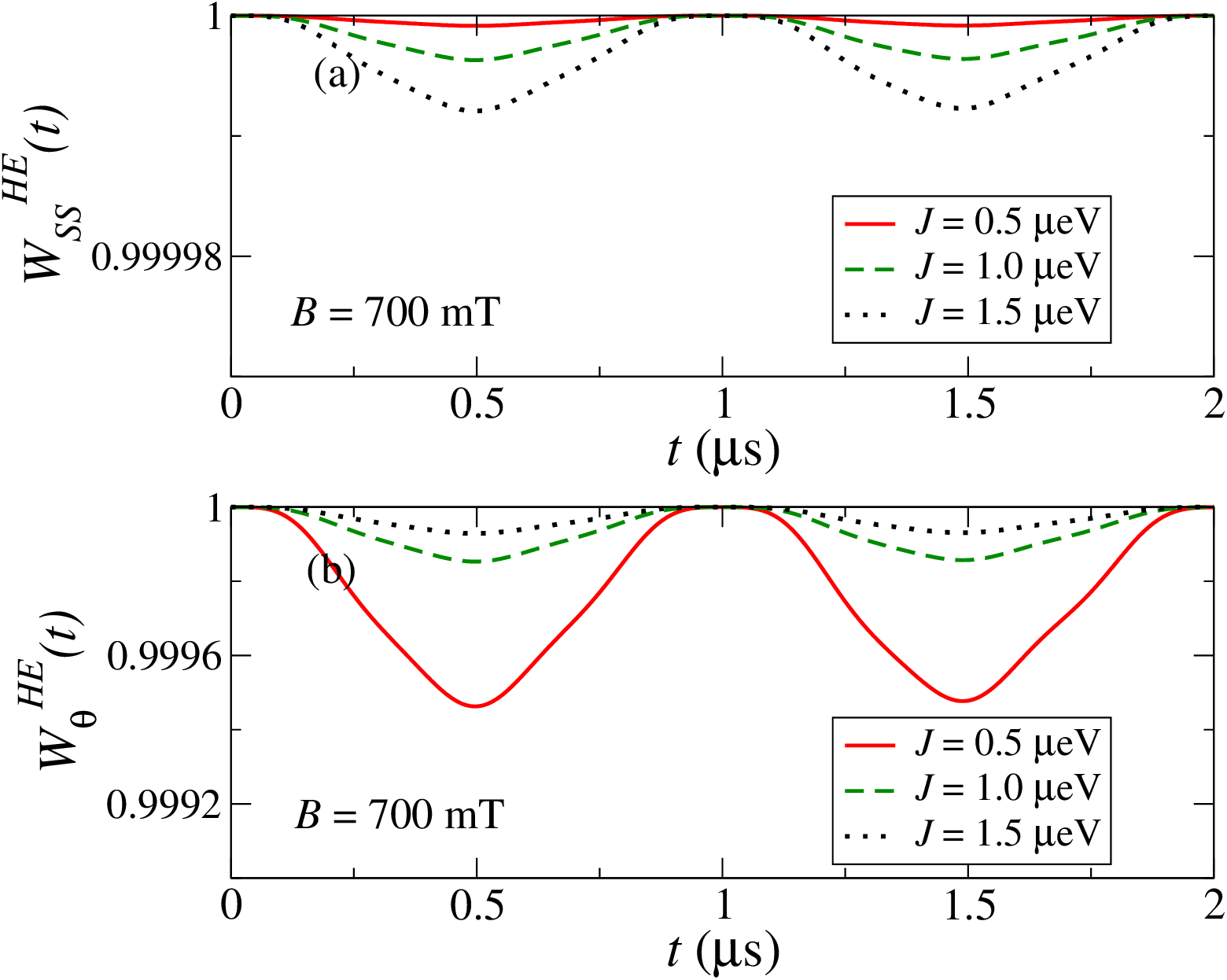}
	\caption{(Color online) (a) $W^{HE}_{SS}(t)$ and (b) $W^{HE}_{\theta}(t)$ for GaAs DQD with $N_{L}=N_{R}=10^{6}$ at $B\!= \! 700$ mT and $\theta_{\rm T}\! = \! 2.5$ mT with various values of $J$. The parameters here are chosen to correspond closely to the ones employed in recent experimental work from Ref.~\onlinecite{Dial_PRL13}.}\label{fig:echoB1}
\end{figure}

For $\tilde{\mathcal{V}}'_{\theta}(t'_{c})$ interaction, we have
\beq
\tilde{\mathcal{V}}'_{\theta}(t'_{c}) = c \sin 2\gamma \! v_{H} \sum_{k,l}(A^{L}_{k}A^{L}_{l} - A^{R}_{k}A^{R}_{l}) (I^{+}_{k}I^{-}_{l} + I^{-}_{k}I^{+}_{l} ) e^{i\omega_{kl}t'} \,\, ,
\eeq
and the filter function is given by Eq.~(\ref{eq:filter}).
Since $\tilde{\mathcal{V}}'_{\theta}$ is a sum of two commuting terms (under our approximation of neglect of the overlap of L and R wavefunctions even at finite $J$), $W^{HE}(t)$ is a product of two functions, each corresponding to one of the dots. Assuming symmetric dots, i.e.,~$N_{L} \! = \! N_{R} \! = \! 2 N_{D}$, we have then
\beq
W^{HE}_{\theta}(t) = \frac{1}{[1+\frac{1}{2}R_{\theta}(t)]^2} \,\, , \label{eq:WHEtheta}
\eeq
with
\beq
R_{\theta}(t) = 32 \sin^{2}2\gamma \,\, \! v_{H}^{2} \sum_{\alpha \neq \beta} a_{\alpha}a_{\beta} n_{\alpha}n_{\beta} \frac{\mathcal{A}^{2}_{\alpha}\mathcal{A}^{2}_{\beta}}{N_{D}^{2}}  \frac{\sin^{4} \frac{\omega_{\alpha\beta}t}{4}}{\omega^{2}_{\alpha\beta}} \,\, . \label{eq:Rtheta}
\eeq
The decoherence functions due to both of the above interactions are shown for GaAs in Figs. \ref{fig:echoB02} and \ref{fig:echoB1}. The Hahn echo signal shows more pronounced oscillations at lower $B$ fields (see Fig.~\ref{fig:echoB02}), when the processes of virtual flip flops between the $S$-$T_{0}$ subspace and the polarized triplets have more importance. Let us mention that in a recent experiment\cite{Dial_PRL13} on Hahn echo in the large $J$ regime, a decay of echo signal on time scale of a few microseconds was seen at $B\! = \! 0.7$ T. Comparison with results presented in Fig.~\ref{fig:echoB1}, which show only visibility loss of less than one percent even at very large $J$, clearly supports the main claim of Ref.~\onlinecite{Dial_PRL13} that charge noise (specifically the fluctuations of $J$) is the dominant source of dephasing in this experiment.

%%%%%%%%%%%%%%%%%%%%%%%
%%% EXCHANGE GATE ERROR%%%%%%
%%%%%%%%%%%%%%%%%%%%%%%
\section{Exchange gate error}  \label{sec:gate}

So far we have focused on hf-induced decoherence for an $S$-$T_0$ qubit with a finite exchange splitting $J$.  In the case of single-spin qubits, exchange interaction plays the crucial role of generating two-qubit gates such as SWAP and Controlled-NOT gates.\cite{Loss_PRA98}  In this section we evaluate how hf-interaction affects the fidelity of such two-qubit gates.  To allow a simple SWAP gate between two single-spin qubits, we assume there is no field gradient between the double dot: $\theta_{\rm T} = 0$.

Let us first make a qualitative examination of the gate error problem.  Recall that the magnitude of the exchange splitting $J$ is generally between 0.1 and 10 $\mu$eV, corresponding to a gate time $T_g$ between 40 and 0.4 ns.  Comparing the short gate time with the decoherence times we have calculated in the previous two sections, the only factor that could significantly impact the fidelity of a SWAP gate within the $S$-$T_0$ subspace is the fluctuations in the Overhauser field difference, $\delta \theta$.  In the absence of a field gradient, $S$-$T_0$ states are affected by $\delta \theta$ only at the second order level through $\HH_A$, as we have shown in Sec.~\ref{sec:W_calculation_0}.  On the other hand, the dephasing Hamiltonians for the polarized triplet states, $\HH_{T_+}$ and $\HH_{T_-}$ [given by Eqs.~(\ref{eq:HT+}) and (\ref{eq:HT-}) in Appendix~\ref{app:other_Heff}, respectively], do contain $\delta \mu$, the fluctuations in the average Overhauser field.  One can reason then that the SWAP gate fidelity would suffer much more severely if polarized triplet states are involved.

We focus on the fidelity of a SWAP gate in a two-electron DQD, which is defined as
\beq
	F \equiv \langle\psi_{\text{in}}|\hat{U}^{\dagger}_{\text{SWAP}}\,  \hat{\rho}_{\text{out}}\, \hat{U}_{\text{SWAP}} |\psi_{\text{in}}\rangle \,\,,
	\label{eq:fidelity}
\eeq
where $\hat{\rho}_{\text{out}}$ refers to the output density matrix and $U_{\text{SWAP}}$ denotes the SWAP operation.  The two qubits are initialized in $\ket{\psi_{\text{in}}}=\ket{\psi_{1}}\otimes\ket{\psi_{2}}$, where $\ket{\psi_{i}}=a_{i}\ket{\uparrow}+b_{i}\ket{\downarrow}$ with $i=1,2$.  Note that the gate fidelity $F$ defined here is input-state-dependent.  Furthermore, it can be easily generalized to the cases where the input state is represented by a density matrix instead of a pure state.  In the current study, we only investigate a few representative examples instead of giving a comprehensive discussion.  However, the results we present should already paint a clear picture on how hyperfine interaction affects the exchange gate fidelity in general.

Here, we focus on three types of input states: (1) a fully polarized state, for example $\ket{\psi_{\text{in}}}=\ket{\uparrow\uparrow}$; (2) an initial state in the $S$-$T_0$ subspace, for example, $\ket{\psi_{\text{in}}}=\ket{\uparrow\downarrow}$; (3) a general initial state that contains both the polarized and unpolarized two-spin eigenstates.

We start with a fully-polarized pure input state, $\ket{\psi_{\text{in}}}=\ket{T_{+}}$ or $\ket{T_{-}}$.  For either of them, the gate fidelity always stays at $1$, since $\HH_{\text{hf}}$ does not mix two spin-up (spin-down) states, and the exchange gate does not change the state at all.  If the input is a superposition of the two polarized triplet states, such as $(\ket{T_{+}} + \ket{T_{-}})/\sqrt{2}$, the exchange gate again does not change either of the triplet states, though the two states do dephase relative to each other because of the $\delta \mu$ term in $\HH_{T_\pm}$, irrespective of whether the exchange gate is turned on or off.

\begin{figure}
\includegraphics[width=3.0in,height=2.0in]{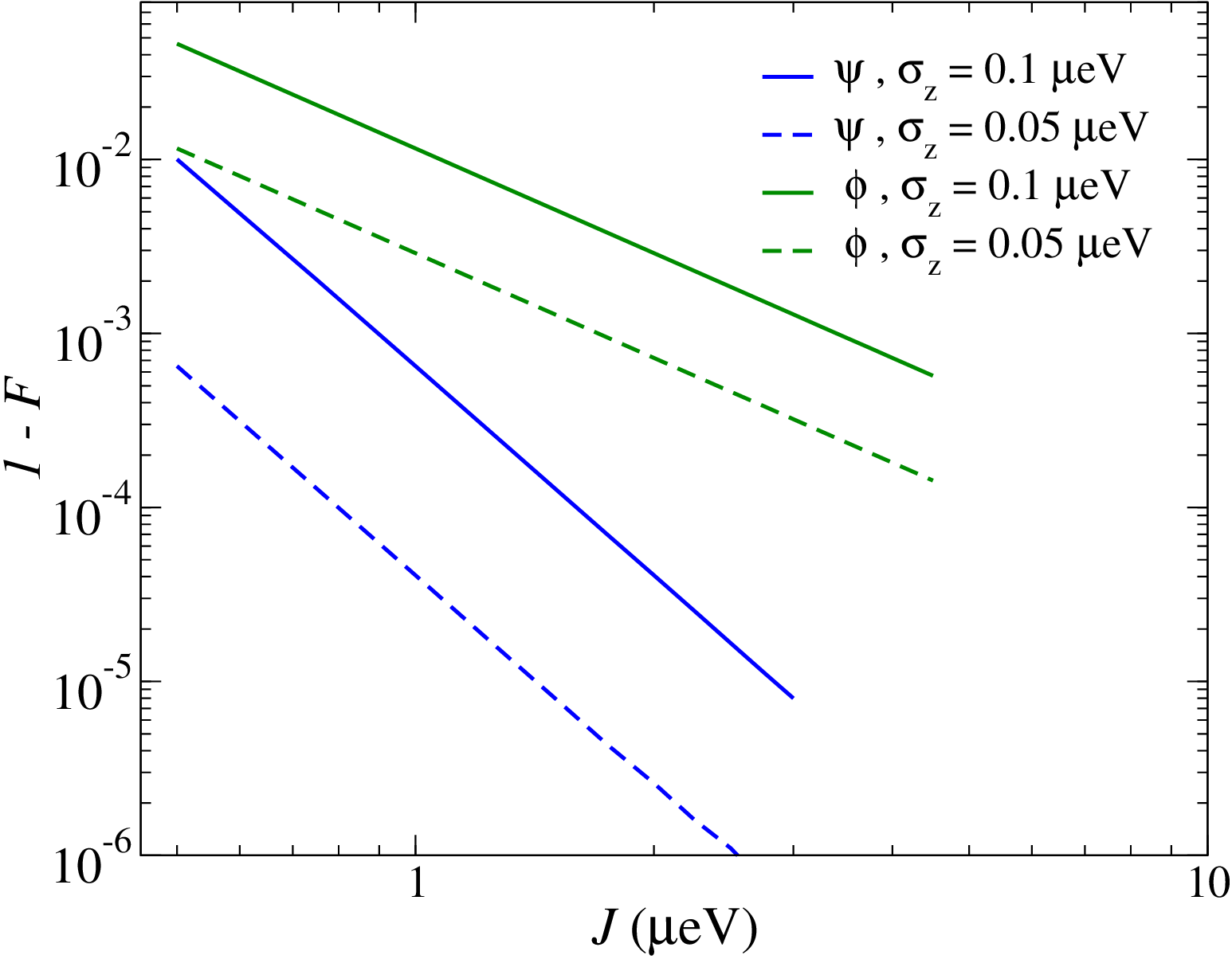}	
\caption{(Color online) The SWAP gate error in GaAs at $B=200$ mT for two input states: (1) $\ket{\psi_{\text{in}}}=\ket{\psi}=\ket{\uparrow\downarrow}$ (blue lines) ; (2) $\ket{\psi_{\text{in}}} = \ket{\phi} = \frac{1}{\sqrt{2}} \left(\ket{\uparrow} + \ket{\downarrow} \right) \otimes \frac{1}{\sqrt{2}} \left(\ket{\uparrow} - \ket{\downarrow} \right)$ (green lines). The blue
lines have steeper slopes than the green ones. Solid lines show the gate errors with the Overhauser fluctuations $\sigma_{z}\approx 0.1$ $\mu$eV, while dashed lines are with $\sigma_{z} \approx 0.05$ $\mu$eV. Here $\sigma_z = n_F' \sigma_\perp$, as discussed in the main text below Eq.~(\ref{eq:T_mu}).}\label{fig:swap_gaas}
\end{figure}

When we have an input state in the $S$-$T_0$ subspace, it can be expressed as a linear combination of $\ket{S}$ and $\ket{T_{0}}$ with appropriate coefficients.  For $\ket{\psi_{\text{in}}}=\ket{\uparrow\downarrow} = (|S\rangle + |T_0\rangle)/\sqrt{2}$, the gate fidelity $F_1$ takes the form
\beq
	F_{1} = \frac{1}{2} + \frac{1}{2}\text{Re}\, W_{ST_{0}}(t)\,\,, \label{eq:F1}
\eeq
which also applies to the case of $\ket{\psi_{\text{in}}}=\ket{\downarrow\uparrow}$.  We have studied the decoherence function $W_{ST_{0}}(t)$ in detail in Sec.~\ref{sec:W_calculation_0}, which decays due to the quasistatic longitudinal Overhauser field $\HH_{A}$ and due to terms proportional to transverse Overhauser fields, $\hat{V}_{SS}$ and $\hat{V}_{\delta\theta}$.  When the nuclear bath is not narrowed, $\HH_{A}=-\Th^{2}/J$ is the leading dephasing mechanism for smaller $J$ (cf. Figs.~\ref{fig:T_GaAs} and \ref{fig:T_Si} for GaAs and Si, respectively).
The resulting gate fidelity can be approximated as (for simplicity, we assume a square pulse of $J$ with a SWAP gate time of $T_g = \pi/J$)
\begin{eqnarray}
	F_{1} & \approx & \frac{1}{2} + \frac{1}{2} \text{Re}\, W_{A}(T_g)
           \approx  1 - \frac{3\pi^2 \sigma_z^4}{16J^4} \,,
\end{eqnarray}
and the gate error is
\beq
1-F_1 = \frac{3\pi^2 \sigma_z^4}{16J^4} = \frac{3\pi^2 n_F^4 \sigma_\perp^4}{4J^4} \,.
\eeq
There are two interesting features here.  First, the gate error is inversely proportional to $J^4$.  Thus the larger the $J$, the smaller the gate error (although the charge noise induced error would more likely be the dominant factor for larger $J$ values, and $J$ should be smaller than $J_{\rm max} \sim 0.64 \mu_{\rm T}$ in order for $\hat{H}_{A}$-induced dephasing to dominate over $\hat{V}_{SS}$-induced one, as shown in Sec.~\ref{sec:dominant}).  Second, narrowing of the nuclear bath (i.e., reducing $n_F$) can help the gate fidelity quite dramatically $1-F_1 \propto n_F^4$.  In Fig.~\ref{fig:swap_gaas}, we plot the SWAP gate error $1 - F_1$ with the input state $\ket{\psi_{\text{in}}}=\ket{\uparrow\downarrow}$ as a function of the exchange splitting $J$.  Both the strong $J$ dependence and the strong $n_F$ dependence are clearly illustrated.

In more general cases, $\ket{\psi_{\text{in}}}$ contains both polarized and unpolarized two-spin states (and it does not have to be a product state either).  Dephasing between the polarized triplet states and the unpolarized states becomes relevant.  Different from the $S_z = 0$ subspace, where $\HH_A$ is the leading cause for dephasing, here it is the average Overhauser field $\delta \mu$ that leads to inhomogeneous broadening and dominates dephasing. In GaAs  this inhomogeneous broadening leads to a $T_2^*$ time in the order of 10 ns, which trumps all other dephasing mechanisms due to higher-order terms in the effective Hamiltonian from Eqs~(\ref{eq:HS})--(\ref{eq:HT-}) .
For example, for an initial state $|\psi_{\rm in}\rangle = \frac{1}{\sqrt{2}} |\uparrow + \downarrow\rangle \otimes \frac{1}{\sqrt{2}} |\uparrow - \downarrow\rangle$, using the classical averaging procedure we employed in previous sections we obtain the SWAP gate fidelity as
\bea
	F_{2} & \approx & \frac{3}{8} +  \frac{ e^{-T_g^{2}/T^{2}_{\mu}} }{2} + \frac{ e^{-4T_g^{2}/T^{2}_{\mu}} }{8}
           \approx  1 - \frac{T_g^2}{T_\mu^2}
\eea
where $T_{\mu}$ is the characteristic time due to the inhomogeneous broadening effect from $\delta \hat{\mu}$,
\beq
	T_{\mu} = \frac{\sqrt{2}}{\sqrt{(\sigma^{2}_{z,L}+\sigma^{2}_{z,R})}} = \frac{\sqrt{2}}{n'_F \sigma_\perp} \ \,\,.
\label{eq:T_mu}
\eeq
Here we have assumed that $\sqrt{\sigma_{z,L}^2 + \sigma_{z,R}^2} = n'_F \sigma_\perp$, i.e.,~we include the possibility of the narrowing of the Overhauser field distribution in each of the dots separately.  Note that this is distinct from the previously considered case of narrowing of Overhauser field {\em difference} between the two dots, leading to our introduction of a narrowing factor $n'_{F}$ instead of $n_{F}$ used before.
Narrowing of this difference (i.e.,~decreasing $\sigma_{z}$, the standard deviation of the field difference, from its natural value by a factor of $n_{F}$) can occur when no narrowing of the field distribution in each of the dots is present --- only a correlation between the $z$ components of the Overhauser field in L and R dots is needed. Keeping this in mind, we see  that the gate error takes a simple form of
\beq
1-F_2 = \frac{T_g^2}{T_\mu^2} \approx \frac{\pi^2 (n'_F)^2 \sigma_\perp^2}{J^2} \,.
\eeq
Since $\sigma_\perp$ is on the order of 0.1 $\mu$eV for GaAs, the exchange coupling $J$ needs to be much larger than 1 $\mu$eV for the gate error to be manageable in the context of quantum information processing.  In Fig.~\ref{fig:swap_gaas}, we show our results of $1-F_2$ as a function of $J$ for the input state given above.  Comparing to the curves for the initial state in the $S$-$T_0$ subspace, the gate error here has a larger magnitude and a weaker $J$ dependence.  This difference comes from the fact that $1-F_1$ is dominated by $\HH_A = \delta \theta^2/J$, while $1-F_2$ is dominated by $\delta \mu$ in $H_{T_\pm}$.

In short, while high-fidelity exchange gate can be obtained when an input state is in the $S$-$T_0$ subspace, gate error is generally larger for an input state containing polarized states, dominated by the inhomogeneous broadening from the quasistatic nuclear spin reservoir.  Since the later type of input states is the norm rather than the exception, the upper two curves in Fig.~\ref{fig:swap_gaas} are more representative of the qualitative behavior of a general input state.  Furthermore, the numerical coefficient for the gate error $1-F$ is input-state-dependent: it is determined by the weight of $|T_{\pm}\rangle$ within the input state.  Finally, it is also important to point out that our consideration in this section assumes the simplest implementation of the SWAP gate.  As pointed out in Refs.~\onlinecite{Wang_NatComm12, Kestner_PRL13}, pulse shaping and other means can be employed to reduce the sensitivity of an exchange gate to the Overhauser field fluctuations.

%%%%%%%%%%%%%%%%
%%% SUMMARY
%%%%%%%%%%%%%%%%
\section{Summary and conclusions}
With the aim of helping to understand experiments related to $S$-$T_{0}$ qubits and single-spin qubits, we have performed a theoretical study of hyperfine-interaction-induced dephasing for two electron spins in a double quantum dot at finite $S$-$T_{0}$ splitting $J$.  For the $S$-$T_0$ qubit, we evaluate various hf-induced dephasing terms between the singlet $S$ and the unpolarized triplet state $T_{0}$, and identify the dominant dephasing channels.  Specifically, we find that in the absence of a magnetic field gradient across the DQD, inhomogeneous broadening for an $S$-$T_0$ qubit is significantly suppressed relative to a single spin qubit, with the most significant source of inhomogeneous broadening being the $\HH_A$ term, proportional to the square of the fluctuations of the longitudinal Overhauser field difference between the dots.  Another important decoherence channel is the $\hat{V}_{SS}$ term, which originates from dressing of the singlet state by the polarized triplets due to transverse Overhauser field.  Throughout a wide range of parameters, these two terms compete for the role of the dominant source of decoherence, leading to a predicted non-monotonic dependence of the $T_{2}$ dephasing time on $J$.

The physical picture changes when there is a finite magnetic field gradient across the double dot.  The gradient dictates that the two-spin eigenstates are now superpositions of the singlet and unpolarized triplet states, and there is a finite electron spin polarization in each of the two quantum dots.  Random longitudinal Overhauser field difference can now cause dephasing between the two-spin eigenstates, in analogy to inhomogeneous broadening in a single spin qubit.  In other words, the $S$-$T_0$ encoding acquires more of a single-spin-qubit decoherence characteristics as the singlet-triplet mixing increases. This increased hf-induced dephasing is the price that one has to pay for having two-axis control (with the $x$ axis rotations provided by the field gradient) over the $S$-$T_0$ qubit.

We have also considered the Hahn echo experiment in which the superposition of $S$ and $T_{0}$ states evolves at finite $J$, while being subject to a $\pi$-pulse (generated by a built-in gradient of the effective magnetic field between the dots, which drives the rotation when $J$ is suppressed for a chosen time) in the middle of the evolution period (for a recent experiment see Ref.~\onlinecite{Dial_PRL13}). The dynamics of the echo signal is due to the transverse Overhauser field terms, which lead to appearance of oscillations (closely related to the ones predicted\cite{Cywinski_PRL09,Cywinski_PRB09} and observed\cite{Bluhm_NP10} in the $J \! \approx \! 0$ regime), the amplitude of which increases as the applied magnetic field is lowered. The observation of such oscillations in a Hahn echo experiment in the large $J$ regime would be a clear signal of achieving a strong suppression of the charge noise (causing fluctuations of $J$) in the system. A nonmonotonic dependence of the characteristic free-evolution decoherence time $T_{2}$ on $J$ (with the maximal coherence time seen at $J_{\text{max}} \! \approx \! 0.64 \mu_{T}$) is another predicted experimental feature signifying the dominance of hyperfine mechanisms of decoherence in the system.

We have also studied hf-induced gate errors in an exchange-gate for two single-spin qubits, with a particular focus on the SWAP gate.
For a general input state, having a non-negligible component of polarized triplet states, inhomogeneous broadening from the longitudinal Overhauser field is the dominant source of gate errors.
By exploring several typical initial states, we are able to obtain the  dependence of the gate errors on $J$ and the standard deviation of the distribution of the Overhauser fields in the two dots. The obtained results show that for realistic multiqubit operations with single-spin qubits, the fluctuations of longitudinal Overhauser field remain a serious source of computation errors and control schemes focused on mitigating these errors by exploiting the quasistatic nature of the Overhauser field fluctuation, similar to the ones presented in Refs.~\onlinecite{Wang_NatComm12, Kestner_PRL13} for single qubit operations, should be seriously considered.  This necessitates a generalization of the currently proposed techniques for noise-resistant single qubit operations to the multiqubit case, which is still very much in its infancy.\cite{Hickman_arXiv13}

While our calculations are focused on two coupled spin qubits, the lessons we have learned should be useful for analysis of decoherence of multiple coupled qubits.  For example, our results on the $S$-$T_0$ qubit in the absence of field gradient indicate that electron interaction and symmetry can help suppress decoherence, in the same manner as decoherence-free subspace, even though contact hyperfine coupling is a completely local interaction.  On the other hand, if the symmetries are broken, whether by intrinsic or extrinsic inhomogeneities, the coherence properties of the overall interacting system is more similar to an ensemble of individual components.  Nevertheless, based on our results for $S$-$T_0$ decoherence, we may speculate that dephasing between two states of a coupled $n$-qubit system should generally be slower than the dephasing of two states of an uncoupled $n$-qubit system.

\section*{Acknowledgements}
We thank R.-B.~Liu, F.~Kuemmeth, and A.~Higginbotham for helpful discussions.
JH and XH acknowledge financial support by NSA/LPS through ARO, DARPA QuEST through AFOSR, and NSF. {\L}C acknowledges funding from the Homing Programme of the Foundation for Polish Science supported by the EEA Financial Mechanism.  SDS acknowledges financial support by LPS-CMTC and IARPA-MQCO.

%%%%%%%%%%%%%%%%
%%% Appendix
%%%%%%%%%%%%%%%%
\appendix

%%%%%%%%%%%%%%%%%%%%%%%%%%%%%%%%%%%%%%%%%%%%%
%%% EXP
%%%%%%%%%%%%%%%%%%%%%%%%%%%%%%%%%%%%%%%%%%%%%
\section{POSSIBLE WAYS TO MEASURE THE $S$-$T_{0}$ COHERENCE AT LARGE EXCHANGE SPLITTING $J$ IN GATED DQDS} \label{app:experiment}
With full control over the $S$-$T_0$ qubit available,\cite{Foletti_NP09} a measurement of coherence between $S$ and $T_{0}$ states can be performed at nonzero $J$ in the following way (which in fact has been recently realized in Ref.~\onlinecite{Dial_PRL13}). Using the $x$ rotations (enabled by a nanomagnet located close to the DQD\cite{Tokura_PRL06,Pioro_NP08,Brunner_PRL11,Petersen_PRL13} or a pre-established gradient of the nuclear polarization between the two dots\cite{Foletti_NP09,Bluhm_PRL10}), one can rotate the $S$ state initialized in a standard way\cite{Hanson_RMP07} into a superposition of $S$ and $T_{0}$ (such as $\ket{\pm Y} \! = \! (\ket{S}\pm i \ket{T_{0}})/\sqrt{2}$ states), and then set $J$ to be much larger than the $\theta_{\rm T}$ term.  At such a large $J$, the influence of the environment is expected to lead only to pure dephasing of the superposition of states --- the diagonal elements $\rho_{_{SS}}$ and $\rho_{_{T_{0}T_{0}}}$ of the reduced density matrix of the qubit are conserved, while the off-diagonal elements, $\rho_{_{ST_{0}}} \! = \! \rho^{*}_{_{T_{0}S}}$ can decay due to fluctuations of the energy splitting of the $S$ and $T_{0}$ states.
After letting the state freely evolve for time $t$, one can then measure the degree of coherence (i.e.,~$\rho_{_{ST_{0}}}$) by performing a state rotation which would bring the qubit back to the $S$ state \emph{provided that the actual state after time $t$ is the same as the initial one}. Subsequent measurement of the singlet return probability $P_{S}(t)$ is a measure of how much dephasing occurred in time $t$.

The coherence-measurement protocol can be formalized as follows, assuming a constant finite $\theta_{\rm T}$.  The qubit is first initialized into the singlet $S$ state.\cite{Petta_Science05}  At $t\! = \! 0$, a $(\pi/2)_x$ pulse is applied (by pulsing the DQD to the $J \ll \theta_{\rm T}$ regime and stay there for a period of time $\tau$, such that $\theta_{\rm T} \tau = \pi/4$), so that the qubit is rotated to the initial state of $|-Y\rangle$.  The DQD is then kept in the large-$J$ regime and evolve freely for a time period $t$.  After this time another $(3\pi/2)_{x}$ rotation is performed, and subsequently, the singlet probability $P_{S}(t)$ is read out by charge sensing using the Pauli spin blockade.\cite{Hanson_RMP07}  We assume that all the gate operations here take negligible time compared to the free-evolution time $t$. The singlet probability $P_{S}(t)$ is defined as
\beq
	P_{S}(t) = \text{Tr} \left [ \ket{S}\bra{S} \hat{\rho}(t) \right ] \,\, ,
\eeq
where the final density matrix is given by
\beq
	\hat{\rho}(t) = \hat{U}_{3\pi/2,x}e^{-i\HH t} \hat{\rho}_{Q}(0)\otimes \hat{\rho}_{I}(0)e^{i\HH t} \hat{U}^{\dagger}_{3\pi/2,x} \,\, ,
\eeq
with $\hat{\rho}_{Q}(0)$ and $\hat{\rho}_{I}(0)$ being the initial density matrices of the qubit and the nuclear bath, respectively, and $\hat{U}_{x,3\pi/2} \! = \! -(1+i\hat{\sigma}_{x})/\sqrt{2}$ being the unitary transformation corresponding to the final rotation of the qubit's state. After simple manipulations we obtain
\beq
	P_{S}(t) = \frac{1}{2} +\frac{1}{4}\tr_{I}\left ( \hat{\rho}_{I}(0) e^{i\hat{H}t} \hat{\sigma}_{y}e^{-i\hat{H}t}\hat{\sigma}_{y} \right ) \,\, .  \label{eq:PS_nonadiabatic}
\eeq
Since the DQD Hamiltonian is a pure dephasing one at large $J$, $\hat{H} \! = \! \HH_{S}\ket{S}\bra{S} + \HH_{T_{0}}\ket{T_{0}}\bra{T_{0}}$ (where $\HH_{S}$, $\HH_{T_{0}}$ are operators in the Hilbert space of the nuclear bath), we obtain
\beq
	P_{S}(t) = \frac{1}{2} +\frac{1}{2}\text{Re}\, W_{ST_{0}}(t) \,\, ,  \label{eq:PSW}
\eeq
where
\beq
	W^{ST_{0}}(t) \equiv  \frac{\rho^{Q}_{ST_{0}}(t)}{\rho^{Q}_{ST_{0}}(0)} = \tr_{I}\left ( \hat{\rho}_{I}(0) e^{i\HH_{T_{0}}t}e^{-i\hat{H}_{S}t} \right ) \,\, .
\eeq
In other words, the time evolution of the singlet probability $P_S(t)$ carries the complete information of the $S$-$T_0$ coherence $W_{ST_0}(t)$.

In the absence of a controlled $\theta_{\rm T}$ gradient field, it is still possible to create an initial state being a superposition of $\ket{S}$ and $\ket{T_{0}}$ by initializing the $S(0,2)$ state, and then \emph{adiabatically} lowering $J$, so that the state obtained at $J\! \ll \! \sigma_{z}$ in the $(1,1)$ charge regime is an eigenstate of the electron-nuclear-spin Hamiltonian.\cite{Higginbotham_arXiv13}  Depending on the sign of the expectation value of $\delta\hat{\theta}$, one of the $\ket{\pm X} \! = \! (\ket{S}\pm \ket{T_{0}})/\sqrt{2}$ states is created.  The $S$-$T_{0}$ splitting can then be rapidly increased to a finite value $J$, and the DQD system is allowed to evolve freely for time $t$.  After this evolution $J$ is again rapidly reduced to very low values, such that the state is projected onto the two-electron product state basis.  Subsequently the DQD is adiabatically swept to the large $J$ regime for the standard readout procedure via spin blockade. A calculation similar to the one given above gives us
\beq
P_{S}(t) = \frac{1}{2} + \frac{1}{4}\tr_{I}\left ( \hat{\rho}_{I}(0) \hat{\sigma}_{x} e^{i\hat{H}t} \hat{\sigma}_{x}e^{-i\hat{H}t} \right ) \,\, .		\label{eq:PS_adiabatic}
\eeq
If we now make the pure-dephasing approximation for the Hamiltonian, we arrive again at Eq.~(\ref{eq:PSW}).

%%%%%%%%%%%%%%%%%%%%%%%%%%%%%%%%%%%%%%%%%%%%%
%%% Derivation of hyperfine coupling Hamiltonian
%%%%%%%%%%%%%%%%%%%%%%%%%%%%%%%%%%%%%%%%%%%%%
\section{DERIVATION OF HYPERFINE COUPLING HAMILTONIAN} \label{app:Hhf}
To focus on the coupled electron-nuclear-spin dynamics, we project the hyperfine interaction onto the lowest-energy two-electron states and obtain Eq.~(\ref{eq:Hhf}) in the main text.  The basis states are the ground singlet and triplet states:
\bea
	&&\ket{S} =\nonumber \frac{\ket{\uparrow\downarrow}-\ket{\downarrow\uparrow}}{\sqrt{2}}
			\otimes\psi_{\text{S}}(\mathbf{r}_1,\mathbf{r}_2)\,, \\
	&&\ket{T_{0}} =\nonumber  \frac{\ket{\uparrow\downarrow}+\ket{\downarrow\uparrow}}{\sqrt{2}}
			\otimes\psi_{\text{AS}}(\mathbf{r}_1,\mathbf{r}_2)\,, \\
	&&\ket{T_{+}} =\nonumber  \ket{\uparrow\uparrow}
			\otimes\psi_{\text{AS}}(\mathbf{r}_1,\mathbf{r}_2)	\,, \\
	&&\ket{T_{-}} = \ket{\downarrow\downarrow}
			\otimes\psi_{\text{AS}}(\mathbf{r}_1,\mathbf{r}_2)	\,.
\eea
Here, $\psi_{\text{(A)S}}(\mathbf{r}_1,\mathbf{r}_2)$ denotes the (anti-)symmetric combination of the orbital envelopes $\Psi_{L}(\mathbf{r})$ and $\Psi_{R}(\mathbf{r})$,
\bea
	&&\psi_{\text{S}}(\mathbf{r}_1,\mathbf{r}_2)=\nonumber\frac{	[\Psi_{L}(\mathbf{r}_{1})\Psi_{R}(\mathbf{r}_{2})+\Psi_{R}(\mathbf{r}_{1})\Psi_{L}(\mathbf{r}_{2})]}{\sqrt{\nu_{0}}\sqrt{2(1+|\chi|^{2})}}
		\,,\\
			&& \\
	&&\psi_{\text{AS}}(\mathbf{r}_1,\mathbf{r}_2) = \nonumber \frac{[\Psi_{L}(\mathbf{r}_{1}) \Psi_{R}(\mathbf{r}_{2}) - \Psi_{R}(\mathbf{r}_{1}) \Psi_{L}(\mathbf{r}_{2})]}{\sqrt{\nu_{0}}\sqrt{2(1-|\chi|^{2})}} \,,\\
			&&
\eea
with the orbital overlap $\chi=\frac{1}{\nu_{0}}\int d\mathbf{r}\,{\Psi^{*}_{L}(\mathbf{r})} {\Psi_{R}(\mathbf{r})}\,$, where $\nu_{0}$ is the volume of the primitive unit cell (note that we use the normalization in which the envelope wavefunctions are dimensionless, and all the dependence of the hf interaction on the periodic parts of the Bloch functions is carried by $\mathcal{A}_{\alpha}$ constants below). Generally, $\chi \ll 1$, since the local orbitals decay rapidly away from the center of each dot.

To calculate the hf matrix element, we need to first evaluate the matrix elements of the $\delta$-functions.  Specifically,
\begin{widetext}
\bea
\bra{\psi_{\text{S}}}\MA_{\alpha[i]}\nu_{0}\delta(\mathbf{r}_{k}-\mathbf{R}_{i}) \ket{\psi_{\text{AS}}} & & \nonumber \\
& & \hspace*{-1.5in}= \int d\mathbf{r}_{1} \int d\mathbf{r}_{2} \frac{\MA_{\alpha[i]}\nu_{0}\delta(\mathbf{r}_{k}-\mathbf{R}_{i})}{2\sqrt{1-|\chi|^{4}}} [\Psi^{*}_{L}(\mathbf{r}_{1})\Psi^{*}_{R}(\mathbf{r}_{2}) + \Psi^{*}_{R}(\mathbf{r}_{1})\Psi^{*}_{L}(\mathbf{r}_{2})] \times  [\Psi_{L}(\mathbf{r}_{1})\Psi_{R}(\mathbf{r}_{2}) - \Psi_{R}(\mathbf{r}_{1})\Psi_{L}(\mathbf{r}_{2})] \nonumber \\
& & \hspace*{-1.5in}= \frac{(-1)^{k+1}}{2\sqrt{1-|\chi|^{4}}}\MA_{\alpha[i]} \left\{  |\Psi_{L}(\mathbf{R}_{i})|^{2} - |\Psi_{R}(\mathbf{R}_{i})|^{2} - 2\text{Im}\left[ \chi^{*}\Psi^{*}_{L}(\mathbf{R}_{i})\Psi_{R}(\mathbf{R}_{i})\right] \right\} \equiv  (-1)^{k+1}B_{i} (\chi) \;, k = 1,2 \,\,. \\
\bra{\psi_{\text{AS}}}\MA_{\alpha[i]}\nu_{0}\delta(\mathbf{r}_{k}-\mathbf{R}_{i}) \ket{\psi_{\text{AS}}} & \!\equiv\! & C_{i} (\chi) \!=\! \frac{1}{2(1-|\chi|^{2})} \MA_{\alpha[i]}\left\{  |\Psi_{L}(\mathbf{R}_{i})|^{2} \!+\! |\Psi_{R}(\mathbf{R}_{i})|^{2} \!-\! 2\text{Re}\left[ \chi^{*}\Psi^{*}_{L}(\mathbf{R}_{i}\Psi_{R}(\mathbf{R}_{i})\right] \right\}, \\
\bra{\psi_{\text{S}}}\MA_{\alpha[i]}\nu_{0}\delta(\mathbf{r}_{k}-\mathbf{R}_{i}) \ket{\psi_{\text{S}}} & \!\equiv\! & D_{i} (\chi) \!=\! \frac{1}{2(1+|\chi|^{2})} \MA_{\alpha[i]}\left\{ |\Psi_{L}(\mathbf{R}_{i})|^{2} \!+\! |\Psi_{R}(\mathbf{R}_{i})|^{2} \!+\! 2\text{Re}\left[ \chi^{*}\Psi^{*}_{L}(\mathbf{R}_{i}\Psi_{R}(\mathbf{R}_{i})\right] \right\}.
% \approx C_{i} (\chi) +\frac{4}{2(1+|\chi|^{2})} \MA_{\alpha[i]}\text{Re}\left[ \chi^{*}\Psi^{*}_{L}(\mathbf{R}_{i})\Psi_{R}(\mathbf{R}_{i})\right]\;.
\eea
\end{widetext}
In contrast to $B_{i} (\chi)$, $C_{i} (\chi)$ and $D_{i} (\chi)$ are both symmetric in $(\mathbf{r}_{1},\mathbf{r}_{2})$.

Writing $\mathbf{S}_{k}\cdot \mathbf{I}_{i} = S^{z}_{k}I^{z}_{i}+\frac{1}{2}(S^{+}_{k} I^{-}_{i}+S^{-}_{k}I^{+}_{i})$, we can now evaluate the hf matrix elements straightforwardly.  For example,
\bea
\bra{S} \HH_{\text{hf}} \ket{T_{0}}
 %	&=&\nonumber \bra{S}  \sum_{i} \MA_{\alpha[i]}\delta(\mathbf{r}_{1}-\mathbf{R}_{i})\, \left[ S^{z}_{1}I^{z}_{i}+\frac{1}{2}(S^{+}_{1}I^{-}_{i}+S^{-}_{1}I^{+}_{i}) \right] + \sum_{i}  \MA_{\alpha[i]}\delta(\mathbf{r}_{2}-\mathbf{R}_{i}) \,\left[ S^{z}_{2}I^{z}_{i}+\frac{1}{2}(S^{+}_{2}I^{-}_{i}+S^{-}_{2}I^{+}_{i}) \right] \ket{T_{0}} \\
	&=&\sum_{k=1,2}\nonumber \bra{\psi_{\text{S}}}\sum_{i}  \MA_{\alpha[i]}\nu_{0}\delta(\mathbf{r}_{k}-\mathbf{R}_{i})\ket{\psi_{\text{AS}}} \\
	&&\nonumber \times  \frac{\bra{\uparrow\downarrow}S^{z}_{k}I^{z}_{i}\ket{\uparrow\downarrow} - \bra{\downarrow\uparrow} S^{z}_{k}I^{z}_{i} \ket{\downarrow\uparrow}}{2} \\
	 %&=& \frac{1}{2} \left[ \sum_{i} B_{i}(\chi) I^{z}_{i} + \sum_{i} B_{i}(\chi) I^{z}_{i} \right] \\
	 &=& \sum_{i} B_{i}(\chi) I^{z}_{i} \,\,.	\nonumber	
\eea
The generalized form of the complete hf coupling matrix is then: 
\bea
 \HH_{\text{hf}}(\chi)&\!\!\!\!\!=\!\!\!\!\!&\!\! \sum_{i} \left( \begin{array}{cccc}
		\!\! 0 & B_{i}(\chi)I^{z}_{i} &\!  \frac{-B_{i}(\chi)}{\sqrt{2}}I^{+}_{i} &\!  \frac{B_{i}(\chi)}{\sqrt{2}}I^{-}_{i} \!\!\!\\
		\!\! B_{i}(\chi)I^{z}_{i} & 0 &\!  \frac{C_{i}(\chi)}{\sqrt{2}}I^{+}_{i}  &\!  \frac{C_{i}(\chi)}{\sqrt{2}} I^{-}_{i} \!\!\!\\
		\!\! \frac{-B_{i}(\chi)}{\sqrt{2}}I^{-}_{i} &\! \frac{C_{i}(\chi)}{\sqrt{2}}I^{-}_{i} &\!  C_{i}(\chi) I^{z}_{i} &\! 0 \!\!\!\\
		\!\! \frac{B_{i}(\chi)}{\sqrt{2}}I^{+}_{i} &\! \frac{C_{i}(\chi)}{\sqrt{2}}I^{+}_{i} &\! 0 &\! -C_{i}(\chi)I^{z}_{i}\!\!\!\end{array}\right). \nonumber \\ \label{eq:appendix_ignore_chi}
\eea

In Eq.~(\ref{eq:appendix_ignore_chi}), we can neglect terms containing $\chi$ and/or $\Psi_{L}(\mathbf{R}_{i}) \Psi^{*}_{R}(\mathbf{R}_{i})\,$.  This approximation is justified because the number of nuclei in the overlapping region is very small, and their interaction with the electron very weak (as they are located in the tail regions of both orbitals).  After some simple algebra we find, for instance,
\bea
		\bra{S}\HH_{\text{hf}}\ket{T_{0}} \approx \sum_{i} \frac{1}{2}(A^{L}_{i}-A^{R}_{i}) I^{z}_{i} = \sum_{i} B_{i} I^{z}_{i}\,.
\eea
The rest of the hf matrix elements can be obtained in a similar manner.  With a negligible $\chi\,$, the hf coupling matrix can be approximated as Eq.~(\ref{eq:Hhf}).

%%%%%%%%%%%%%%%%%%%%%%%%%%%%%%%%%%%%%%%%%%%%%
%%% Classical approach
%%%%%%%%%%%%%%%%%%%%%%%%%%%%%%%%%%%%%%%%%%%%%
\section{SEMICLASSICAL APPROACH TO THE OVERHAUSER FIELDS AND THE DEPHASING INDUCED BY AVERAGING OVER THEM} \label{app:classical}

The slow dynamics and the large numbers of nuclear spins in a quantum dot makes a semiclassical description of their average effect possible.  Indeed, it has been pointed out\cite{Neder_PRB11} that at the short-time limit, before the electron-mediated nuclear dynamics causes any significant effects on the nuclear spin dynamics, a semiclassical description that account for only the Larmor precession of the nuclear spins is sufficient to explain most experimental observations.  Here we give a brief summary of the semiclassical picture of the Overhauser fields, and
estimate the magnitude of various hf-related terms appearing in our effective Hamiltonians.

We write the Overhauser operator as $\hat{\mathbf h}_{L/R} = \sum_{i}A^{L/R}_{i} \hat{\mathbf I}_{i}$ and treat its mean field average as a classical field $\mathbf{h}_{L/R}$.  The transverse Overhauser operator is given by $h^{\pm}_{L/R} = h^{x}_{L/R} \pm i h^{y}_{L/R}$.  In the classical limit, neglecting the possibility of a finite bath polarization, $h^{+}_{L/R}h^{-}_{L/R} \approx (h^{x}_{L/R})^2 + (h^{y}_{L/R})^2$.  The classical expressions for the hf-related terms in the effective Hamiltonian are then:
\bea
	\theta &=&  \frac{1}{2}(h^{z}_{L}-h^{z}_{R})  \,\,, \label{eq:theta_classical} \\
	V_{SS} &=& \frac{1}{4}\frac{J}{\EZT^{2}-J^{2}} \left ( \mathbf{h}_{L}^{\perp} - \mathbf{h}_{R}^{\perp} \right )^2 = v_{ss} \left ( \mathbf{h}_{L}^{\perp} - \mathbf{h}_{R}^{\perp} \right )^2 \,,\label{eq:VSS_classical}\\
	V_{S,T_{0}} &=&\nonumber -\frac{1}{8}(\frac{1}{\EZT}+\frac{\EZT}{\EZT^{2}-J^{2}}) \left [ (\mathbf{h}_{L}^{\perp})^2 - (\mathbf{h}_{R}^{\perp})^2 \right ] \\
			&& -\frac{i}{4}\frac{J}{\EZT^{2}-J^{2}}\, \mathbf{z}\cdot\left( \mathbf{h}^{\perp}_{R}\times\mathbf{h}^{\perp}_{L} \right ) \,\,, \label{eq:VST_classical}\\
	H_{A} &=& -\frac{1}{4J}(h^{z}_{L}-h^{z}_{R})^2\,\,\label{eq:HA_classical} \,\,, \\
	V_{\delta\theta} &=&\nonumber \frac{1}{8J}(\frac{1}{\EZT}+\frac{\EZT}{\EZT^{2}-J^{2}})(h^{z}_{L}-h^{z}_{R})
			\left[ (\mathbf{h}_{L}^{\perp})^2 - (\mathbf{h}_{R}^{\perp})^2 \right ] \,\, \nonumber\\
			& = & v_{\theta} \Delta h^{z} 	\left[ (\mathbf{h}_{L}^{\perp})^2 - (\mathbf{h}_{R}^{\perp})^2 \right ] \,\, ,
				\label{eq:Vtheta_classical}
\eea
Replacing $(h^{z}_{L}-h^{z}_{R})/2$ by $\sigma_{z}$, and $|\mathbf{h}^{\perp}|$ by $\sigma_{\perp}$, we can estimate the typical magnitude of these terms.

The semiclassical calculation of dephasing due to the above terms amounts to Gaussian averages [with respect to distribution from Eq.~(\ref{eq:PO})] of the phases corresponding to various terms given above. For example,  the $S$-$T_{0}$ dephasing function due to $V_{SS}$ is given by
\bea
W^{\text{cl}}_{SS}(t) & = & \int \frac{d^{2}h^{\perp}_{L}}{2\pi \sigma_{\perp,L}^2} e^{-\frac{(h^{\perp}_{L})^2}{2\sigma^{2}_{\perp,L}}} \int \frac{d^{2}h^{\perp}_{R}}{2\pi \sigma_{\perp,R}^2} e^{-\frac{(h^{\perp}_{R})^2}{2\sigma^{2}_{\perp,R}}} \nonumber\\
& & \times \exp \left[ - iv_{ss} t (\mathbf{h}_{L}^{\perp} - \mathbf{h}_{R}^{\perp} )^2 \right ] \,\, .
\eea
Keeping in mind that $\sigma^{2}_{\perp} \! = \! \sigma^{2}_{\perp,L}+\sigma^{2}_{\perp,R}$, we calculate the Gaussian integrals and obtain the expression for $W^{\text{cl}}_{SS}(t)$, which turns out to be equal to the one given in Eq.~(\ref{eq:WSS_short}).  In other words, the semiclassical calculation of dephasing due to $\hat{V}_{SS}$ is equivalent to neglecting the Larmor precession of nuclear spins in the RDT calculation given in the main text.

Similarly, the dephasing function due to $V_{\delta\theta}$ is given by
\bea
W^{\text{cl}}_{\delta\theta}(t) & = & \int \frac{dh^{z}_{L}}{\sqrt{2\pi}\sigma_{z,L}} e^{-\frac{(h^{z}_{L})^2}{2(\sigma_{z,L})^2}} \int \frac{dh^{z}_{R}}{\sqrt{2\pi}\sigma_{z,R}} e^{-\frac{(h^{z}_{R})^2}{2(\sigma_{z,R})^2}} \nonumber\\
& & \times W_{L}(h^{z}_{L},h^{z}_{R}) W_{R}(h^{z}_{L},h^{z}_{R})
\eea
with
\bea
W_{L}(h^{z}_{L},h^{z}_{R}) & = & \int \frac{dh^{x}_{L}dh^{y}_{L}}{2\pi \sigma^{2}_{\perp,L}} e^{-(h^{x}_{L})^2/2\sigma^{2}_{\perp,L}} e^{-(h^{y}_{L})^2/2\sigma^{2}_{\perp,L}} \nonumber\\
& & \times e^{-2iv_{\theta}\Delta h^{z} ( (h^{x}_{L})^2+(h^{y}_{L})^2 )}  \,\,.
\eea
The expression for $W_{R}$ is analogous, only with the sign in front of $v_{\theta}$ inverted.  After carrying out the integrals in $W_{L}$ and $W_{R}$, we obtain
\beq
W_{L/R} (h^{z}_{L},h^{z}_{R}) =  \frac{1}{1\pm 4iv_{\theta}\Delta h^{z} \sigma^{2}_{L/R}t} \,\, .
\eeq
The decoherence function for a fixed value of $\Delta h^{z} \! = \! 2\theta$ can be obtained by multiplying the two expressions above.  Since the coherence decay is faster when $\Delta h^{z}$ is larger, we find the lower bound for the decay time by replacing $\Delta h^{z}$ with $2\sigma_{z}$. Assuming dots of equal sizes, we obtain the expression for $W^{\text{cl}}_{\delta\theta}(t)$ that is identical to the short-time (or single-isotope) result from Eq.~(\ref{eq:Wthetashort}).

In both calculations here, we have neglected any nontrivial effects that the bath polarization can have on the free induction decay signal.  These effects, discussed in Ref.~\onlinecite{Barnes_PRL12}, cannot be captured by the semiclassical approximation. However, at small nuclear polarizations considered in this paper, they amount to a rather small quantitative change in decay time scale,\cite{Barnes_PRL12} which justifies our approximation of neglecting them.

%%% WAVEFUNCTION-SHAPE DEPENDENT CALCULATIONS
\section{THD DISTRIBUTION OF HYPERFINE COUPLINGS DEPENDENT ON THE WAVE-FUNCTION SHAPE} \label{app:rho}

In the study of coupled electron-nuclear spin dynamics, we often need to evaluate moments of the hyperfine interaction.  Here we discuss the wave-function dependence of these moments.  For example, the $n$th moment can be written as a sum over the nuclear index $i$ as
\begin{eqnarray}
\sum_{i} A^{n}_{i} & = & \sum_{\alpha}n_{\alpha} \mathcal{A}^{n}_{\alpha} \sum_{u} |\Psi(\mathbf{r}_{u})|^{2n} \,\, , \nonumber \\
& = & \sum_{\alpha}n_{\alpha} \mathcal{A}^{n}_{\alpha} \int \xi^{n} \rho(\xi) \text{d}\xi \,\, ,
\end{eqnarray}
where the index $u$ labels the Wigner-Seitz (WS) unit cells, and $\nu_{0}$ is the volume of the WS unit cell.  The envelope wavefunction is normalized as $\int |\Psi(\mathbf{r})|^2 \! = \! \nu_{0}$, and the function $\rho(\xi)$ parametrizes the ``density of states'' of hf couplings\cite{Cywinski_PRB09}
\beq
\rho(\xi) = \frac{1}{\nu_{0}} \int \delta \left (\xi-|\Psi(\mathbf{r})|^2 \right ) \text{d}^{3}r \,\, .
\eeq
As an example, if we assume that the envelope wavefunction is a two-dimensional Gaussian:
\beq
\Psi(\mathbf{r}) = \frac{\sqrt{\nu_{0}}}{\sqrt{\pi a} L} e^{-(x^{2}+y^{2})/2L^2} \Theta(a/2- |z| ) \,\, ,
\eeq
we have
\beq
\rho(\xi) = \frac{N_{\Psi}}{2\xi} \Theta\left( \frac{2}{N_{\Psi}} - \xi \right ) \,\, , \label{eq:rho2D}
\eeq
where
\beq
N_{\Psi} \equiv \frac{1}{\sum_{u} |\Psi(\mathbf{r}_{u})|^4} =  \frac{2\pi a L^2}{\nu_{0}} \,\, .
\eeq
As an example application, let us use the above $\rho(\xi)$ to calculate the quantity appearing in Sec.~\ref{sec:HBC}:
\beq
\sum_{u} |\Psi(\mathbf{r}_{u})|^8 = \int \xi^{4}\rho(\xi) \text{d}\xi = \frac{2}{N^{3}_{\Psi}} \,\, .
\eeq
The prefactor 2 here depends on the exact shape of the wavefunction, but the $\sim \! N_{\Psi}^{-3}$ scaling is a general result.

%%%%%%%%%%%%%%%%%%%%%%%%%%%%%%%%%%%%%%%%%%%%%
%%% Effective pure dephasing H beyond the S-T0 subspace
%%%%%%%%%%%%%%%%%%%%%%%%%%%%%%%%%%%%%%%%%%%%%
\section{EFFECTIVE DEPHASING HAMILTONIANS BEYOND THE $S$-$T_0$ SUBSPACE} \label{app:other_Heff}

For single-spin qubits, exchange interaction is used to perform two-qubit gates and to transport spin states.\cite{Loss_PRA98}  In these operations, generally, all four two-spin states are involved.  For simplicity, we assume there is no intentional field gradient between the neighboring quantum dots ($\theta_{\rm T} = 0$).  The presence of a field gradient would complicate gates such as a simple SWAP.  Under these conditions, the effective dephasing Hamiltonians in the two-spin Hilbert space can be written as
\bea
\tilde{H}_{ST} & = & \hat{H}_{S}\ket{S}\bra{S}+\hat{H}_{T_{+}}\ket{T_{+}}\bra{T_{+}} \nonumber \\
& & + \hat{H}_{T_{0}}\ket{T_{0}}\bra{T_{0}} + \hat{H}_{T_{-}}\ket{T_{-}}\bra{T_{-}} \,,
\eea
where
\bea
	\hat{H}_{S} &=&\nonumber -J + \HH_{A} - \frac{1}{2(\EZT+J)}\sum_{i,j} B_{i}B_{j} I^{-}_{i} I^{+}_{j} \\
			&&+\, \frac{1}{2(\EZT-J)}\sum_{i,j} B_{i}B_{j} I^{+}_{i} I^{-}_{j} \nonumber \\
			& = & -J + \HH_{A} + \hat{V}_{SS} + \hat{H}_{B} \,\,,\label{eq:HS}\\
	\hat{H}_{T_{+}} &=&\nonumber -\EZT + \delta \hat{\mu} - \frac{1}{2\EZT}\sum_{i,j} C_{i}C_{j} I^{-}_{i} I^{+}_{j} \\
			&&-\, \frac{1}{2(\EZT-J)}\sum_{i,j} B_{i}B_{j} I^{+}_{i} I^{-}_{j} \label{eq:HT+}\\
	\hat{H}_{T_{0}} &=& -\HH_{A} - \frac{1}{2\EZT}\sum_{i,j} C_{i}C_{j} I^{-}_{i} I^{+}_{j} \nonumber \\
			& & + \frac{1}{2\EZT}\sum_{i,j} C_{i}C_{j} I^{+}_{i} I^{-}_{j} \nonumber \\
			& = & -\HH_{A} + \hat{H}_{C}\,\,, \label{eq:HT0}\\
	\hat{H}_{T_{-}} &=&\nonumber \EZT -\delta \hat{\mu} + \frac{1}{2(\EZT+J)}\sum_{i,j} B_{i}B_{j} I^{+}_{i} I^{-}_{j} \\
			&&+\, \frac{1}{2\EZT}\sum_{i,j} C_{i}C_{j} I^{-}_{i} I^{+}_{j} \,\,,\label{eq:HT-} \\
    \HH_{A} & = & -\frac{\hat{\theta}^2}{J}\,\,.
\eea
This is the Hamiltonian we use to calculate the exchange gate errors due to hyperfine interaction with nuclear spins in Sec.~\ref{sec:gate}.

\end{document}